\documentstyle{article}
\begin{document}
\title{Addendum to ``The Dirac-Coulomb Sturmians and the series
expansion of the Dirac-Coulomb Green function: application to the
relativistic polarizability of the hydrogen-like atom'' \\
{[J.~Phys.~B:~At.~Mol.~Opt.~Phys.\/ {\bf 30} (1997) 825--61, \\
(E) {\bf 30} (1997) 2747]}}
\author{Rados{\l}aw Szmytkowski \\*[8mm]
Atomic Physics Division \\
Faculty of Applied Physics and Mathematics \\
Technical University of Gda{\'n}sk \\
ul.\ Gabriela Narutowicza 11/12, PL 80-952 Gda{\'n}sk, Poland \\
E-mail: radek@mif.pg.gda.pl}
\date{19th February 1999}
\maketitle
\begin{abstract}
Closure relations satisfied by the radial Dirac-Coulomb Sturmians are
proved analytically. The Sturmian expansion of the Dirac-Coulomb
Green function is transformed to the form containing only series with
summations running over non-negative indices. The main paper was
published in {\em J.~Phys.~B:~At.~Mol.~Opt.~Phys.\/} {\bf 30} (1997)
825--61 [Erratum: {\bf 30} (1997) 2747], see also {\em
J.~Phys.~A:~Math.~Gen.\/} {\bf 31} (1998) 4963--90 [Erratum: {\bf 31}
(1998) 7415--6]. \\*[3mm]
\end{abstract}
\newpage
\section{\rm Proofs of the closure relations}
In this section it is our goal to prove correctness of the following
closure relations satisfied by the Dirac-Coulomb Sturmians (equations
(91) and (26) of Ref.~\cite{Szmy97})
\begin{equation}
\frac{\alpha^{-1}}{2}\sum_{n=-\infty}^{\infty}
\left(
\begin{array}{c}
\varepsilon S_{n\kappa}(x) \\
\varepsilon^{-1} T_{n\kappa}(x)
\end{array}
\right)
\left(
\begin{array}{cc}
S_{n\kappa}(x') & T_{n\kappa}(x')
\end{array}
\right)
=\delta(x-x') 
\left(
\begin{array}{cc}
1 & 0 \\
0 & 1
\end{array}
\right)
\label{1}
\end{equation}
and
\begin{equation}
\frac{Z}{x}\sum_{n=-\infty}^{\infty}
\left(
\begin{array}{c}
S_{n\kappa}(x) \\
-\mu_{n\kappa}^{-1}T_{n\kappa}(x)
\end{array}
\right)
\left(
\begin{array}{cc}
\mu_{n\kappa}S_{n\kappa}(x') & T_{n\kappa}(x')
\end{array}
\right)
=\delta(x-x')
\left(
\begin{array}{cc}
1 & 0 \\
0 & 1
\end{array}
\right).
\label{2}
\end{equation}
In the proofs we shall use the following forms of the upper and lower
components of the radial Sturmians (equations (60) and (61) of
Ref.~\cite{Szmy97})
\begin{equation}
S_{n\kappa}(x)=\sqrt{\frac{\alpha(|n|+2\gamma_{\kappa})|n|!}
{2\varepsilon N_{n\kappa}(N_{n\kappa}\mp\kappa)
\Gamma(|n|+2\gamma_{\kappa})}}\:
x^{\gamma_{\kappa}}{\rm e}^{-x/2}
\left[L_{|n|-1}^{(2\gamma_{\kappa})}(x)
+\frac{\kappa\mp N_{n\kappa}}{|n|+2\gamma_{\kappa}}
L_{|n|}^{(2\gamma_{\kappa})}(x)\right],
\label{3}
\end{equation}
\begin{equation}
T_{n\kappa}(x)=
\sqrt{\frac{\alpha\varepsilon(|n|+2\gamma_{\kappa})|n|!}
{2N_{n\kappa}(N_{n\kappa}\mp\kappa)\Gamma(|n|+2\gamma_{\kappa})}}\:
x^{\gamma_{\kappa}}{\rm e}^{-x/2}
\left[L_{|n|-1}^{(2\gamma_{\kappa})}(x)-
\frac{\kappa\mp N_{n\kappa}}{|n|+2\gamma_{\kappa}}
L_{|n|}^{(2\gamma_{\kappa})}(x)\right]
\label{4}
\end{equation}
(the upper signs should be taken for $n>0$ and the lower signs for
$n<0$; for $n=0$ one takes the upper sign if $\kappa<0$ and the lower
one if $\kappa>0$), the expression for the eigenvalue $\mu_{n\kappa}$
associated with the Sturmian $\left(\begin{array}{cc}S_{n\kappa}(x)&
T_{n\kappa}(x)\end{array}\right)^{\top}$ (equation (52) of
Ref.~\cite{Szmy97}) 
\begin{equation}
\mu_{n\kappa}=\varepsilon\zeta(|n|+\gamma_{\kappa}\pm N_{n\kappa})
\label{5}
\end{equation}
and the well-known closure relation for the generalized Laguerre
polynomials 
\begin{equation}
x^{\gamma_{\kappa}}x'^{\gamma_{\kappa}}{\rm e}^{-(x+x')/2}
\sum_{n=0}^{\infty}\frac{n!}{\Gamma(n+2\gamma_{\kappa}+1)}
L_{n}^{(2\gamma_{\kappa})}(x)L_{n}^{(2\gamma_{\kappa})}(x')
=\delta(x-x').
\label{6}
\end{equation}

Consider equation (\ref{1}). It is equivalent to four relations
\begin{equation}
\frac{\alpha^{-1}\varepsilon}{2}\sum_{n=-\infty}^{\infty}
S_{n\kappa}(x)S_{n\kappa}(x')=\delta(x-x'),
\label{7}
\end{equation}
\begin{equation}
\frac{\alpha^{-1}\varepsilon^{-1}}{2}\sum_{n=-\infty}^{\infty}
T_{n\kappa}(x)T_{n\kappa}(x')=\delta(x-x'),
\label{8}
\end{equation}
\begin{equation}
\frac{\alpha^{-1}\varepsilon}{2}\sum_{n=-\infty}^{\infty}
S_{n\kappa}(x)T_{n\kappa}(x')=0
\label{9}
\end{equation}
and
\begin{equation}
\frac{\alpha^{-1}\varepsilon^{-1}}{2}\sum_{n=-\infty}^{\infty}
T_{n\kappa}(x)S_{n\kappa}(x')=0.
\label{10}
\end{equation}
It is to be noticed that it is sufficient to prove relations
(\ref{7})--(\ref{9}) since equation (\ref{10}) is an immediate
consequence of equation (\ref{9}). We start with the relation
(\ref{7}). Utilizing equations (\ref{3})--(\ref{5}) we may
succesively transform its left-hand side as follows
\begin{eqnarray}
&& \frac{\alpha^{-1}\varepsilon}{2}\sum_{n=-\infty}^{\infty}
S_{n\kappa}(x)S_{n\kappa}(x')
\nonumber \\*[5mm]
&=& \frac{\alpha^{-1}\varepsilon}{2}\sum_{n=-\infty}^{\infty}
\frac{\alpha(|n|+2\gamma_{\kappa})|n|!}{2\varepsilon N_{n\kappa}
(N_{n\kappa}\mp\kappa)\Gamma(|n|+2\gamma_{\kappa})}
x^{\gamma_{\kappa}}x'^{\gamma_{\kappa}}{\rm e}^{-(x+x')/2}
\nonumber \\
&\times& \left[L_{|n|-1}^{(2\gamma_{\kappa})}(x)
+\frac{\kappa\mp N_{n\kappa}}{|n|+2\gamma_{\kappa}}
L_{|n|}^{(2\gamma_{\kappa})}(x)\right]
\left[L_{|n|-1}^{(2\gamma_{\kappa})}(x')
+\frac{\kappa\mp N_{n\kappa}}{|n|+2\gamma_{\kappa}}
L_{|n|}^{(2\gamma_{\kappa})}(x')\right]
\nonumber \\*[5mm]
&=& \frac{1}{4}x^{\gamma_{\kappa}}x'^{\gamma_{\kappa}}
{\rm e}^{-(x+x')/2}
\sum_{n=-\infty}^{\infty}\frac{(|n|+2\gamma_{\kappa})|n|!}
{N_{n\kappa}(N_{n\kappa}\mp\kappa)\Gamma(|n|+2\gamma_{\kappa})}
\nonumber \\
&\times& \left[L_{|n|-1}^{(2\gamma_{\kappa})}(x)
+\frac{\kappa\mp N_{n\kappa}}{|n|+2\gamma_{\kappa}}
L_{|n|}^{(2\gamma_{\kappa})}(x)\right]
\left[L_{|n|-1}^{(2\gamma_{\kappa})}(x')
+\frac{\kappa\mp N_{n\kappa}}{|n|+2\gamma_{\kappa}}
L_{|n|}^{(2\gamma_{\kappa})}(x')\right]
\nonumber \\*[5mm]
&=& \frac{1}{4}x^{\gamma_{\kappa}}x'^{\gamma_{\kappa}}
{\rm e}^{-(x+x')/2}\left[
\sum_{n=-\infty}^{\infty}\frac{(|n|+2\gamma_{\kappa})|n|!}
{N_{n\kappa}(N_{n\kappa}\mp\kappa)\Gamma(|n|+2\gamma_{\kappa})}
L_{|n|-1}^{(2\gamma_{\kappa})}(x)L_{|n|-1}^{(2\gamma_{\kappa})}(x')
\right. 
\nonumber \\
&+& \sum_{n=-\infty}^{\infty}\frac{(|n|+2\gamma_{\kappa})|n|!}
{N_{n\kappa}(N_{n\kappa}\mp\kappa)\Gamma(|n|+2\gamma_{\kappa})}
\frac{\kappa\mp N_{n\kappa}}{|n|+2\gamma_{\kappa}}
L_{|n|-1}^{(2\gamma_{\kappa})}(x)L_{|n|}^{(2\gamma_{\kappa})}(x') 
\nonumber \\
&+& \sum_{n=-\infty}^{\infty}\frac{(|n|+2\gamma_{\kappa})|n|!}
{N_{n\kappa}(N_{n\kappa}\mp\kappa)\Gamma(|n|+2\gamma_{\kappa})}
\frac{\kappa\mp N_{n\kappa}}{|n|+2\gamma_{\kappa}}
L_{|n|}^{(2\gamma_{\kappa})}(x)L_{|n|-1}^{(2\gamma_{\kappa})}(x') 
\nonumber \\
&+& \left.
\sum_{n=-\infty}^{\infty}\frac{(|n|+2\gamma_{\kappa})|n|!}
{N_{n\kappa}(N_{n\kappa}\mp\kappa)\Gamma(|n|+2\gamma_{\kappa})}
\left(\frac{\kappa\mp N_{n\kappa}}{|n|+2\gamma_{\kappa}}\right)^{2}
L_{|n|}^{(2\gamma_{\kappa})}(x)L_{|n|}^{(2\gamma_{\kappa})}(x')
\right] 
\nonumber \\*[5mm]
&=& \frac{1}{4}x^{\gamma_{\kappa}}x'^{\gamma_{\kappa}}
{\rm e}^{-(x+x')/2}\left[
\sum_{n=-\infty}^{\infty}\frac{(|n|+2\gamma_{\kappa})|n|!}
{N_{n\kappa}(N_{n\kappa}\mp\kappa)\Gamma(|n|+2\gamma_{\kappa})}
L_{|n|-1}^{(2\gamma_{\kappa})}(x)L_{|n|-1}^{(2\gamma_{\kappa})}(x')
\right. 
\nonumber \\
&+& \sum_{n=-\infty}^{\infty}\frac{\mp|n|!}
{N_{n\kappa}\Gamma(|n|+2\gamma_{\kappa})}
L_{|n|-1}^{(2\gamma_{\kappa})}(x)L_{|n|}^{(2\gamma_{\kappa})}(x') 
\nonumber \\
&+& \sum_{n=-\infty}^{\infty}\frac{\mp|n|!}
{N_{n\kappa}\Gamma(|n|+2\gamma_{\kappa})}
L_{|n|}^{(2\gamma_{\kappa})}(x)L_{|n|-1}^{(2\gamma_{\kappa})}(x') 
\nonumber \\
&+& \left.
\sum_{n=-\infty}^{\infty}\frac{(N_{n\kappa}\mp\kappa)|n|!}
{N_{n\kappa}\Gamma(|n|+2\gamma_{\kappa}+1)}
L_{|n|}^{(2\gamma_{\kappa})}(x)L_{|n|}^{(2\gamma_{\kappa})}(x')
\right] 
\nonumber \\*[5mm]
&=& \frac{1}{4}x^{\gamma_{\kappa}}x'^{\gamma_{\kappa}}
{\rm e}^{-(x+x')/2}\left[
\sum_{n=1}^{\infty}\frac{(|n|+2\gamma_{\kappa})|n|!}
{N_{n\kappa}(N_{n\kappa}-\kappa)\Gamma(|n|+2\gamma_{\kappa})}
L_{|n|-1}^{(2\gamma_{\kappa})}(x)L_{|n|-1}^{(2\gamma_{\kappa})}(x')
\right. 
\nonumber \\
&+& \sum_{n=-\infty}^{-1}\frac{(|n|+2\gamma_{\kappa})|n|!}
{N_{n\kappa}(N_{n\kappa}+\kappa)\Gamma(|n|+2\gamma_{\kappa})}
L_{|n|-1}^{(2\gamma_{\kappa})}(x)L_{|n|-1}^{(2\gamma_{\kappa})}(x')
\nonumber \\
&-& \sum_{n=1}^{\infty}\frac{|n|!}
{N_{n\kappa}\Gamma(|n|+2\gamma_{\kappa})}
L_{|n|-1}^{(2\gamma_{\kappa})}(x)L_{|n|}^{(2\gamma_{\kappa})}(x') 
\nonumber \\
&+& \sum_{n=-\infty}^{-1}\frac{|n|!}
{N_{n\kappa}\Gamma(|n|+2\gamma_{\kappa})}
L_{|n|-1}^{(2\gamma_{\kappa})}(x)L_{|n|}^{(2\gamma_{\kappa})}(x') 
\nonumber \\
&-& \sum_{n=1}^{\infty}\frac{|n|!}
{N_{n\kappa}\Gamma(|n|+2\gamma_{\kappa})}
L_{|n|}^{(2\gamma_{\kappa})}(x)L_{|n|-1}^{(2\gamma_{\kappa})}(x') 
\nonumber \\
&+& \sum_{n=-\infty}^{-1}\frac{|n|!}
{N_{n\kappa}\Gamma(|n|+2\gamma_{\kappa})}
L_{|n|}^{(2\gamma_{\kappa})}(x)L_{|n|-1}^{(2\gamma_{\kappa})}(x') 
\nonumber \\
&+& \sum_{n=0}^{\infty}\frac{(N_{n\kappa}-\kappa)|n|!}
{N_{n\kappa}\Gamma(|n|+2\gamma_{\kappa}+1)}
L_{|n|}^{(2\gamma_{\kappa})}(x)L_{|n|}^{(2\gamma_{\kappa})}(x')
\nonumber \\
&+& \left.
\sum_{n=-\infty}^{0}\frac{(N_{n\kappa}+\kappa)|n|!}
{N_{n\kappa}\Gamma(|n|+2\gamma_{\kappa}+1)}
L_{|n|}^{(2\gamma_{\kappa})}(x)L_{|n|}^{(2\gamma_{\kappa})}(x')
\right] 
\nonumber \\*[5mm]
&=& \frac{1}{4}x^{\gamma_{\kappa}}x'^{\gamma_{\kappa}}
{\rm e}^{-(x+x')/2}\left[
\sum_{n=1}^{\infty}\frac{(n+2\gamma_{\kappa})n!}
{N_{n\kappa}(N_{n\kappa}-\kappa)\Gamma(n+2\gamma_{\kappa})}
L_{n-1}^{(2\gamma_{\kappa})}(x)L_{n-1}^{(2\gamma_{\kappa})}(x')
\right. 
\nonumber \\
&+& \sum_{n=1}^{\infty}\frac{(n+2\gamma_{\kappa})n!}
{N_{n\kappa}(N_{n\kappa}+\kappa)\Gamma(n+2\gamma_{\kappa})}
L_{n-1}^{(2\gamma_{\kappa})}(x)L_{n-1}^{(2\gamma_{\kappa})}(x')
\nonumber \\
&-& \sum_{n=1}^{\infty}\frac{n!}
{N_{n\kappa}\Gamma(n+2\gamma_{\kappa})}
L_{n-1}^{(2\gamma_{\kappa})}(x)L_{n}^{(2\gamma_{\kappa})}(x') 
\nonumber \\
&+& \sum_{n=1}^{\infty}\frac{n!}
{N_{n\kappa}\Gamma(n+2\gamma_{\kappa})}
L_{n-1}^{(2\gamma_{\kappa})}(x)L_{n}^{(2\gamma_{\kappa})}(x') 
\nonumber \\
&-& \sum_{n=1}^{\infty}\frac{n!}
{N_{n\kappa}\Gamma(n+2\gamma_{\kappa})}
L_{n}^{(2\gamma_{\kappa})}(x)L_{n-1}^{(2\gamma_{\kappa})}(x') 
\nonumber \\
&+& \sum_{n=1}^{\infty}\frac{n!}
{N_{n\kappa}\Gamma(n+2\gamma_{\kappa})}
L_{n}^{(2\gamma_{\kappa})}(x)L_{n-1}^{(2\gamma_{\kappa})}(x') 
\nonumber \\
&+& \sum_{n=0}^{\infty}\frac{(N_{n\kappa}-\kappa)n!}
{N_{n\kappa}\Gamma(n+2\gamma_{\kappa}+1)}
L_{n}^{(2\gamma_{\kappa})}(x)L_{n}^{(2\gamma_{\kappa})}(x')
\nonumber \\
&+& \left.
\sum_{n=0}^{\infty}\frac{(N_{n\kappa}+\kappa)n!}
{N_{n\kappa}\Gamma(n+2\gamma_{\kappa}+1)}
L_{n}^{(2\gamma_{\kappa})}(x)L_{n}^{(2\gamma_{\kappa})}(x')
\right] 
\nonumber \\*[5mm]
&=& \frac{1}{4}x^{\gamma_{\kappa}}x'^{\gamma_{\kappa}}
{\rm e}^{-(x+x')/2}\left[
\sum_{n=1}^{\infty}\frac{(n+2\gamma_{\kappa})n!}
{N_{n\kappa}(N_{n\kappa}-\kappa)\Gamma(n+2\gamma_{\kappa})}
L_{n-1}^{(2\gamma_{\kappa})}(x)L_{n-1}^{(2\gamma_{\kappa})}(x')
\right. 
\nonumber \\
&+& \sum_{n=1}^{\infty}\frac{(n+2\gamma_{\kappa})n!}
{N_{n\kappa}(N_{n\kappa}+\kappa)\Gamma(n+2\gamma_{\kappa})}
L_{n-1}^{(2\gamma_{\kappa})}(x)L_{n-1}^{(2\gamma_{\kappa})}(x')
\nonumber \\
&+& \sum_{n=0}^{\infty}\frac{(N_{n\kappa}-\kappa)n!}
{N_{n\kappa}\Gamma(n+2\gamma_{\kappa}+1)}
L_{n}^{(2\gamma_{\kappa})}(x)L_{n}^{(2\gamma_{\kappa})}(x')
\nonumber \\
&+& \left.
\sum_{n=0}^{\infty}\frac{(N_{n\kappa}+\kappa)n!}
{N_{n\kappa}\Gamma(n+2\gamma_{\kappa}+1)}
L_{n}^{(2\gamma_{\kappa})}(x)L_{n}^{(2\gamma_{\kappa})}(x')
\right] 
\nonumber \\*[5mm]
&=& \frac{1}{4} x^{\gamma_{\kappa}}x'^{\gamma_{\kappa}}
{\rm e}^{-(x+x')/2}\left[
\sum_{n=1}^{\infty}\frac{(N_{n\kappa}+\kappa)(n-1)!}
{N_{n\kappa}\Gamma(n+2\gamma_{\kappa})}
L_{n-1}^{(2\gamma_{\kappa})}(x)L_{n-1}^{(2\gamma_{\kappa})}(x')
\right.
\nonumber \\
&+& \sum_{n=1}^{\infty} \frac{(N_{n\kappa}-\kappa)(n-1)!}
{N_{n\kappa}\Gamma(n+2\gamma_{\kappa})}
L_{n-1}^{(2\gamma_{\kappa})}(x)L_{n-1}^{(2\gamma_{\kappa})}(x')
\nonumber \\
&+& \sum_{n=0}^{\infty} \frac{(N_{n\kappa}-\kappa)n!}
{N_{n\kappa}\Gamma(n+2\gamma_{\kappa}+1)}
L_{n}^{(2\gamma_{\kappa})}(x)L_{n}^{(2\gamma_{\kappa})}(x')
\nonumber \\
&+& \left. \sum_{n=0}^{\infty} \frac{(N_{n\kappa}+\kappa)n!}
{N_{n\kappa}\Gamma(n+2\gamma_{\kappa}+1)}
L_{n}^{(2\gamma_{\kappa})}(x)L_{n}^{(2\gamma_{\kappa})}(x')
\right]
\nonumber \\*[5mm]
&=& \frac{1}{2} x^{\gamma_{\kappa}}x'^{\gamma_{\kappa}}
{\rm e}^{-(x+x')/2}\left[
\sum_{n=1}^{\infty}\frac{(n-1)!}{\Gamma(n+2\gamma_{\kappa})}
L_{n-1}^{(2\gamma_{\kappa})}(x)L_{n-1}^{(2\gamma_{\kappa})}(x')
\right.
\nonumber \\
&+& \left.\sum_{n=0}^{\infty} \frac{n!}{\Gamma(n+2\gamma_{\kappa}+1)}
L_{n}^{(2\gamma_{\kappa})}(x)L_{n}^{(2\gamma_{\kappa})}(x') \right]
\nonumber \\*[5mm]
&=& \frac{1}{2} x^{\gamma_{\kappa}}x'^{\gamma_{\kappa}}
{\rm e}^{-(x+x')/2}\left[
\sum_{n=0}^{\infty}\frac{n!}{\Gamma(n+2\gamma_{\kappa}+1)}
L_{n}^{(2\gamma_{\kappa})}(x)L_{n}^{(2\gamma_{\kappa})}(x')
\right.
\nonumber \\
&+& \left.\sum_{n=0}^{\infty} \frac{n!}{\Gamma(n+2\gamma_{\kappa}+1)}
L_{n}^{(2\gamma_{\kappa})}(x)L_{n}^{(2\gamma_{\kappa})}(x') \right]
\nonumber \\*[5mm]
&=& x^{\gamma_{\kappa}}x'^{\gamma_{\kappa}}{\rm e}^{-(x+x')/2}
\sum_{n=0}^{\infty}\frac{n!}{\Gamma(n+2\gamma_{\kappa}+1)}
L_{n}^{(2\gamma_{\kappa})}(x)L_{n}^{(2\gamma_{\kappa})}(x')
=\delta(x-x'),
\label{11}
\end{eqnarray}
where the last equality stems from the closure relation (\ref{6}).
Thus we have proved the relation (\ref{7}). 

Consider now the relation (\ref{8}). Successive transformations of
its left-hand side yield
\begin{eqnarray}
&& \frac{\alpha^{-1}\varepsilon^{-1}}{2}\sum_{n=-\infty}^{\infty}
T_{n\kappa}(x)T_{n\kappa}(x')
\nonumber \\*[5mm]
&=& \frac{\alpha^{-1}\varepsilon^{-1}}{2}\sum_{n=-\infty}^{\infty}
\frac{\varepsilon\alpha(|n|+2\gamma_{\kappa})|n|!}{2N_{n\kappa}
(N_{n\kappa}\mp\kappa)\Gamma(|n|+2\gamma_{\kappa})}
x^{\gamma_{\kappa}}x'^{\gamma_{\kappa}}{\rm e}^{-(x+x')/2}
\nonumber \\
&\times& \left[L_{|n|-1}^{(2\gamma_{\kappa})}(x)
-\frac{\kappa\mp N_{n\kappa}}{|n|+2\gamma_{\kappa}}
L_{|n|}^{(2\gamma_{\kappa})}(x)\right]
\left[L_{|n|-1}^{(2\gamma_{\kappa})}(x')
-\frac{\kappa\mp N_{n\kappa}}{|n|+2\gamma_{\kappa}}
L_{|n|}^{(2\gamma_{\kappa})}(x')\right]
\nonumber \\*[5mm]
&=& \frac{1}{4}x^{\gamma_{\kappa}}x'^{\gamma_{\kappa}}
{\rm e}^{-(x+x')/2}
\sum_{n=-\infty}^{\infty}\frac{(|n|+2\gamma_{\kappa})|n|!}
{N_{n\kappa}(N_{n\kappa}\mp\kappa)\Gamma(|n|+2\gamma_{\kappa})}
\nonumber \\
&\times& \left[L_{|n|-1}^{(2\gamma_{\kappa})}(x)
-\frac{\kappa\mp N_{n\kappa}}{|n|+2\gamma_{\kappa}}
L_{|n|}^{(2\gamma_{\kappa})}(x)\right]
\left[L_{|n|-1}^{(2\gamma_{\kappa})}(x')
-\frac{\kappa\mp N_{n\kappa}}{|n|+2\gamma_{\kappa}}
L_{|n|}^{(2\gamma_{\kappa})}(x')\right]
\nonumber \\*[5mm]
&=& \frac{1}{4}x^{\gamma_{\kappa}}x'^{\gamma_{\kappa}}
{\rm e}^{-(x+x')/2}\left[
\sum_{n=-\infty}^{\infty}\frac{(|n|+2\gamma_{\kappa})|n|!}
{N_{n\kappa}(N_{n\kappa}\mp\kappa)\Gamma(|n|+2\gamma_{\kappa})}
L_{|n|-1}^{(2\gamma_{\kappa})}(x)L_{|n|-1}^{(2\gamma_{\kappa})}(x')
\right. 
\nonumber \\
&-& \sum_{n=-\infty}^{\infty}\frac{(|n|+2\gamma_{\kappa})|n|!}
{N_{n\kappa}(N_{n\kappa}\mp\kappa)\Gamma(|n|+2\gamma_{\kappa})}
\frac{\kappa\mp N_{n\kappa}}{|n|+2\gamma_{\kappa}}
L_{|n|-1}^{(2\gamma_{\kappa})}(x)L_{|n|}^{(2\gamma_{\kappa})}(x') 
\nonumber \\
&-& \sum_{n=-\infty}^{\infty}\frac{(|n|+2\gamma_{\kappa})|n|!}
{N_{n\kappa}(N_{n\kappa}\mp\kappa)\Gamma(|n|+2\gamma_{\kappa})}
\frac{\kappa\mp N_{n\kappa}}{|n|+2\gamma_{\kappa}}
L_{|n|}^{(2\gamma_{\kappa})}(x)L_{|n|-1}^{(2\gamma_{\kappa})}(x') 
\nonumber \\
&+& \left.
\sum_{n=-\infty}^{\infty}\frac{(|n|+2\gamma_{\kappa})|n|!}
{N_{n\kappa}(N_{n\kappa}\mp\kappa)\Gamma(|n|+2\gamma_{\kappa})}
\left(\frac{\kappa\mp N_{n\kappa}}{|n|+2\gamma_{\kappa}}\right)^{2}
L_{|n|}^{(2\gamma_{\kappa})}(x)L_{|n|}^{(2\gamma_{\kappa})}(x')
\right] 
\nonumber \\*[5mm]
&=& \frac{1}{4}x^{\gamma_{\kappa}}x'^{\gamma_{\kappa}}
{\rm e}^{-(x+x')/2}\left[
\sum_{n=-\infty}^{\infty}\frac{(|n|+2\gamma_{\kappa})|n|!}
{N_{n\kappa}(N_{n\kappa}\mp\kappa)\Gamma(|n|+2\gamma_{\kappa})}
L_{|n|-1}^{(2\gamma_{\kappa})}(x)L_{|n|-1}^{(2\gamma_{\kappa})}(x')
\right. 
\nonumber \\
&-& \sum_{n=-\infty}^{\infty}\frac{\mp|n|!}
{N_{n\kappa}\Gamma(|n|+2\gamma_{\kappa})}
L_{|n|-1}^{(2\gamma_{\kappa})}(x)L_{|n|}^{(2\gamma_{\kappa})}(x') 
\nonumber \\
&-& \sum_{n=-\infty}^{\infty}\frac{\mp|n|!}
{N_{n\kappa}\Gamma(|n|+2\gamma_{\kappa})}
L_{|n|}^{(2\gamma_{\kappa})}(x)L_{|n|-1}^{(2\gamma_{\kappa})}(x') 
\nonumber \\
&+& \left.
\sum_{n=-\infty}^{\infty}\frac{(N_{n\kappa}\mp\kappa)|n|!}
{N_{n\kappa}\Gamma(|n|+2\gamma_{\kappa}+1)}
L_{|n|}^{(2\gamma_{\kappa})}(x)L_{|n|}^{(2\gamma_{\kappa})}(x')
\right] 
\nonumber \\*[5mm]
&=& \frac{1}{4}x^{\gamma_{\kappa}}x'^{\gamma_{\kappa}}
{\rm e}^{-(x+x')/2}\left[
\sum_{n=1}^{\infty}\frac{(|n|+2\gamma_{\kappa})|n|!}
{N_{n\kappa}(N_{n\kappa}-\kappa)\Gamma(|n|+2\gamma_{\kappa})}
L_{|n|-1}^{(2\gamma_{\kappa})}(x)L_{|n|-1}^{(2\gamma_{\kappa})}(x')
\right. 
\nonumber \\
&+& \sum_{n=-\infty}^{-1}\frac{(|n|+2\gamma_{\kappa})|n|!}
{N_{n\kappa}(N_{n\kappa}+\kappa)\Gamma(|n|+2\gamma_{\kappa})}
L_{|n|-1}^{(2\gamma_{\kappa})}(x)L_{|n|-1}^{(2\gamma_{\kappa})}(x')
\nonumber \\
&+& \sum_{n=1}^{\infty}\frac{|n|!}
{N_{n\kappa}\Gamma(|n|+2\gamma_{\kappa})}
L_{|n|-1}^{(2\gamma_{\kappa})}(x)L_{|n|}^{(2\gamma_{\kappa})}(x') 
\nonumber \\
&-& \sum_{n=-\infty}^{-1}\frac{|n|!}
{N_{n\kappa}\Gamma(|n|+2\gamma_{\kappa})}
L_{|n|-1}^{(2\gamma_{\kappa})}(x)L_{|n|}^{(2\gamma_{\kappa})}(x') 
\nonumber \\
&+& \sum_{n=1}^{\infty}\frac{|n|!}
{N_{n\kappa}\Gamma(|n|+2\gamma_{\kappa})}
L_{|n|}^{(2\gamma_{\kappa})}(x)L_{|n|-1}^{(2\gamma_{\kappa})}(x') 
\nonumber \\
&-& \sum_{n=-\infty}^{-1}\frac{|n|!}
{N_{n\kappa}\Gamma(|n|+2\gamma_{\kappa})}
L_{|n|}^{(2\gamma_{\kappa})}(x)L_{|n|-1}^{(2\gamma_{\kappa})}(x') 
\nonumber \\
&+& \sum_{n=0}^{\infty}\frac{(N_{n\kappa}-\kappa)|n|!}
{N_{n\kappa}\Gamma(|n|+2\gamma_{\kappa}+1)}
L_{|n|}^{(2\gamma_{\kappa})}(x)L_{|n|}^{(2\gamma_{\kappa})}(x')
\nonumber \\
&+& \left.
\sum_{n=-\infty}^{0}\frac{(N_{n\kappa}+\kappa)|n|!}
{N_{n\kappa}\Gamma(|n|+2\gamma_{\kappa}+1)}
L_{|n|}^{(2\gamma_{\kappa})}(x)L_{|n|}^{(2\gamma_{\kappa})}(x')
\right] 
\nonumber \\*[5mm]
&=& \frac{1}{4}x^{\gamma_{\kappa}}x'^{\gamma_{\kappa}}
{\rm e}^{-(x+x')/2}\left[
\sum_{n=1}^{\infty}\frac{(n+2\gamma_{\kappa})n!}
{N_{n\kappa}(N_{n\kappa}-\kappa)\Gamma(n+2\gamma_{\kappa})}
L_{n-1}^{(2\gamma_{\kappa})}(x)L_{n-1}^{(2\gamma_{\kappa})}(x')
\right. 
\nonumber \\
&+& \sum_{n=1}^{\infty}\frac{(n+2\gamma_{\kappa})n!}
{N_{n\kappa}(N_{n\kappa}+\kappa)\Gamma(n+2\gamma_{\kappa})}
L_{n-1}^{(2\gamma_{\kappa})}(x)L_{n-1}^{(2\gamma_{\kappa})}(x')
\nonumber \\
&+& \sum_{n=1}^{\infty}\frac{n!}
{N_{n\kappa}\Gamma(n+2\gamma_{\kappa})}
L_{n-1}^{(2\gamma_{\kappa})}(x)L_{n}^{(2\gamma_{\kappa})}(x') 
\nonumber \\
&-& \sum_{n=1}^{\infty}\frac{n!}
{N_{n\kappa}\Gamma(n+2\gamma_{\kappa})}
L_{n-1}^{(2\gamma_{\kappa})}(x)L_{n}^{(2\gamma_{\kappa})}(x') 
\nonumber \\
&+& \sum_{n=1}^{\infty}\frac{n!}
{N_{n\kappa}\Gamma(n+2\gamma_{\kappa})}
L_{n}^{(2\gamma_{\kappa})}(x)L_{n-1}^{(2\gamma_{\kappa})}(x') 
\nonumber \\
&-& \sum_{n=1}^{\infty}\frac{n!}
{N_{n\kappa}\Gamma(n+2\gamma_{\kappa})}
L_{n}^{(2\gamma_{\kappa})}(x)L_{n-1}^{(2\gamma_{\kappa})}(x') 
\nonumber \\
&+& \sum_{n=0}^{\infty}\frac{(N_{n\kappa}-\kappa)n!}
{N_{n\kappa}\Gamma(n+2\gamma_{\kappa}+1)}
L_{n}^{(2\gamma_{\kappa})}(x)L_{n}^{(2\gamma_{\kappa})}(x')
\nonumber \\
&+& \left.
\sum_{n=0}^{\infty}\frac{(N_{n\kappa}+\kappa)n!}
{N_{n\kappa}\Gamma(n+2\gamma_{\kappa}+1)}
L_{n}^{(2\gamma_{\kappa})}(x)L_{n}^{(2\gamma_{\kappa})}(x')
\right] 
\nonumber \\*[5mm]
&=& \frac{1}{4}x^{\gamma_{\kappa}}x'^{\gamma_{\kappa}}
{\rm e}^{-(x+x')/2}\left[
\sum_{n=1}^{\infty}\frac{(n+2\gamma_{\kappa})n!}
{N_{n\kappa}(N_{n\kappa}-\kappa)\Gamma(n+2\gamma_{\kappa})}
L_{n-1}^{(2\gamma_{\kappa})}(x)L_{n-1}^{(2\gamma_{\kappa})}(x')
\right. 
\nonumber \\
&+& \sum_{n=1}^{\infty}\frac{(n+2\gamma_{\kappa})n!}
{N_{n\kappa}(N_{n\kappa}+\kappa)\Gamma(n+2\gamma_{\kappa})}
L_{n-1}^{(2\gamma_{\kappa})}(x)L_{n-1}^{(2\gamma_{\kappa})}(x')
\nonumber \\
&+& \sum_{n=0}^{\infty}\frac{(N_{n\kappa}-\kappa)n!}
{N_{n\kappa}\Gamma(n+2\gamma_{\kappa}+1)}
L_{n}^{(2\gamma_{\kappa})}(x)L_{n}^{(2\gamma_{\kappa})}(x')
\nonumber \\
&+& \left.
\sum_{n=0}^{\infty}\frac{(N_{n\kappa}+\kappa)n!}
{N_{n\kappa}\Gamma(n+2\gamma_{\kappa}+1)}
L_{n}^{(2\gamma_{\kappa})}(x)L_{n}^{(2\gamma_{\kappa})}(x')
\right] 
\nonumber \\*[5mm]
&=& \frac{1}{4} x^{\gamma_{\kappa}}x'^{\gamma_{\kappa}}
{\rm e}^{-(x+x')/2}\left[
\sum_{n=1}^{\infty}\frac{(N_{n\kappa}+\kappa)(n-1)!}
{N_{n\kappa}\Gamma(n+2\gamma_{\kappa})}
L_{n-1}^{(2\gamma_{\kappa})}(x)L_{n-1}^{(2\gamma_{\kappa})}(x')
\right.
\nonumber \\
&+& \sum_{n=1}^{\infty} \frac{(N_{n\kappa}-\kappa)(n-1)!}
{N_{n\kappa}\Gamma(n+2\gamma_{\kappa})}
L_{n-1}^{(2\gamma_{\kappa})}(x)L_{n-1}^{(2\gamma_{\kappa})}(x')
\nonumber \\
&+& \sum_{n=0}^{\infty} \frac{(N_{n\kappa}-\kappa)n!}
{N_{n\kappa}\Gamma(n+2\gamma_{\kappa}+1)}
L_{n}^{(2\gamma_{\kappa})}(x)L_{n}^{(2\gamma_{\kappa})}(x')
\nonumber \\
&+& \left. \sum_{n=0}^{\infty} \frac{(N_{n\kappa}+\kappa)n!}
{N_{n\kappa}\Gamma(n+2\gamma_{\kappa}+1)}
L_{n}^{(2\gamma_{\kappa})}(x)L_{n}^{(2\gamma_{\kappa})}(x')
\right]
\nonumber \\*[5mm]
&=& \frac{1}{2} x^{\gamma_{\kappa}}x'^{\gamma_{\kappa}}
{\rm e}^{-(x+x')/2}\left[
\sum_{n=1}^{\infty}\frac{(n-1)!}{\Gamma(n+2\gamma_{\kappa})}
L_{n-1}^{(2\gamma_{\kappa})}(x)L_{n-1}^{(2\gamma_{\kappa})}(x')
\right.
\nonumber \\
&+& \left.\sum_{n=0}^{\infty} \frac{n!}{\Gamma(n+2\gamma_{\kappa}+1)}
L_{n}^{(2\gamma_{\kappa})}(x)L_{n}^{(2\gamma_{\kappa})}(x') \right]
\nonumber \\*[5mm]
&=& \frac{1}{2} x^{\gamma_{\kappa}}x'^{\gamma_{\kappa}}
{\rm e}^{-(x+x')/2}\left[
\sum_{n=0}^{\infty}\frac{n!}{\Gamma(n+2\gamma_{\kappa}+1)}
L_{n}^{(2\gamma_{\kappa})}(x)L_{n}^{(2\gamma_{\kappa})}(x')
\right.
\nonumber \\
&+& \left.\sum_{n=0}^{\infty} \frac{n!}{\Gamma(n+2\gamma_{\kappa}+1)}
L_{n}^{(2\gamma_{\kappa})}(x)L_{n}^{(2\gamma_{\kappa})}(x') \right]
\nonumber \\*[5mm]
&=& x^{\gamma_{\kappa}}x'^{\gamma_{\kappa}}{\rm e}^{-(x+x')/2}
\sum_{n=0}^{\infty}\frac{n!}{\Gamma(n+2\gamma_{\kappa}+1)}
L_{n}^{(2\gamma_{\kappa})}(x)L_{n}^{(2\gamma_{\kappa})}(x')
=\delta(x-x'),
\label{12}
\end{eqnarray}
where again the last equality follows from the closure relation
(\ref{6}). Thus we have proved correctness of the relation (\ref{8}).

Finally, successive transformations of the left-hand side of the
relation (\ref{9}) give
\begin{eqnarray}
&& \frac{\alpha^{-1}\varepsilon}{2}\sum_{n=-\infty}^{\infty}
S_{n\kappa}(x)T_{n\kappa}(x')
\nonumber \\*[5mm]
&=& \frac{\alpha^{-1}\varepsilon}{2}\sum_{n=-\infty}^{\infty}
\frac{\alpha(|n|+2\gamma_{\kappa})|n|!}{2N_{n\kappa}
(N_{n\kappa}\mp\kappa)\Gamma(|n|+2\gamma_{\kappa})}
x^{\gamma_{\kappa}}x'^{\gamma_{\kappa}}{\rm e}^{-(x+x')/2}
\nonumber \\
&\times& \left[L_{|n|-1}^{(2\gamma_{\kappa})}(x)
+\frac{\kappa\mp N_{n\kappa}}{|n|+2\gamma_{\kappa}}
L_{|n|}^{(2\gamma_{\kappa})}(x)\right]
\left[L_{|n|-1}^{(2\gamma_{\kappa})}(x')
-\frac{\kappa\mp N_{n\kappa}}{|n|+2\gamma_{\kappa}}
L_{|n|}^{(2\gamma_{\kappa})}(x')\right]
\nonumber \\*[5mm]
&=& \frac{\varepsilon}{4}x^{\gamma_{\kappa}}x'^{\gamma_{\kappa}}
{\rm e}^{-(x+x')/2}
\sum_{n=-\infty}^{\infty}\frac{(|n|+2\gamma_{\kappa})|n|!}
{N_{n\kappa}(N_{n\kappa}\mp\kappa)\Gamma(|n|+2\gamma_{\kappa})}
\nonumber \\
&\times& \left[L_{|n|-1}^{(2\gamma_{\kappa})}(x)
+\frac{\kappa\mp N_{n\kappa}}{|n|+2\gamma_{\kappa}}
L_{|n|}^{(2\gamma_{\kappa})}(x)\right]
\left[L_{|n|-1}^{(2\gamma_{\kappa})}(x')
-\frac{\kappa\mp N_{n\kappa}}{|n|+2\gamma_{\kappa}}
L_{|n|}^{(2\gamma_{\kappa})}(x')\right]
\nonumber \\*[5mm]
&=& \frac{\varepsilon}{4}x^{\gamma_{\kappa}}x'^{\gamma_{\kappa}}
{\rm e}^{-(x+x')/2}\left[
\sum_{n=-\infty}^{\infty}\frac{(|n|+2\gamma_{\kappa})|n|!}
{N_{n\kappa}(N_{n\kappa}\mp\kappa)\Gamma(|n|+2\gamma_{\kappa})}
L_{|n|-1}^{(2\gamma_{\kappa})}(x)L_{|n|-1}^{(2\gamma_{\kappa})}(x')
\right. 
\nonumber \\
&-& \sum_{n=-\infty}^{\infty}\frac{(|n|+2\gamma_{\kappa})|n|!}
{N_{n\kappa}(N_{n\kappa}\mp\kappa)\Gamma(|n|+2\gamma_{\kappa})}
\frac{\kappa\mp N_{n\kappa}}{|n|+2\gamma_{\kappa}}
L_{|n|-1}^{(2\gamma_{\kappa})}(x)L_{|n|}^{(2\gamma_{\kappa})}(x') 
\nonumber \\
&+& \sum_{n=-\infty}^{\infty}\frac{(|n|+2\gamma_{\kappa})|n|!}
{N_{n\kappa}(N_{n\kappa}\mp\kappa)\Gamma(|n|+2\gamma_{\kappa})}
\frac{\kappa\mp N_{n\kappa}}{|n|+2\gamma_{\kappa}}
L_{|n|}^{(2\gamma_{\kappa})}(x)L_{|n|-1}^{(2\gamma_{\kappa})}(x') 
\nonumber \\
&-& \left.
\sum_{n=-\infty}^{\infty}\frac{(|n|+2\gamma_{\kappa})|n|!}
{N_{n\kappa}(N_{n\kappa}\mp\kappa)\Gamma(|n|+2\gamma_{\kappa})}
\left(\frac{\kappa\mp N_{n\kappa}}{|n|+2\gamma_{\kappa}}\right)^{2}
L_{|n|}^{(2\gamma_{\kappa})}(x)L_{|n|}^{(2\gamma_{\kappa})}(x')
\right] 
\nonumber \\*[5mm]
&=& \frac{\varepsilon}{4}x^{\gamma_{\kappa}}x'^{\gamma_{\kappa}}
{\rm e}^{-(x+x')/2}\left[
\sum_{n=-\infty}^{\infty}\frac{(|n|+2\gamma_{\kappa})|n|!}
{N_{n\kappa}(N_{n\kappa}\mp\kappa)\Gamma(|n|+2\gamma_{\kappa})}
L_{|n|-1}^{(2\gamma_{\kappa})}(x)L_{|n|-1}^{(2\gamma_{\kappa})}(x')
\right. 
\nonumber \\
&-& \sum_{n=-\infty}^{\infty}\frac{\mp|n|!}
{N_{n\kappa}\Gamma(|n|+2\gamma_{\kappa})}
L_{|n|-1}^{(2\gamma_{\kappa})}(x)L_{|n|}^{(2\gamma_{\kappa})}(x') 
\nonumber \\
&+& \sum_{n=-\infty}^{\infty}\frac{\mp|n|!}
{N_{n\kappa}\Gamma(|n|+2\gamma_{\kappa})}
L_{|n|}^{(2\gamma_{\kappa})}(x)L_{|n|-1}^{(2\gamma_{\kappa})}(x') 
\nonumber \\
&-& \left.
\sum_{n=-\infty}^{\infty}\frac{(N_{n\kappa}\mp\kappa)|n|!}
{N_{n\kappa}\Gamma(|n|+2\gamma_{\kappa}+1)}
L_{|n|}^{(2\gamma_{\kappa})}(x)L_{|n|}^{(2\gamma_{\kappa})}(x')
\right] 
\nonumber \\*[5mm]
&=& \frac{\varepsilon}{4}x^{\gamma_{\kappa}}x'^{\gamma_{\kappa}}
{\rm e}^{-(x+x')/2}\left[
\sum_{n=1}^{\infty}\frac{(|n|+2\gamma_{\kappa})|n|!}
{N_{n\kappa}(N_{n\kappa}-\kappa)\Gamma(|n|+2\gamma_{\kappa})}
L_{|n|-1}^{(2\gamma_{\kappa})}(x)L_{|n|-1}^{(2\gamma_{\kappa})}(x')
\right. 
\nonumber \\
&+& \sum_{n=-\infty}^{-1}\frac{(|n|+2\gamma_{\kappa})|n|!}
{N_{n\kappa}(N_{n\kappa}+\kappa)\Gamma(|n|+2\gamma_{\kappa})}
L_{|n|-1}^{(2\gamma_{\kappa})}(x)L_{|n|-1}^{(2\gamma_{\kappa})}(x')
\nonumber \\
&+& \sum_{n=1}^{\infty}\frac{|n|!}
{N_{n\kappa}\Gamma(|n|+2\gamma_{\kappa})}
L_{|n|-1}^{(2\gamma_{\kappa})}(x)L_{|n|}^{(2\gamma_{\kappa})}(x') 
\nonumber \\
&-& \sum_{n=-\infty}^{-1}\frac{|n|!}
{N_{n\kappa}\Gamma(|n|+2\gamma_{\kappa})}
L_{|n|-1}^{(2\gamma_{\kappa})}(x)L_{|n|}^{(2\gamma_{\kappa})}(x') 
\nonumber \\
&-& \sum_{n=1}^{\infty}\frac{|n|!}
{N_{n\kappa}\Gamma(|n|+2\gamma_{\kappa})}
L_{|n|}^{(2\gamma_{\kappa})}(x)L_{|n|-1}^{(2\gamma_{\kappa})}(x') 
\nonumber \\
&+& \sum_{n=-\infty}^{-1}\frac{|n|!}
{N_{n\kappa}\Gamma(|n|+2\gamma_{\kappa})}
L_{|n|}^{(2\gamma_{\kappa})}(x)L_{|n|-1}^{(2\gamma_{\kappa})}(x') 
\nonumber \\
&-& \sum_{n=0}^{\infty}\frac{(N_{n\kappa}-\kappa)|n|!}
{N_{n\kappa}\Gamma(|n|+2\gamma_{\kappa}+1)}
L_{|n|}^{(2\gamma_{\kappa})}(x)L_{|n|}^{(2\gamma_{\kappa})}(x')
\nonumber \\
&-& \left.
\sum_{n=-\infty}^{0}\frac{(N_{n\kappa}+\kappa)|n|!}
{N_{n\kappa}\Gamma(|n|+2\gamma_{\kappa}+1)}
L_{|n|}^{(2\gamma_{\kappa})}(x)L_{|n|}^{(2\gamma_{\kappa})}(x')
\right] 
\nonumber \\*[5mm]
&=& \frac{\varepsilon}{4}x^{\gamma_{\kappa}}x'^{\gamma_{\kappa}}
{\rm e}^{-(x+x')/2}\left[
\sum_{n=1}^{\infty}\frac{(n+2\gamma_{\kappa})n!}
{N_{n\kappa}(N_{n\kappa}-\kappa)\Gamma(n+2\gamma_{\kappa})}
L_{n-1}^{(2\gamma_{\kappa})}(x)L_{n-1}^{(2\gamma_{\kappa})}(x')
\right. 
\nonumber \\
&+& \sum_{n=1}^{\infty}\frac{(n+2\gamma_{\kappa})n!}
{N_{n\kappa}(N_{n\kappa}+\kappa)\Gamma(n+2\gamma_{\kappa})}
L_{n-1}^{(2\gamma_{\kappa})}(x)L_{n-1}^{(2\gamma_{\kappa})}(x')
\nonumber \\
&+& \sum_{n=1}^{\infty}\frac{n!}
{N_{n\kappa}\Gamma(n+2\gamma_{\kappa})}
L_{n-1}^{(2\gamma_{\kappa})}(x)L_{n}^{(2\gamma_{\kappa})}(x') 
\nonumber \\
&-& \sum_{n=1}^{\infty}\frac{n!}
{N_{n\kappa}\Gamma(n+2\gamma_{\kappa})}
L_{n-1}^{(2\gamma_{\kappa})}(x)L_{n}^{(2\gamma_{\kappa})}(x') 
\nonumber \\
&-& \sum_{n=1}^{\infty}\frac{n!}
{N_{n\kappa}\Gamma(n+2\gamma_{\kappa})}
L_{n}^{(2\gamma_{\kappa})}(x)L_{n-1}^{(2\gamma_{\kappa})}(x') 
\nonumber \\
&+& \sum_{n=1}^{\infty}\frac{n!}
{N_{n\kappa}\Gamma(n+2\gamma_{\kappa})}
L_{n}^{(2\gamma_{\kappa})}(x)L_{n-1}^{(2\gamma_{\kappa})}(x') 
\nonumber \\
&-& \sum_{n=0}^{\infty}\frac{(N_{n\kappa}-\kappa)n!}
{N_{n\kappa}\Gamma(n+2\gamma_{\kappa}+1)}
L_{n}^{(2\gamma_{\kappa})}(x)L_{n}^{(2\gamma_{\kappa})}(x')
\nonumber \\
&-& \left.
\sum_{n=0}^{\infty}\frac{(N_{n\kappa}+\kappa)n!}
{N_{n\kappa}\Gamma(n+2\gamma_{\kappa}+1)}
L_{n}^{(2\gamma_{\kappa})}(x)L_{n}^{(2\gamma_{\kappa})}(x')
\right] 
\nonumber \\*[5mm]
&=& \frac{\varepsilon}{4}x^{\gamma_{\kappa}}x'^{\gamma_{\kappa}}
{\rm e}^{-(x+x')/2}\left[
\sum_{n=1}^{\infty}\frac{(n+2\gamma_{\kappa})n!}
{N_{n\kappa}(N_{n\kappa}-\kappa)\Gamma(n+2\gamma_{\kappa})}
L_{n-1}^{(2\gamma_{\kappa})}(x)L_{n-1}^{(2\gamma_{\kappa})}(x')
\right. 
\nonumber \\
&+& \sum_{n=1}^{\infty}\frac{(n+2\gamma_{\kappa})n!}
{N_{n\kappa}(N_{n\kappa}+\kappa)\Gamma(n+2\gamma_{\kappa})}
L_{n-1}^{(2\gamma_{\kappa})}(x)L_{n-1}^{(2\gamma_{\kappa})}(x')
\nonumber \\
&-& \sum_{n=0}^{\infty}\frac{(N_{n\kappa}-\kappa)n!}
{N_{n\kappa}\Gamma(n+2\gamma_{\kappa}+1)}
L_{n}^{(2\gamma_{\kappa})}(x)L_{n}^{(2\gamma_{\kappa})}(x')
\nonumber \\
&-& \left.
\sum_{n=0}^{\infty}\frac{(N_{n\kappa}+\kappa)n!}
{N_{n\kappa}\Gamma(n+2\gamma_{\kappa}+1)}
L_{n}^{(2\gamma_{\kappa})}(x)L_{n}^{(2\gamma_{\kappa})}(x')
\right] 
\nonumber \\*[5mm]
&=& \frac{\varepsilon}{4} x^{\gamma_{\kappa}}x'^{\gamma_{\kappa}}
{\rm e}^{-(x+x')/2}\left[
\sum_{n=1}^{\infty}\frac{(N_{n\kappa}+\kappa)(n-1)!}
{N_{n\kappa}\Gamma(n+2\gamma_{\kappa})}
L_{n-1}^{(2\gamma_{\kappa})}(x)L_{n-1}^{(2\gamma_{\kappa})}(x')
\right.
\nonumber \\
&+& \sum_{n=1}^{\infty} \frac{(N_{n\kappa}-\kappa)(n-1)!}
{N_{n\kappa}\Gamma(n+2\gamma_{\kappa})}
L_{n-1}^{(2\gamma_{\kappa})}(x)L_{n-1}^{(2\gamma_{\kappa})}(x')
\nonumber \\
&-& \sum_{n=0}^{\infty} \frac{(N_{n\kappa}-\kappa)n!}
{N_{n\kappa}\Gamma(n+2\gamma_{\kappa}+1)}
L_{n}^{(2\gamma_{\kappa})}(x)L_{n}^{(2\gamma_{\kappa})}(x')
\nonumber \\
&-& \left. \sum_{n=0}^{\infty} \frac{(N_{n\kappa}+\kappa)n!}
{N_{n\kappa}\Gamma(n+2\gamma_{\kappa}+1)}
L_{n}^{(2\gamma_{\kappa})}(x)L_{n}^{(2\gamma_{\kappa})}(x')
\right]
\nonumber \\*[5mm]
&=& \frac{\varepsilon}{2} x^{\gamma_{\kappa}}x'^{\gamma_{\kappa}}
{\rm e}^{-(x+x')/2}\left[
\sum_{n=1}^{\infty}\frac{(n-1)!}{\Gamma(n+2\gamma_{\kappa})}
L_{n-1}^{(2\gamma_{\kappa})}(x)L_{n-1}^{(2\gamma_{\kappa})}(x')
\right.
\nonumber \\
&-& \left.\sum_{n=0}^{\infty} \frac{n!}{\Gamma(n+2\gamma_{\kappa}+1)}
L_{n}^{(2\gamma_{\kappa})}(x)L_{n}^{(2\gamma_{\kappa})}(x') \right]
\nonumber \\*[5mm]
&=& \frac{\varepsilon}{2} x^{\gamma_{\kappa}}x'^{\gamma_{\kappa}}
{\rm e}^{-(x+x')/2}\left[
\sum_{n=0}^{\infty}\frac{n!}{\Gamma(n+2\gamma_{\kappa}+1)}
L_{n}^{(2\gamma_{\kappa})}(x)L_{n}^{(2\gamma_{\kappa})}(x')
\right.
\nonumber \\
&-& \left.\sum_{n=0}^{\infty} \frac{n!}{\Gamma(n+2\gamma_{\kappa}+1)}
L_{n}^{(2\gamma_{\kappa})}(x)L_{n}^{(2\gamma_{\kappa})}(x') \right]=0,
\label{13}
\end{eqnarray}
where the last equality is obvious. This completes the proof of
correctness of the closure relation (\ref{1}).

Consider now the closure relation (\ref{2}). It is equivalent to four
relations
\begin{equation}
\frac{Z}{x}\sum_{n=-\infty}^{\infty}
\mu_{n\kappa}S_{n\kappa}(x)S_{n\kappa}(x')=\delta(x-x'),
\label{14}
\end{equation}
\begin{equation}
-\frac{Z}{x}\sum_{n=-\infty}^{\infty}
\mu_{n\kappa}^{-1}T_{n\kappa}(x)T_{n\kappa}(x')=\delta(x-x'),
\label{15}
\end{equation}
\begin{equation}
\frac{Z}{x}\sum_{n=-\infty}^{\infty}
S_{n\kappa}(x)T_{n\kappa}(x')=0
\label{16}
\end{equation}
and
\begin{equation}
-\frac{Z}{x}\sum_{n=-\infty}^{\infty}
T_{n\kappa}(x)S_{n\kappa}(x')=0.
\label{17}
\end{equation}
We observe that the `off-diagonal' relations (\ref{16}) and
(\ref{17}) follow immediately from the relations (\ref{9}) and
(\ref{10}) and therefore it is sufficient to prove the `diagonal'
relations (\ref{14}) and (\ref{15}), a task that is a little bit
more difficult than proofs of the relations (\ref{7}) and (\ref{8}).
Upon transforming the left-hand side of equation (\ref{14}) we have
\begin{eqnarray}
&& \frac{Z}{x} \sum_{n=-\infty}^{\infty} \mu_{n\kappa}
S_{n\kappa}(x)S_{n\kappa}(x')
\nonumber \\*[5mm]
&=& \frac{Z}{x} \sum_{n=-\infty}^{\infty}\frac{\varepsilon}{\alpha Z}
(|n|+\gamma_{\kappa}\pm N_{n\kappa})
\frac{\alpha(|n|+2\gamma_{\kappa})|n|!}{2\varepsilon N_{n\kappa}
(N_{n\kappa}\mp\kappa)\Gamma(|n|+2\gamma_{\kappa})}
x^{\gamma_{\kappa}}x'^{\gamma_{\kappa}}{\rm e}^{-(x+x')/2}
\nonumber \\
&\times& \left[L_{|n|-1}^{(2\gamma_{\kappa})}(x)
+\frac{\kappa\mp N_{n\kappa}}{|n|+2\gamma_{\kappa}}
L_{|n|}^{(2\gamma_{\kappa})}(x)\right]
\left[L_{|n|-1}^{(2\gamma_{\kappa})}(x')
+\frac{\kappa\mp N_{n\kappa}}{|n|+2\gamma_{\kappa}}
L_{|n|}^{(2\gamma_{\kappa})}(x')\right]
\nonumber \\*[5mm]
&=& \frac{1}{2}x^{\gamma_{\kappa}-1}x'^{\gamma_{\kappa}}
{\rm e}^{-(x+x')/2} \sum_{n=-\infty}^{\infty}
\frac{(|n|+\gamma_{\kappa}\pm N_{n\kappa})(|n|+2\gamma_{\kappa})|n|!}
{N_{n\kappa}(N_{n\kappa}\mp\kappa)\Gamma(|n|+2\gamma_{\kappa})}
\nonumber \\
&\times& \left[L_{|n|-1}^{(2\gamma_{\kappa})}(x)
+\frac{\kappa\mp N_{n\kappa}}{|n|+2\gamma_{\kappa}}
L_{|n|}^{(2\gamma_{\kappa})}(x)\right]
\left[L_{|n|-1}^{(2\gamma_{\kappa})}(x')
+\frac{\kappa\mp N_{n\kappa}}{|n|+2\gamma_{\kappa}}
L_{|n|}^{(2\gamma_{\kappa})}(x')\right]
\nonumber \\*[5mm]
&=& \frac{1}{2}x^{\gamma_{\kappa}-1}x'^{\gamma_{\kappa}}
{\rm e}^{-(x+x')/2} \left[\sum_{n=-\infty}^{\infty}
\frac{(|n|+\gamma_{\kappa}\pm N_{n\kappa})(|n|+2\gamma_{\kappa})|n|!}
{N_{n\kappa}(N_{n\kappa}\mp\kappa)\Gamma(|n|+2\gamma_{\kappa})}
L_{|n|-1}^{(2\gamma_{\kappa})}(x)L_{|n|-1}^{(2\gamma_{\kappa})}(x')
\right.
\nonumber \\
&+& \sum_{n=-\infty}^{\infty}
\frac{(|n|+\gamma_{\kappa}\pm N_{n\kappa})(|n|+2\gamma_{\kappa})|n|!}
{N_{n\kappa}(N_{n\kappa}\mp\kappa)\Gamma(|n|+2\gamma_{\kappa})}
\frac{\kappa\mp N_{n\kappa}}{|n|+2\gamma_{\kappa}}
L_{|n|-1}^{(2\gamma_{\kappa})}(x)L_{|n|}^{(2\gamma_{\kappa})}(x')
\nonumber \\
&+& \sum_{n=-\infty}^{\infty}
\frac{(|n|+\gamma_{\kappa}\pm N_{n\kappa})(|n|+2\gamma_{\kappa})|n|!}
{N_{n\kappa}(N_{n\kappa}\mp\kappa)\Gamma(|n|+2\gamma_{\kappa})}
\frac{\kappa\mp N_{n\kappa}}{|n|+2\gamma_{\kappa}}
L_{|n|}^{(2\gamma_{\kappa})}(x)L_{|n|-1}^{(2\gamma_{\kappa})}(x')
\nonumber \\
&+& \left. \sum_{n=-\infty}^{\infty}
\frac{(|n|+\gamma_{\kappa}\pm N_{n\kappa})(|n|+2\gamma_{\kappa})|n|!}
{N_{n\kappa}(N_{n\kappa}\mp\kappa)\Gamma(|n|+2\gamma_{\kappa})}
\left(\frac{\kappa\mp N_{n\kappa}}{|n|+2\gamma_{\kappa}}\right)^{2}
L_{|n|}^{(2\gamma_{\kappa})}(x)L_{|n|}^{(2\gamma_{\kappa})}(x')
\right]
\nonumber \\*[5mm]
&=& \frac{1}{2}x^{\gamma_{\kappa}-1}x'^{\gamma_{\kappa}}
{\rm e}^{-(x+x')/2} \left[\sum_{n=-\infty}^{\infty}
\frac{(|n|+\gamma_{\kappa}\pm N_{n\kappa})(|n|+2\gamma_{\kappa})|n|!}
{N_{n\kappa}(N_{n\kappa}\mp\kappa)\Gamma(|n|+2\gamma_{\kappa})}
L_{|n|-1}^{(2\gamma_{\kappa})}(x)L_{|n|-1}^{(2\gamma_{\kappa})}(x')
\right.
\nonumber \\
&+& \sum_{n=-\infty}^{\infty}
\frac{\mp(|n|+\gamma_{\kappa}\pm N_{n\kappa})|n|!}
{N_{n\kappa}\Gamma(|n|+2\gamma_{\kappa})}
L_{|n|-1}^{(2\gamma_{\kappa})}(x)L_{|n|}^{(2\gamma_{\kappa})}(x')
\nonumber \\
&+& \sum_{n=-\infty}^{\infty}
\frac{\mp(|n|+\gamma_{\kappa}\pm N_{n\kappa})|n|!}
{N_{n\kappa}\Gamma(|n|+2\gamma_{\kappa})}
L_{|n|}^{(2\gamma_{\kappa})}(x)L_{|n|-1}^{(2\gamma_{\kappa})}(x')
\nonumber \\
&+& \left. \sum_{n=-\infty}^{\infty}
\frac{(|n|+\gamma_{\kappa}\pm N_{n\kappa})(N_{n\kappa}\mp\kappa)|n|!}
{N_{n\kappa}\Gamma(|n|+2\gamma_{\kappa}+1)}
L_{|n|}^{(2\gamma_{\kappa})}(x)L_{|n|}^{(2\gamma_{\kappa})}(x')
\right]
\nonumber \\*[5mm]
&=& \frac{1}{2}x^{\gamma_{\kappa}-1}x'^{\gamma_{\kappa}}
{\rm e}^{-(x+x')/2} \left[\sum_{n=1}^{\infty}
\frac{(|n|+\gamma_{\kappa}+N_{n\kappa})(|n|+2\gamma_{\kappa})|n|!}
{N_{n\kappa}(N_{n\kappa}-\kappa)\Gamma(|n|+2\gamma_{\kappa})}
L_{|n|-1}^{(2\gamma_{\kappa})}(x)L_{|n|-1}^{(2\gamma_{\kappa})}(x')
\right.
\nonumber \\
&+& \sum_{n=-\infty}^{-1}
\frac{(|n|+\gamma_{\kappa}-N_{n\kappa})(|n|+2\gamma_{\kappa})|n|!} 
{N_{n\kappa}(N_{n\kappa}+\kappa)\Gamma(|n|+2\gamma_{\kappa})}
L_{|n|-1}^{(2\gamma_{\kappa})}(x)L_{|n|-1}^{(2\gamma_{\kappa})}(x')
\nonumber \\
&-& \sum_{n=1}^{\infty}
\frac{(|n|+\gamma_{\kappa}+N_{n\kappa})|n|!}
{N_{n\kappa}\Gamma(|n|+2\gamma_{\kappa})}
L_{|n|-1}^{(2\gamma_{\kappa})}(x)L_{|n|}^{(2\gamma_{\kappa})}(x')
\nonumber \\
&+& \sum_{n=-\infty}^{-1}
\frac{(|n|+\gamma_{\kappa}-N_{n\kappa})|n|!}
{N_{n\kappa}\Gamma(|n|+2\gamma_{\kappa})}
L_{|n|-1}^{(2\gamma_{\kappa})}(x)L_{|n|}^{(2\gamma_{\kappa})}(x')
\nonumber \\
&-& \sum_{n=1}^{\infty}
\frac{(|n|+\gamma_{\kappa}+N_{n\kappa})|n|!}
{N_{n\kappa}\Gamma(|n|+2\gamma_{\kappa})}
L_{|n|}^{(2\gamma_{\kappa})}(x)L_{|n|-1}^{(2\gamma_{\kappa})}(x')
\nonumber \\
&+& \sum_{n=-\infty}^{-1}
\frac{(|n|+\gamma_{\kappa}-N_{n\kappa})|n|!}
{N_{n\kappa}\Gamma(|n|+2\gamma_{\kappa})}
L_{|n|}^{(2\gamma_{\kappa})}(x)L_{|n|-1}^{(2\gamma_{\kappa})}(x')
\nonumber \\
&+& \sum_{n=0}^{\infty}
\frac{(|n|+\gamma_{\kappa}+N_{n\kappa})(N_{n\kappa}-\kappa)|n|!}
{N_{n\kappa}\Gamma(|n|+2\gamma_{\kappa}+1)}
L_{|n|}^{(2\gamma_{\kappa})}(x)L_{|n|}^{(2\gamma_{\kappa})}(x')
\nonumber \\
&+& \left. \sum_{n=-\infty}^{0}
\frac{(|n|+\gamma_{\kappa}-N_{n\kappa})(N_{n\kappa}+\kappa)|n|!}
{N_{n\kappa}\Gamma(|n|+2\gamma_{\kappa}+1)}
L_{|n|}^{(2\gamma_{\kappa})}(x)L_{|n|}^{(2\gamma_{\kappa})}(x')
\right]
\nonumber \\*[5mm]
&=& \frac{1}{2}x^{\gamma_{\kappa}-1}x'^{\gamma_{\kappa}}
{\rm e}^{-(x+x')/2} \left[\sum_{n=1}^{\infty}
\frac{(n+\gamma_{\kappa}+N_{n\kappa})(n+2\gamma_{\kappa})n!}
{N_{n\kappa}(N_{n\kappa}-\kappa)\Gamma(n+2\gamma_{\kappa})}
L_{n-1}^{(2\gamma_{\kappa})}(x)L_{n-1}^{(2\gamma_{\kappa})}(x')
\right.
\nonumber \\
&+& \sum_{n=1}^{\infty}
\frac{(n+\gamma_{\kappa}-N_{n\kappa})(n+2\gamma_{\kappa})n!} 
{N_{n\kappa}(N_{n\kappa}+\kappa)\Gamma(n+2\gamma_{\kappa})}
L_{n-1}^{(2\gamma_{\kappa})}(x)L_{n-1}^{(2\gamma_{\kappa})}(x')
\nonumber \\
&-& \sum_{n=1}^{\infty}
\frac{(n+\gamma_{\kappa}+N_{n\kappa})n!}
{N_{n\kappa}\Gamma(n+2\gamma_{\kappa})}
L_{n-1}^{(2\gamma_{\kappa})}(x)L_{n}^{(2\gamma_{\kappa})}(x')
\nonumber \\
&+& \sum_{n=1}^{\infty}
\frac{(n+\gamma_{\kappa}-N_{n\kappa})n!}
{N_{n\kappa}\Gamma(n+2\gamma_{\kappa})}
L_{n-1}^{(2\gamma_{\kappa})}(x)L_{n}^{(2\gamma_{\kappa})}(x')
\nonumber \\
&-& \sum_{n=1}^{\infty}
\frac{(n+\gamma_{\kappa}+N_{n\kappa})n!}
{N_{n\kappa}\Gamma(n+2\gamma_{\kappa})}
L_{n}^{(2\gamma_{\kappa})}(x)L_{n-1}^{(2\gamma_{\kappa})}(x')
\nonumber \\
&+& \sum_{n=1}^{\infty}
\frac{(n+\gamma_{\kappa}-N_{n\kappa})n!}
{N_{n\kappa}\Gamma(n+2\gamma_{\kappa})}
L_{n}^{(2\gamma_{\kappa})}(x)L_{n-1}^{(2\gamma_{\kappa})}(x')
\nonumber \\
&+& \sum_{n=0}^{\infty}
\frac{(n+\gamma_{\kappa}+N_{n\kappa})(N_{n\kappa}-\kappa)n!}
{N_{n\kappa}\Gamma(n+2\gamma_{\kappa}+1)}
L_{n}^{(2\gamma_{\kappa})}(x)L_{n}^{(2\gamma_{\kappa})}(x')
\nonumber \\
&+& \left. \sum_{n=0}^{\infty}
\frac{(n+\gamma_{\kappa}-N_{n\kappa})(N_{n\kappa}+\kappa)n!}
{N_{n\kappa}\Gamma(n+2\gamma_{\kappa}+1)}
L_{n}^{(2\gamma_{\kappa})}(x)L_{n}^{(2\gamma_{\kappa})}(x')
\right]
\nonumber \\*[5mm]
&=& \frac{1}{2}x^{\gamma_{\kappa}-1}x'^{\gamma_{\kappa}}
{\rm e}^{-(x+x')/2} \left[\sum_{n=1}^{\infty}
\frac{(n+\gamma_{\kappa}+N_{n\kappa})(N_{n\kappa}+\kappa)(n-1)!}
{N_{n\kappa}\Gamma(n+2\gamma_{\kappa})}
L_{n-1}^{(2\gamma_{\kappa})}(x)L_{n-1}^{(2\gamma_{\kappa})}(x')
\right.
\nonumber \\
&+& \sum_{n=1}^{\infty}
\frac{(n+\gamma_{\kappa}-N_{n\kappa})(N_{n\kappa}-\kappa)(n-1)!} 
{N_{n\kappa}\Gamma(n+2\gamma_{\kappa})}
L_{n-1}^{(2\gamma_{\kappa})}(x)L_{n-1}^{(2\gamma_{\kappa})}(x')
\nonumber \\
&-& \sum_{n=1}^{\infty}
\frac{(n+\gamma_{\kappa}+N_{n\kappa})n!}
{N_{n\kappa}\Gamma(n+2\gamma_{\kappa})}
L_{n-1}^{(2\gamma_{\kappa})}(x)L_{n}^{(2\gamma_{\kappa})}(x')
\nonumber \\
&+& \sum_{n=1}^{\infty}
\frac{(n+\gamma_{\kappa}-N_{n\kappa})n!}
{N_{n\kappa}\Gamma(n+2\gamma_{\kappa})}
L_{n-1}^{(2\gamma_{\kappa})}(x)L_{n}^{(2\gamma_{\kappa})}(x')
\nonumber \\
&-& \sum_{n=1}^{\infty}
\frac{(n+\gamma_{\kappa}+N_{n\kappa})n!}
{N_{n\kappa}\Gamma(n+2\gamma_{\kappa})}
L_{n}^{(2\gamma_{\kappa})}(x)L_{n-1}^{(2\gamma_{\kappa})}(x')
\nonumber \\
&+& \sum_{n=1}^{\infty}
\frac{(n+\gamma_{\kappa}-N_{n\kappa})n!}
{N_{n\kappa}\Gamma(n+2\gamma_{\kappa})}
L_{n}^{(2\gamma_{\kappa})}(x)L_{n-1}^{(2\gamma_{\kappa})}(x')
\nonumber \\
&+& \sum_{n=0}^{\infty}
\frac{(n+\gamma_{\kappa}+N_{n\kappa})(N_{n\kappa}-\kappa)n!}
{N_{n\kappa}\Gamma(n+2\gamma_{\kappa}+1)}
L_{n}^{(2\gamma_{\kappa})}(x)L_{n}^{(2\gamma_{\kappa})}(x')
\nonumber \\
&+& \left. \sum_{n=0}^{\infty}
\frac{(n+\gamma_{\kappa}-N_{n\kappa})(N_{n\kappa}+\kappa)n!}
{N_{n\kappa}\Gamma(n+2\gamma_{\kappa}+1)}
L_{n}^{(2\gamma_{\kappa})}(x)L_{n}^{(2\gamma_{\kappa})}(x')
\right]
\nonumber \\*[5mm]
&=& x^{\gamma_{\kappa}-1}x'^{\gamma_{\kappa}}{\rm e}^{-(x+x')/2} 
\left[\sum_{n=1}^{\infty}
\frac{(n+\gamma_{\kappa}+\kappa)(n-1)!}{\Gamma(n+2\gamma_{\kappa})}
L_{n-1}^{(2\gamma_{\kappa})}(x)L_{n-1}^{(2\gamma_{\kappa})}(x')
\right.
\nonumber \\
&-& \sum_{n=1}^{\infty}\frac{n!}{\Gamma(n+2\gamma_{\kappa})}
L_{n-1}^{(2\gamma_{\kappa})}(x)L_{n}^{(2\gamma_{\kappa})}(x')
\nonumber \\
&-& \sum_{n=1}^{\infty}\frac{n!}{\Gamma(n+2\gamma_{\kappa})}
L_{n}^{(2\gamma_{\kappa})}(x)L_{n-1}^{(2\gamma_{\kappa})}(x')
\nonumber \\
&+& \left. \sum_{n=0}^{\infty}
\frac{(n+\gamma_{\kappa}-\kappa)n!}{\Gamma(n+2\gamma_{\kappa}+1)}
L_{n}^{(2\gamma_{\kappa})}(x)L_{n}^{(2\gamma_{\kappa})}(x')
\right]
\nonumber \\*[5mm]
&=& x^{\gamma_{\kappa}-1}x'^{\gamma_{\kappa}}{\rm e}^{-(x+x')/2} 
\left[\sum_{n=0}^{\infty}
\frac{(n+\gamma_{\kappa}+\kappa+1)n!}{\Gamma(n+2\gamma_{\kappa}+1)}
L_{n}^{(2\gamma_{\kappa})}(x)L_{n}^{(2\gamma_{\kappa})}(x')
\right.
\nonumber \\
&-& \sum_{n=1}^{\infty}\frac{n!}{\Gamma(n+2\gamma_{\kappa})}
L_{n-1}^{(2\gamma_{\kappa})}(x)L_{n}^{(2\gamma_{\kappa})}(x')
\nonumber \\
&-& \sum_{n=1}^{\infty}\frac{n!}{\Gamma(n+2\gamma_{\kappa})}
L_{n}^{(2\gamma_{\kappa})}(x)L_{n-1}^{(2\gamma_{\kappa})}(x')
\nonumber \\
&+& \left. \sum_{n=0}^{\infty}
\frac{(n+\gamma_{\kappa}-\kappa)n!}{\Gamma(n+2\gamma_{\kappa}+1)}
L_{n}^{(2\gamma_{\kappa})}(x)L_{n}^{(2\gamma_{\kappa})}(x')
\right]
\nonumber \\*[5mm]
&=& x^{\gamma_{\kappa}-1}x'^{\gamma_{\kappa}}{\rm e}^{-(x+x')/2} 
\left[\sum_{n=0}^{\infty}
\frac{(2n+2\gamma_{\kappa}+1)n!}{\Gamma(n+2\gamma_{\kappa}+1)}
L_{n}^{(2\gamma_{\kappa})}(x)L_{n}^{(2\gamma_{\kappa})}(x')
\right.
\nonumber \\
&-& \sum_{n=1}^{\infty}\frac{n!}{\Gamma(n+2\gamma_{\kappa})}
L_{n-1}^{(2\gamma_{\kappa})}(x)L_{n}^{(2\gamma_{\kappa})}(x')
\nonumber \\
&-& \left. \sum_{n=1}^{\infty}\frac{n!}{\Gamma(n+2\gamma_{\kappa})}
L_{n}^{(2\gamma_{\kappa})}(x)L_{n-1}^{(2\gamma_{\kappa})}(x')
\right].
\label{18}
\end{eqnarray}
Before we shall identify the obtained expression, we shall transform
to the same form the left-hand side of the relation (\ref{15})
\begin{eqnarray}
&& -\frac{Z}{x} \sum_{n=-\infty}^{\infty} \mu_{n\kappa}^{-1}
T_{n\kappa}(x)T_{n\kappa}(x')
\nonumber \\*[5mm]
&=& -\frac{Z}{x} \sum_{n=-\infty}^{\infty}
\frac{-\varepsilon^{-1}}{\alpha Z}(|n|+\gamma_{\kappa}\mp N_{n\kappa})
\frac{\alpha\varepsilon(|n|+2\gamma_{\kappa})|n|!}{2N_{n\kappa}
(N_{n\kappa}\mp\kappa)\Gamma(|n|+2\gamma_{\kappa})}
x^{\gamma_{\kappa}}x'^{\gamma_{\kappa}}{\rm e}^{-(x+x')/2}
\nonumber \\
&\times& \left[L_{|n|-1}^{(2\gamma_{\kappa})}(x)
-\frac{\kappa\mp N_{n\kappa}}{|n|+2\gamma_{\kappa}}
L_{|n|}^{(2\gamma_{\kappa})}(x)\right]
\left[L_{|n|-1}^{(2\gamma_{\kappa})}(x')
-\frac{\kappa\mp N_{n\kappa}}{|n|+2\gamma_{\kappa}}
L_{|n|}^{(2\gamma_{\kappa})}(x')\right]
\nonumber \\*[5mm]
&=& \frac{1}{2}x^{\gamma_{\kappa}-1}x'^{\gamma_{\kappa}}
{\rm e}^{-(x+x')/2} \sum_{n=-\infty}^{\infty}
\frac{(|n|+\gamma_{\kappa}\mp N_{n\kappa})(|n|+2\gamma_{\kappa})|n|!}
{N_{n\kappa}(N_{n\kappa}\mp\kappa)\Gamma(|n|+2\gamma_{\kappa})}
\nonumber \\
&\times& \left[L_{|n|-1}^{(2\gamma_{\kappa})}(x)
-\frac{\kappa\mp N_{n\kappa}}{|n|+2\gamma_{\kappa}}
L_{|n|}^{(2\gamma_{\kappa})}(x)\right]
\left[L_{|n|-1}^{(2\gamma_{\kappa})}(x')
-\frac{\kappa\mp N_{n\kappa}}{|n|+2\gamma_{\kappa}}
L_{|n|}^{(2\gamma_{\kappa})}(x')\right]
\nonumber \\*[5mm]
&=& \frac{1}{2}x^{\gamma_{\kappa}-1}x'^{\gamma_{\kappa}}
{\rm e}^{-(x+x')/2} \left[\sum_{n=-\infty}^{\infty}
\frac{(|n|+\gamma_{\kappa}\mp N_{n\kappa})(|n|+2\gamma_{\kappa})|n|!}
{N_{n\kappa}(N_{n\kappa}\mp\kappa)\Gamma(|n|+2\gamma_{\kappa})}
L_{|n|-1}^{(2\gamma_{\kappa})}(x)L_{|n|-1}^{(2\gamma_{\kappa})}(x')
\right.
\nonumber \\
&-& \sum_{n=-\infty}^{\infty}
\frac{(|n|+\gamma_{\kappa}\mp N_{n\kappa})(|n|+2\gamma_{\kappa})|n|!}
{N_{n\kappa}(N_{n\kappa}\mp\kappa)\Gamma(|n|+2\gamma_{\kappa})}
\frac{\kappa\mp N_{n\kappa}}{|n|+2\gamma_{\kappa}}
L_{|n|-1}^{(2\gamma_{\kappa})}(x)L_{|n|}^{(2\gamma_{\kappa})}(x')
\nonumber \\
&-& \sum_{n=-\infty}^{\infty}
\frac{(|n|+\gamma_{\kappa}\mp N_{n\kappa})(|n|+2\gamma_{\kappa})|n|!}
{N_{n\kappa}(N_{n\kappa}\mp\kappa)\Gamma(|n|+2\gamma_{\kappa})}
\frac{\kappa\mp N_{n\kappa}}{|n|+2\gamma_{\kappa}}
L_{|n|}^{(2\gamma_{\kappa})}(x)L_{|n|-1}^{(2\gamma_{\kappa})}(x')
\nonumber \\
&+& \left. \sum_{n=-\infty}^{\infty}
\frac{(|n|+\gamma_{\kappa}\mp N_{n\kappa})(|n|+2\gamma_{\kappa})|n|!}
{N_{n\kappa}(N_{n\kappa}\mp\kappa)\Gamma(|n|+2\gamma_{\kappa})}
\left(\frac{\kappa\mp N_{n\kappa}}{|n|+2\gamma_{\kappa}}\right)^{2}
L_{|n|}^{(2\gamma_{\kappa})}(x)L_{|n|}^{(2\gamma_{\kappa})}(x')
\right]
\nonumber \\*[5mm]
&=& \frac{1}{2}x^{\gamma_{\kappa}-1}x'^{\gamma_{\kappa}}
{\rm e}^{-(x+x')/2} \left[\sum_{n=-\infty}^{\infty}
\frac{(|n|+\gamma_{\kappa}\mp N_{n\kappa})(|n|+2\gamma_{\kappa})|n|!}
{N_{n\kappa}(N_{n\kappa}\mp\kappa)\Gamma(|n|+2\gamma_{\kappa})}
L_{|n|-1}^{(2\gamma_{\kappa})}(x)L_{|n|-1}^{(2\gamma_{\kappa})}(x')
\right.
\nonumber \\
&-& \sum_{n=-\infty}^{\infty}
\frac{\mp(|n|+\gamma_{\kappa}\mp N_{n\kappa})|n|!}
{N_{n\kappa}\Gamma(|n|+2\gamma_{\kappa})}
L_{|n|-1}^{(2\gamma_{\kappa})}(x)L_{|n|}^{(2\gamma_{\kappa})}(x')
\nonumber \\
&-& \sum_{n=-\infty}^{\infty}
\frac{\mp(|n|+\gamma_{\kappa}\mp N_{n\kappa})|n|!}
{N_{n\kappa}\Gamma(|n|+2\gamma_{\kappa})}
L_{|n|}^{(2\gamma_{\kappa})}(x)L_{|n|-1}^{(2\gamma_{\kappa})}(x')
\nonumber \\
&+& \left. \sum_{n=-\infty}^{\infty}
\frac{(|n|+\gamma_{\kappa}\mp N_{n\kappa})(N_{n\kappa}\mp\kappa)|n|!}
{N_{n\kappa}\Gamma(|n|+2\gamma_{\kappa}+1)}
L_{|n|}^{(2\gamma_{\kappa})}(x)L_{|n|}^{(2\gamma_{\kappa})}(x')
\right]
\nonumber \\*[5mm]
&=& \frac{1}{2}x^{\gamma_{\kappa}-1}x'^{\gamma_{\kappa}}
{\rm e}^{-(x+x')/2} \left[\sum_{n=1}^{\infty}
\frac{(|n|+\gamma_{\kappa}-N_{n\kappa})(|n|+2\gamma_{\kappa})|n|!}
{N_{n\kappa}(N_{n\kappa}-\kappa)\Gamma(|n|+2\gamma_{\kappa})}
L_{|n|-1}^{(2\gamma_{\kappa})}(x)L_{|n|-1}^{(2\gamma_{\kappa})}(x')
\right.
\nonumber \\
&+& \sum_{n=-\infty}^{-1}
\frac{(|n|+\gamma_{\kappa}+N_{n\kappa})(|n|+2\gamma_{\kappa})|n|!} 
{N_{n\kappa}(N_{n\kappa}+\kappa)\Gamma(|n|+2\gamma_{\kappa})}
L_{|n|-1}^{(2\gamma_{\kappa})}(x)L_{|n|-1}^{(2\gamma_{\kappa})}(x')
\nonumber \\
&+& \sum_{n=1}^{\infty}
\frac{(|n|+\gamma_{\kappa}-N_{n\kappa})|n|!}
{N_{n\kappa}\Gamma(|n|+2\gamma_{\kappa})}
L_{|n|-1}^{(2\gamma_{\kappa})}(x)L_{|n|}^{(2\gamma_{\kappa})}(x')
\nonumber \\
&-& \sum_{n=-\infty}^{-1}
\frac{(|n|+\gamma_{\kappa}+N_{n\kappa})|n|!}
{N_{n\kappa}\Gamma(|n|+2\gamma_{\kappa})}
L_{|n|-1}^{(2\gamma_{\kappa})}(x)L_{|n|}^{(2\gamma_{\kappa})}(x')
\nonumber \\
&+& \sum_{n=1}^{\infty}
\frac{(|n|+\gamma_{\kappa}-N_{n\kappa})|n|!}
{N_{n\kappa}\Gamma(|n|+2\gamma_{\kappa})}
L_{|n|}^{(2\gamma_{\kappa})}(x)L_{|n|-1}^{(2\gamma_{\kappa})}(x')
\nonumber \\
&-& \sum_{n=-\infty}^{-1}
\frac{(|n|+\gamma_{\kappa}+N_{n\kappa})|n|!}
{N_{n\kappa}\Gamma(|n|+2\gamma_{\kappa})}
L_{|n|}^{(2\gamma_{\kappa})}(x)L_{|n|-1}^{(2\gamma_{\kappa})}(x')
\nonumber \\
&+& \sum_{n=0}^{\infty}
\frac{(|n|+\gamma_{\kappa}-N_{n\kappa})(N_{n\kappa}-\kappa)|n|!}
{N_{n\kappa}\Gamma(|n|+2\gamma_{\kappa}+1)}
L_{|n|}^{(2\gamma_{\kappa})}(x)L_{|n|}^{(2\gamma_{\kappa})}(x')
\nonumber \\
&+& \left. \sum_{n=-\infty}^{0}
\frac{(|n|+\gamma_{\kappa}+N_{n\kappa})(N_{n\kappa}+\kappa)|n|!}
{N_{n\kappa}\Gamma(|n|+2\gamma_{\kappa}+1)}
L_{|n|}^{(2\gamma_{\kappa})}(x)L_{|n|}^{(2\gamma_{\kappa})}(x')
\right]
\nonumber \\*[5mm]
&=& \frac{1}{2}x^{\gamma_{\kappa}-1}x'^{\gamma_{\kappa}}
{\rm e}^{-(x+x')/2} \left[\sum_{n=1}^{\infty}
\frac{(n+\gamma_{\kappa}-N_{n\kappa})(n+2\gamma_{\kappa})n!}
{N_{n\kappa}(N_{n\kappa}-\kappa)\Gamma(n+2\gamma_{\kappa})}
L_{n-1}^{(2\gamma_{\kappa})}(x)L_{n-1}^{(2\gamma_{\kappa})}(x')
\right.
\nonumber \\
&+& \sum_{n=1}^{\infty}
\frac{(n+\gamma_{\kappa}+N_{n\kappa})(n+2\gamma_{\kappa})n!} 
{N_{n\kappa}(N_{n\kappa}+\kappa)\Gamma(n+2\gamma_{\kappa})}
L_{n-1}^{(2\gamma_{\kappa})}(x)L_{n-1}^{(2\gamma_{\kappa})}(x')
\nonumber \\
&+& \sum_{n=1}^{\infty}
\frac{(n+\gamma_{\kappa}-N_{n\kappa})n!}
{N_{n\kappa}\Gamma(n+2\gamma_{\kappa})}
L_{n-1}^{(2\gamma_{\kappa})}(x)L_{n}^{(2\gamma_{\kappa})}(x')
\nonumber \\
&-& \sum_{n=1}^{\infty}
\frac{(n+\gamma_{\kappa}+N_{n\kappa})n!}
{N_{n\kappa}\Gamma(n+2\gamma_{\kappa})}
L_{n-1}^{(2\gamma_{\kappa})}(x)L_{n}^{(2\gamma_{\kappa})}(x')
\nonumber \\
&+& \sum_{n=1}^{\infty}
\frac{(n+\gamma_{\kappa}-N_{n\kappa})n!}
{N_{n\kappa}\Gamma(n+2\gamma_{\kappa})}
L_{n}^{(2\gamma_{\kappa})}(x)L_{n-1}^{(2\gamma_{\kappa})}(x')
\nonumber \\
&-& \sum_{n=1}^{\infty}
\frac{(n+\gamma_{\kappa}+N_{n\kappa})n!}
{N_{n\kappa}\Gamma(n+2\gamma_{\kappa})}
L_{n}^{(2\gamma_{\kappa})}(x)L_{n-1}^{(2\gamma_{\kappa})}(x')
\nonumber \\
&+& \sum_{n=0}^{\infty}
\frac{(n+\gamma_{\kappa}-N_{n\kappa})(N_{n\kappa}-\kappa)n!}
{N_{n\kappa}\Gamma(n+2\gamma_{\kappa}+1)}
L_{n}^{(2\gamma_{\kappa})}(x)L_{n}^{(2\gamma_{\kappa})}(x')
\nonumber \\
&+& \left. \sum_{n=0}^{\infty}
\frac{(n+\gamma_{\kappa}+N_{n\kappa})(N_{n\kappa}+\kappa)n!}
{N_{n\kappa}\Gamma(n+2\gamma_{\kappa}+1)}
L_{n}^{(2\gamma_{\kappa})}(x)L_{n}^{(2\gamma_{\kappa})}(x')
\right]
\nonumber \\*[5mm]
&=& \frac{1}{2}x^{\gamma_{\kappa}-1}x'^{\gamma_{\kappa}}
{\rm e}^{-(x+x')/2} \left[\sum_{n=1}^{\infty}
\frac{(n+\gamma_{\kappa}-N_{n\kappa})(N_{n\kappa}+\kappa)(n-1)!}
{N_{n\kappa}\Gamma(n+2\gamma_{\kappa})}
L_{n-1}^{(2\gamma_{\kappa})}(x)L_{n-1}^{(2\gamma_{\kappa})}(x')
\right.
\nonumber \\
&+& \sum_{n=1}^{\infty}
\frac{(n+\gamma_{\kappa}+N_{n\kappa})(N_{n\kappa}-\kappa)(n-1)!} 
{N_{n\kappa}\Gamma(n+2\gamma_{\kappa})}
L_{n-1}^{(2\gamma_{\kappa})}(x)L_{n-1}^{(2\gamma_{\kappa})}(x')
\nonumber \\
&+& \sum_{n=1}^{\infty}
\frac{(n+\gamma_{\kappa}-N_{n\kappa})n!}
{N_{n\kappa}\Gamma(n+2\gamma_{\kappa})}
L_{n-1}^{(2\gamma_{\kappa})}(x)L_{n}^{(2\gamma_{\kappa})}(x')
\nonumber \\
&-& \sum_{n=1}^{\infty}
\frac{(n+\gamma_{\kappa}+N_{n\kappa})n!}
{N_{n\kappa}\Gamma(n+2\gamma_{\kappa})}
L_{n-1}^{(2\gamma_{\kappa})}(x)L_{n}^{(2\gamma_{\kappa})}(x')
\nonumber \\
&+& \sum_{n=1}^{\infty}
\frac{(n+\gamma_{\kappa}-N_{n\kappa})n!}
{N_{n\kappa}\Gamma(n+2\gamma_{\kappa})}
L_{n}^{(2\gamma_{\kappa})}(x)L_{n-1}^{(2\gamma_{\kappa})}(x')
\nonumber \\
&-& \sum_{n=1}^{\infty}
\frac{(n+\gamma_{\kappa}+N_{n\kappa})n!}
{N_{n\kappa}\Gamma(n+2\gamma_{\kappa})}
L_{n}^{(2\gamma_{\kappa})}(x)L_{n-1}^{(2\gamma_{\kappa})}(x')
\nonumber \\
&+& \sum_{n=0}^{\infty}
\frac{(n+\gamma_{\kappa}-N_{n\kappa})(N_{n\kappa}-\kappa)n!}
{N_{n\kappa}\Gamma(n+2\gamma_{\kappa}+1)}
L_{n}^{(2\gamma_{\kappa})}(x)L_{n}^{(2\gamma_{\kappa})}(x')
\nonumber \\
&+& \left. \sum_{n=0}^{\infty}
\frac{(n+\gamma_{\kappa}+N_{n\kappa})(N_{n\kappa}+\kappa)n!}
{N_{n\kappa}\Gamma(n+2\gamma_{\kappa}+1)}
L_{n}^{(2\gamma_{\kappa})}(x)L_{n}^{(2\gamma_{\kappa})}(x')
\right]
\nonumber \\*[5mm]
&=& x^{\gamma_{\kappa}-1}x'^{\gamma_{\kappa}}{\rm e}^{-(x+x')/2} 
\left[\sum_{n=1}^{\infty}
\frac{(n+\gamma_{\kappa}-\kappa)(n-1)!}{\Gamma(n+2\gamma_{\kappa})}
L_{n-1}^{(2\gamma_{\kappa})}(x)L_{n-1}^{(2\gamma_{\kappa})}(x')
\right.
\nonumber \\
&-& \sum_{n=1}^{\infty}\frac{n!}{\Gamma(n+2\gamma_{\kappa})}
L_{n-1}^{(2\gamma_{\kappa})}(x)L_{n}^{(2\gamma_{\kappa})}(x')
\nonumber \\
&-& \sum_{n=1}^{\infty}\frac{n!}{\Gamma(n+2\gamma_{\kappa})}
L_{n}^{(2\gamma_{\kappa})}(x)L_{n-1}^{(2\gamma_{\kappa})}(x')
\nonumber \\
&+& \left. \sum_{n=0}^{\infty}
\frac{(n+\gamma_{\kappa}+\kappa)n!}{\Gamma(n+2\gamma_{\kappa}+1)}
L_{n}^{(2\gamma_{\kappa})}(x)L_{n}^{(2\gamma_{\kappa})}(x')
\right]
\nonumber \\*[5mm]
&=& x^{\gamma_{\kappa}-1}x'^{\gamma_{\kappa}}{\rm e}^{-(x+x')/2} 
\left[\sum_{n=0}^{\infty}
\frac{(n+\gamma_{\kappa}-\kappa+1)n!}{\Gamma(n+2\gamma_{\kappa}+1)}
L_{n}^{(2\gamma_{\kappa})}(x)L_{n}^{(2\gamma_{\kappa})}(x')
\right.
\nonumber \\
&-& \sum_{n=1}^{\infty}\frac{n!}{\Gamma(n+2\gamma_{\kappa})}
L_{n-1}^{(2\gamma_{\kappa})}(x)L_{n}^{(2\gamma_{\kappa})}(x')
\nonumber \\
&-& \sum_{n=1}^{\infty}\frac{n!}{\Gamma(n+2\gamma_{\kappa})}
L_{n}^{(2\gamma_{\kappa})}(x)L_{n-1}^{(2\gamma_{\kappa})}(x')
\nonumber \\
&+& \left. \sum_{n=0}^{\infty}
\frac{(n+\gamma_{\kappa}+\kappa)n!}{\Gamma(n+2\gamma_{\kappa}+1)}
L_{n}^{(2\gamma_{\kappa})}(x)L_{n}^{(2\gamma_{\kappa})}(x')
\right]
\nonumber \\*[5mm]
&=& x^{\gamma_{\kappa}-1}x'^{\gamma_{\kappa}}{\rm e}^{-(x+x')/2} 
\left[\sum_{n=0}^{\infty}
\frac{(2n+2\gamma_{\kappa}+1)n!}{\Gamma(n+2\gamma_{\kappa}+1)}
L_{n}^{(2\gamma_{\kappa})}(x)L_{n}^{(2\gamma_{\kappa})}(x')
\right.
\nonumber \\
&-& \sum_{n=1}^{\infty}\frac{n!}{\Gamma(n+2\gamma_{\kappa})}
L_{n-1}^{(2\gamma_{\kappa})}(x)L_{n}^{(2\gamma_{\kappa})}(x')
\nonumber \\
&-& \left. \sum_{n=1}^{\infty}\frac{n!}{\Gamma(n+2\gamma_{\kappa})}
L_{n}^{(2\gamma_{\kappa})}(x)L_{n-1}^{(2\gamma_{\kappa})}(x')
\right].
\label{19}
\end{eqnarray}
Consider now the closure relation (\ref{6}) in the case
$\alpha=2\gamma_{\kappa}-1$. Upon utilizing the recurrence relation 
\begin{equation}
L_{n}^{(\alpha)}(x)=L_{n}^{(\alpha+1)}(x)-L_{n-1}^{(\alpha+1)}(x)
\label{20}
\end{equation}
we may transform it as follows
\begin{eqnarray}
&&\delta(x-x')=x^{\gamma_{\kappa}-1/2}x'^{\gamma_{\kappa}-1/2}
{\rm e}^{-(x+x')/2}
\sum_{n=0}^{\infty}\frac{n!}{\Gamma(n+2\gamma_{\kappa})}
L_{n}^{(2\gamma_{\kappa}-1)}(x)L_{n}^{(2\gamma_{\kappa}-1)}(x')
\nonumber \\*[5mm]
&&=x^{\gamma_{\kappa}-1/2}x'^{\gamma_{\kappa}-1/2}{\rm e}^{-(x+x')/2}
\sum_{n=0}^{\infty}\frac{n!}{\Gamma(n+2\gamma_{\kappa})}
\left[L_{n}^{(2\gamma_{\kappa})}(x)-
L_{n-1}^{(2\gamma_{\kappa})}(x)\right]
\left[L_{n}^{(2\gamma_{\kappa})}(x')-
L_{n-1}^{(2\gamma_{\kappa})}(x')\right]
\nonumber \\*[5mm]
&&=x^{\gamma_{\kappa}-1/2}x'^{\gamma_{\kappa}-1/2}{\rm e}^{-(x+x')/2}
\left[\sum_{n=0}^{\infty}\frac{n!}{\Gamma(n+2\gamma_{\kappa})}
L_{n}^{(2\gamma_{\kappa})}(x)L_{n}^{(2\gamma_{\kappa})}(x')\right.
\nonumber \\
&&-\sum_{n=0}^{\infty}\frac{n!}{\Gamma(n+2\gamma_{\kappa})}
L_{n}^{(2\gamma_{\kappa})}(x)L_{n-1}^{(2\gamma_{\kappa})}(x')
-\sum_{n=0}^{\infty}\frac{n!}{\Gamma(n+2\gamma_{\kappa})}
L_{n-1}^{(2\gamma_{\kappa})}(x)L_{n}^{(2\gamma_{\kappa})}(x')
\nonumber \\
&&\left.+\sum_{n=0}^{\infty}\frac{n!}{\Gamma(n+2\gamma_{\kappa})}
L_{n-1}^{(2\gamma_{\kappa})}(x)L_{n-1}^{(2\gamma_{\kappa})}(x')\right]
\nonumber \\*[5mm]
&&=x^{\gamma_{\kappa}-1/2}x'^{\gamma_{\kappa}-1/2}{\rm e}^{-(x+x')/2}
\left[\sum_{n=0}^{\infty}\frac{n!}{\Gamma(n+2\gamma_{\kappa})}
L_{n}^{(2\gamma_{\kappa})}(x)L_{n}^{(2\gamma_{\kappa})}(x')\right.
\nonumber \\
&&-\sum_{n=1}^{\infty}\frac{n!}{\Gamma(n+2\gamma_{\kappa})}
L_{n}^{(2\gamma_{\kappa})}(x)L_{n-1}^{(2\gamma_{\kappa})}(x')
-\sum_{n=1}^{\infty}\frac{n!}{\Gamma(n+2\gamma_{\kappa})}
L_{n-1}^{(2\gamma_{\kappa})}(x)L_{n}^{(2\gamma_{\kappa})}(x')
\nonumber \\
&&\left.+\sum_{n=0}^{\infty}
\frac{(n+1)!}{\Gamma(n+2\gamma_{\kappa}+1)}
L_{n}^{(2\gamma_{\kappa})}(x)L_{n}^{(2\gamma_{\kappa})}(x')\right]
\nonumber \\*[5mm]
&&=x^{\gamma_{\kappa}-1/2}x'^{\gamma_{\kappa}-1/2}{\rm e}^{-(x+x')/2}
\left[\sum_{n=0}^{\infty}
\frac{(2n+2\gamma_{\kappa}+1)n!}{\Gamma(n+2\gamma_{\kappa}+1)}
L_{n}^{(2\gamma_{\kappa})}(x)L_{n}^{(2\gamma_{\kappa})}(x')\right.
\nonumber \\
&&\left.-\sum_{n=1}^{\infty}\frac{n!}{\Gamma(n+2\gamma_{\kappa})}
L_{n}^{(2\gamma_{\kappa})}(x)L_{n-1}^{(2\gamma_{\kappa})}(x')
-\sum_{n=1}^{\infty}\frac{n!}{\Gamma(n+2\gamma_{\kappa})}
L_{n-1}^{(2\gamma_{\kappa})}(x)L_{n}^{(2\gamma_{\kappa})}(x')\right].
\label{21}
\end{eqnarray}
Comparison of equations (\ref{18}), (\ref{19}) and (\ref{21}) yields
\begin{equation}
\frac{Z}{x}\sum_{n=-\infty}^{\infty}
\mu_{n\kappa}S_{n\kappa}(x)S_{n\kappa}(x')=
\sqrt{x'/x}\:\delta(x-x')=\delta(x-x')
\label{22}
\end{equation}
and
\begin{equation}
-\frac{Z}{x}\sum_{n=-\infty}^{\infty}
\mu_{n\kappa}^{-1}T_{n\kappa}(x)T_{n\kappa}(x')=
\sqrt{x'/x}\:\delta(x-x')=\delta(x-x')
\label{23}
\end{equation}
which completes the proofs of the relations (\ref{14}) and
(\ref{15}) and consequently of the closure relation (\ref{2}).
\section{\rm A transformed form of the radial Dirac-Coulomb Green
function}
In the Sturmian expansions of the radial Dirac-Coulomb Green function
(equations (134) and (135) of Ref.~\cite{Szmy97}), the elements
$g_{\kappa}^{(ij)}(r,r';E)$ have been expressed as infinite series
with summations running over all, positive and negative, integers. By
reducing terms with the same $|n|$ to a common denominator, the
expansions may be transformed to forms containing only summations
over non-negative integers. Definning
\begin{equation}
\epsilon=\frac{E}{mc^{2}},
\hspace*{1cm} \omega=\frac{\zeta\epsilon}{\sqrt{1-\epsilon^{2}}},
\label{24}
\end{equation}
one has
\begin{eqnarray*}
&& g_{\kappa}^{(11)}(r,r';E)
=\sum_{n=-\infty}^{\infty}\frac{\mu_{n\kappa}}{\mu_{n\kappa}-1}
S_{n\kappa}(x)S_{n\kappa}(x')
\nonumber \\*[5mm]
&=& \sum_{n=-\infty}^{\infty}
\frac{1}{1+\frac{|n|+\gamma_{\kappa}\mp N_{n\kappa}}
{\varepsilon\zeta}}\:\frac{\alpha(|n|+2\gamma_{\kappa})|n|!}
{2\varepsilon N_{n\kappa}(N_{n\kappa}\mp\kappa)
\Gamma(|n|+2\gamma_{\kappa})}x^{\gamma_{\kappa}}x'^{\gamma_{\kappa}}
{\rm e}^{-(x+x')/2}
\nonumber \\
&\times& \left[L_{|n|-1}^{(2\gamma_{\kappa})}(x)
+\frac{\kappa\mp N_{n\kappa}}{|n|+2\gamma_{\kappa}}
L_{|n|}^{(2\gamma_{\kappa})}(x)\right]
\left[L_{|n|-1}^{(2\gamma_{\kappa})}(x')
+\frac{\kappa\mp N_{n\kappa}}{|n|+2\gamma_{\kappa}}
L_{|n|}^{(2\gamma_{\kappa})}(x')\right]
\nonumber \\*[5mm]
&=& x^{\gamma_{\kappa}}x'^{\gamma_{\kappa}}
{\rm e}^{-(x+x')/2}\sum_{n=-\infty}^{\infty}
\frac{\alpha\zeta(|n|+2\gamma_{\kappa})|n|!}
{2N_{n\kappa}(N_{n\kappa}\mp\kappa)\Gamma(|n|+2\gamma_{\kappa})}
\frac{|n|+\gamma_{\kappa}+\varepsilon\zeta\pm N_{n\kappa}}
{(|n|+\gamma_{\kappa}+\varepsilon\zeta)^{2}-N_{n\kappa}^{2}}
\nonumber \\
&\times& \left[L_{|n|-1}^{(2\gamma_{\kappa})}(x)
+\frac{\kappa\mp N_{n\kappa}}{|n|+2\gamma_{\kappa}}
L_{|n|}^{(2\gamma_{\kappa})}(x)\right]
\left[L_{|n|-1}^{(2\gamma_{\kappa})}(x')
+\frac{\kappa\mp N_{n\kappa}}{|n|+2\gamma_{\kappa}}
L_{|n|}^{(2\gamma_{\kappa})}(x')\right]
\nonumber \\*[5mm]
&=& x^{\gamma_{\kappa}}x'^{\gamma_{\kappa}}
{\rm e}^{-(x+x')/2}\sum_{n=-\infty}^{\infty}
\frac{\alpha\zeta(|n|+2\gamma_{\kappa})|n|!}
{2N_{n\kappa}(N_{n\kappa}\mp\kappa)\Gamma(|n|+2\gamma_{\kappa})}
\frac{|n|+\gamma_{\kappa}+\varepsilon\zeta\pm N_{n\kappa}}
{2\varepsilon\zeta(|n|+\gamma_{\kappa}-\omega)}
\nonumber \\
&\times& \left[L_{|n|-1}^{(2\gamma_{\kappa})}(x)
+\frac{\kappa\mp N_{n\kappa}}{|n|+2\gamma_{\kappa}}
L_{|n|}^{(2\gamma_{\kappa})}(x)\right]
\left[L_{|n|-1}^{(2\gamma_{\kappa})}(x')
+\frac{\kappa\mp N_{n\kappa}}{|n|+2\gamma_{\kappa}}
L_{|n|}^{(2\gamma_{\kappa})}(x')\right]
\nonumber \\*[5mm]
&=& \frac{\alpha}{4\varepsilon}
x^{\gamma_{\kappa}}x'^{\gamma_{\kappa}}
{\rm e}^{-(x+x')/2}\sum_{n=-\infty}^{\infty}
\frac{(|n|+\gamma_{\kappa}+\varepsilon\zeta\pm N_{n\kappa})
(|n|+2\gamma_{\kappa})|n|!}{N_{n\kappa}(N_{n\kappa}\mp\kappa)
(|n|+\gamma_{\kappa}-\omega)\Gamma(|n|+2\gamma_{\kappa})}
\nonumber \\
&\times& \left[L_{|n|-1}^{(2\gamma_{\kappa})}(x)
+\frac{\kappa\mp N_{n\kappa}}{|n|+2\gamma_{\kappa}}
L_{|n|}^{(2\gamma_{\kappa})}(x)\right]
\left[L_{|n|-1}^{(2\gamma_{\kappa})}(x')
+\frac{\kappa\mp N_{n\kappa}}{|n|+2\gamma_{\kappa}}
L_{|n|}^{(2\gamma_{\kappa})}(x')\right]
\nonumber \\*[5mm]
&=& \frac{\alpha}{4\varepsilon}x^{\gamma_{\kappa}}
x'^{\gamma_{\kappa}}{\rm e}^{-(x+x')/2}\left[
\sum_{n=-\infty}^{\infty}
\frac{(|n|+\gamma_{\kappa}+\varepsilon\zeta\pm N_{n\kappa})
(|n|+2\gamma_{\kappa})|n|!}{N_{n\kappa}(N_{n\kappa}\mp\kappa)
(|n|+\gamma_{\kappa}-\omega)\Gamma(|n|+2\gamma_{\kappa})}
L_{|n|-1}^{(2\gamma_{\kappa})}(x)L_{|n|-1}^{(2\gamma_{\kappa})}(x')
\right.
\nonumber \\
&+& \sum_{n=-\infty}^{\infty}
\frac{(|n|+\gamma_{\kappa}+\varepsilon\zeta\pm N_{n\kappa})
(|n|+2\gamma_{\kappa})|n|!}{N_{n\kappa}(N_{n\kappa}\mp\kappa)
(|n|+\gamma_{\kappa}-\omega)\Gamma(|n|+2\gamma_{\kappa})}
\frac{\kappa\mp N_{n\kappa}}{|n|+2\gamma_{\kappa}}
L_{|n|-1}^{(2\gamma_{\kappa})}(x)L_{|n|}^{(2\gamma_{\kappa})}(x')
\nonumber \\
&+& \sum_{n=-\infty}^{\infty}
\frac{(|n|+\gamma_{\kappa}+\varepsilon\zeta\pm N_{n\kappa})
(|n|+2\gamma_{\kappa})|n|!}{N_{n\kappa}(N_{n\kappa}\mp\kappa)
(|n|+\gamma_{\kappa}-\omega)\Gamma(|n|+2\gamma_{\kappa})}
\frac{\kappa\mp N_{n\kappa}}{|n|+2\gamma_{\kappa}}
L_{|n|}^{(2\gamma_{\kappa})}(x)L_{|n|-1}^{(2\gamma_{\kappa})}(x')
\nonumber \\
&+& \left.\sum_{n=-\infty}^{\infty}
\frac{(|n|+\gamma_{\kappa}+\varepsilon\zeta\pm N_{n\kappa})
(|n|+2\gamma_{\kappa})|n|!}{N_{n\kappa}(N_{n\kappa}\mp\kappa)
(|n|+\gamma_{\kappa}-\omega)\Gamma(|n|+2\gamma_{\kappa})}
\left(\frac{\kappa\mp N_{n\kappa}}{|n|+2\gamma_{\kappa}}\right)^{2}
L_{|n|}^{(2\gamma_{\kappa})}(x)L_{|n|}^{(2\gamma_{\kappa})}(x')
\right]
\nonumber \\*[5mm]
&=& \frac{\alpha}{4\varepsilon}x^{\gamma_{\kappa}}
x'^{\gamma_{\kappa}}{\rm e}^{-(x+x')/2}\left[
\sum_{n=-\infty}^{\infty}
\frac{(|n|+\gamma_{\kappa}+\varepsilon\zeta\pm N_{n\kappa})
(|n|+2\gamma_{\kappa})|n|!}{N_{n\kappa}(N_{n\kappa}\mp\kappa)
(|n|+\gamma_{\kappa}-\omega)\Gamma(|n|+2\gamma_{\kappa})}
L_{|n|-1}^{(2\gamma_{\kappa})}(x)L_{|n|-1}^{(2\gamma_{\kappa})}(x')
\right.
\nonumber \\
&+& \sum_{n=-\infty}^{\infty}
\frac{\mp(|n|+\gamma_{\kappa}+\varepsilon\zeta\pm N_{n\kappa})|n|!}
{N_{n\kappa}(|n|+\gamma_{\kappa}-\omega)\Gamma(|n|+2\gamma_{\kappa})}
L_{|n|-1}^{(2\gamma_{\kappa})}(x)L_{|n|}^{(2\gamma_{\kappa})}(x')
\nonumber \\
&+& \sum_{n=-\infty}^{\infty}
\frac{\mp(|n|+\gamma_{\kappa}+\varepsilon\zeta\pm N_{n\kappa})|n|!}
{N_{n\kappa}(|n|+\gamma_{\kappa}-\omega)\Gamma(|n|+2\gamma_{\kappa})}
L_{|n|}^{(2\gamma_{\kappa})}(x)L_{|n|-1}^{(2\gamma_{\kappa})}(x')
\nonumber \\
&+& \left.\sum_{n=-\infty}^{\infty}
\frac{(|n|+\gamma_{\kappa}+\varepsilon\zeta\pm N_{n\kappa})
(N_{n\kappa}\mp\kappa)|n|!}{N_{n\kappa}
(|n|+\gamma_{\kappa}-\omega)\Gamma(|n|+2\gamma_{\kappa}+1)}
L_{|n|}^{(2\gamma_{\kappa})}(x)L_{|n|}^{(2\gamma_{\kappa})}(x')
\right]
\nonumber \\*[5mm]
&=& \frac{\alpha}{4\varepsilon}x^{\gamma_{\kappa}}
x'^{\gamma_{\kappa}}{\rm e}^{-(x+x')/2}\left[
\sum_{n=1}^{\infty}
\frac{(|n|+\gamma_{\kappa}+\varepsilon\zeta+N_{n\kappa})
(|n|+2\gamma_{\kappa})|n|!}{N_{n\kappa}(N_{n\kappa}-\kappa)
(|n|+\gamma_{\kappa}-\omega)\Gamma(|n|+2\gamma_{\kappa})}
L_{|n|-1}^{(2\gamma_{\kappa})}(x)L_{|n|-1}^{(2\gamma_{\kappa})}(x')
\right.
\nonumber \\
&+& \sum_{n=-\infty}^{-1}
\frac{(|n|+\gamma_{\kappa}+\varepsilon\zeta-N_{n\kappa})
(|n|+2\gamma_{\kappa})|n|!}{N_{n\kappa}(N_{n\kappa}+\kappa)
(|n|+\gamma_{\kappa}-\omega)\Gamma(|n|+2\gamma_{\kappa})}
L_{|n|-1}^{(2\gamma_{\kappa})}(x)L_{|n|-1}^{(2\gamma_{\kappa})}(x')
\nonumber \\
&-& \sum_{n=1}^{\infty}
\frac{(|n|+\gamma_{\kappa}+\varepsilon\zeta+N_{n\kappa})|n|!}
{N_{n\kappa}(|n|+\gamma_{\kappa}-\omega)\Gamma(|n|+2\gamma_{\kappa})}
L_{|n|-1}^{(2\gamma_{\kappa})}(x)L_{|n|}^{(2\gamma_{\kappa})}(x')
\nonumber \\
&+& \sum_{n=-\infty}^{-1}
\frac{(|n|+\gamma_{\kappa}+\varepsilon\zeta-N_{n\kappa})|n|!}
{N_{n\kappa}(|n|+\gamma_{\kappa}-\omega)\Gamma(|n|+2\gamma_{\kappa})}
L_{|n|-1}^{(2\gamma_{\kappa})}(x)L_{|n|}^{(2\gamma_{\kappa})}(x')
\nonumber \\
&-& \sum_{n=1}^{\infty}
\frac{(|n|+\gamma_{\kappa}+\varepsilon\zeta+N_{n\kappa})|n|!}
{N_{n\kappa}(|n|+\gamma_{\kappa}-\omega)\Gamma(|n|+2\gamma_{\kappa})}
L_{|n|}^{(2\gamma_{\kappa})}(x)L_{|n|-1}^{(2\gamma_{\kappa})}(x')
\nonumber \\
&+& \sum_{n=-\infty}^{-1}
\frac{(|n|+\gamma_{\kappa}+\varepsilon\zeta-N_{n\kappa})|n|!}
{N_{n\kappa}(|n|+\gamma_{\kappa}-\omega)\Gamma(|n|+2\gamma_{\kappa})}
L_{|n|}^{(2\gamma_{\kappa})}(x)L_{|n|-1}^{(2\gamma_{\kappa})}(x')
\nonumber \\
&+& \sum_{n=0}^{\infty}
\frac{(|n|+\gamma_{\kappa}+\varepsilon\zeta+N_{n\kappa})
(N_{n\kappa}-\kappa)|n|!}{N_{n\kappa}
(|n|+\gamma_{\kappa}-\omega)\Gamma(|n|+2\gamma_{\kappa}+1)}
L_{|n|}^{(2\gamma_{\kappa})}(x)L_{|n|}^{(2\gamma_{\kappa})}(x')
\nonumber \\
&+& \left.\sum_{n=-\infty}^{0}
\frac{(|n|+\gamma_{\kappa}+\varepsilon\zeta-N_{n\kappa})
(N_{n\kappa}+\kappa)|n|!}{N_{n\kappa}
(|n|+\gamma_{\kappa}-\omega)\Gamma(|n|+2\gamma_{\kappa}+1)}
L_{|n|}^{(2\gamma_{\kappa})}(x)L_{|n|}^{(2\gamma_{\kappa})}(x')
\right]
\nonumber \\*[5mm]
&=& \frac{\alpha}{4\varepsilon}x^{\gamma_{\kappa}}
x'^{\gamma_{\kappa}}{\rm e}^{-(x+x')/2}\left[
\sum_{n=1}^{\infty}
\frac{(n+\gamma_{\kappa}+\varepsilon\zeta+N_{n\kappa})
(n+2\gamma_{\kappa})n!}{N_{n\kappa}(N_{n\kappa}-\kappa)
(n+\gamma_{\kappa}-\omega)\Gamma(n+2\gamma_{\kappa})}
L_{n-1}^{(2\gamma_{\kappa})}(x)L_{n-1}^{(2\gamma_{\kappa})}(x')
\right.
\nonumber \\
&+& \sum_{n=1}^{\infty}
\frac{(n+\gamma_{\kappa}+\varepsilon\zeta-N_{n\kappa})
(n+2\gamma_{\kappa})n!}{N_{n\kappa}(N_{n\kappa}+\kappa)
(n+\gamma_{\kappa}-\omega)\Gamma(n+2\gamma_{\kappa})}
L_{n-1}^{(2\gamma_{\kappa})}(x)L_{n-1}^{(2\gamma_{\kappa})}(x')
\nonumber \\
&-& \sum_{n=1}^{\infty}
\frac{(n+\gamma_{\kappa}+\varepsilon\zeta+N_{n\kappa})n!}
{N_{n\kappa}(n+\gamma_{\kappa}-\omega)\Gamma(n+2\gamma_{\kappa})}
L_{n-1}^{(2\gamma_{\kappa})}(x)L_{n}^{(2\gamma_{\kappa})}(x')
\nonumber \\
&+& \sum_{n=1}^{\infty}
\frac{(n+\gamma_{\kappa}+\varepsilon\zeta-N_{n\kappa})n!}
{N_{n\kappa}(n+\gamma_{\kappa}-\omega)\Gamma(n+2\gamma_{\kappa})}
L_{n-1}^{(2\gamma_{\kappa})}(x)L_{n}^{(2\gamma_{\kappa})}(x')
\nonumber \\
&-& \sum_{n=1}^{\infty}
\frac{(n+\gamma_{\kappa}+\varepsilon\zeta+N_{n\kappa})n!}
{N_{n\kappa}(n+\gamma_{\kappa}-\omega)\Gamma(n+2\gamma_{\kappa})}
L_{n}^{(2\gamma_{\kappa})}(x)L_{n-1}^{(2\gamma_{\kappa})}(x')
\nonumber \\
&+& \sum_{n=1}^{\infty}
\frac{(n+\gamma_{\kappa}+\varepsilon\zeta-N_{n\kappa})n!}
{N_{n\kappa}(n+\gamma_{\kappa}-\omega)\Gamma(n+2\gamma_{\kappa})}
L_{n}^{(2\gamma_{\kappa})}(x)L_{n-1}^{(2\gamma_{\kappa})}(x')
\nonumber \\
&+& \sum_{n=0}^{\infty}
\frac{(n+\gamma_{\kappa}+\varepsilon\zeta+N_{n\kappa})
(N_{n\kappa}-\kappa)n!}{N_{n\kappa}
(n+\gamma_{\kappa}-\omega)\Gamma(n+2\gamma_{\kappa}+1)}
L_{n}^{(2\gamma_{\kappa})}(x)L_{n}^{(2\gamma_{\kappa})}(x')
\nonumber \\
&+& \left.\sum_{n=0}^{\infty}
\frac{(n+\gamma_{\kappa}+\varepsilon\zeta-N_{n\kappa})
(N_{n\kappa}+\kappa)n!}{N_{n\kappa}
(n+\gamma_{\kappa}-\omega)\Gamma(n+2\gamma_{\kappa}+1)}
L_{n}^{(2\gamma_{\kappa})}(x)L_{n}^{(2\gamma_{\kappa})}(x')
\right],
\end{eqnarray*}
\begin{eqnarray}
&=& \frac{\alpha}{4\varepsilon}x^{\gamma_{\kappa}}
x'^{\gamma_{\kappa}}{\rm e}^{-(x+x')/2}\left[
\sum_{n=1}^{\infty}\frac{(n+\gamma_{\kappa}+\varepsilon\zeta
+N_{n\kappa})(N_{n\kappa}+\kappa)(n-1)!}{N_{n\kappa}
(n+\gamma_{\kappa}-\omega)\Gamma(n+2\gamma_{\kappa})}
L_{n-1}^{(2\gamma_{\kappa})}(x)L_{n-1}^{(2\gamma_{\kappa})}(x')
\right.
\nonumber \\
&+& \sum_{n=1}^{\infty}\frac{(n+\gamma_{\kappa}+\varepsilon\zeta
-N_{n\kappa})(N_{n\kappa}-\kappa)(n-1)!}{N_{n\kappa}
(n+\gamma_{\kappa}-\omega)\Gamma(n+2\gamma_{\kappa})}
L_{n-1}^{(2\gamma_{\kappa})}(x)L_{n-1}^{(2\gamma_{\kappa})}(x')
\nonumber \\
&-& 2\sum_{n=1}^{\infty}\frac{n!}{(n+\gamma_{\kappa}-\omega)
\Gamma(n+2\gamma_{\kappa})}
L_{n-1}^{(2\gamma_{\kappa})}(x)L_{n}^{(2\gamma_{\kappa})}(x')
\nonumber \\
&-& 2\sum_{n=1}^{\infty}\frac{n!}{(n+\gamma_{\kappa}-\omega)
\Gamma(n+2\gamma_{\kappa})}
L_{n}^{(2\gamma_{\kappa})}(x)L_{n-1}^{(2\gamma_{\kappa})}(x')
\nonumber \\
&+& \sum_{n=0}^{\infty}
\frac{(n+\gamma_{\kappa}+\varepsilon\zeta+N_{n\kappa})
(N_{n\kappa}-\kappa)n!}{N_{n\kappa}(n+\gamma_{\kappa}-\omega)
\Gamma(n+2\gamma_{\kappa}+1)}
L_{n}^{(2\gamma_{\kappa})}(x)L_{n}^{(2\gamma_{\kappa})}(x')
\nonumber \\
&+& \left.\sum_{n=0}^{\infty}
\frac{(n+\gamma_{\kappa}+\varepsilon\zeta-N_{n\kappa})
(N_{n\kappa}+\kappa)n!}{N_{n\kappa}(n+\gamma_{\kappa}-\omega)
\Gamma(n+2\gamma_{\kappa}+1)}
L_{n}^{(2\gamma_{\kappa})}(x)L_{n}^{(2\gamma_{\kappa})}(x')
\right]
\nonumber \\*[5mm]
&=& \frac{\alpha}{2\varepsilon}x^{\gamma_{\kappa}}
x'^{\gamma_{\kappa}}{\rm e}^{-(x+x')/2}\left[
\sum_{n=1}^{\infty}\frac{(n+\gamma_{\kappa}+\varepsilon\zeta+\kappa)
(n-1)!}{(n+\gamma_{\kappa}-\omega)\Gamma(n+2\gamma_{\kappa})}
L_{n-1}^{(2\gamma_{\kappa})}(x)L_{n-1}^{(2\gamma_{\kappa})}(x')
\right.
\nonumber \\
&-& \sum_{n=0}^{\infty}\frac{n!}{(n+\gamma_{\kappa}-\omega)
\Gamma(n+2\gamma_{\kappa})}L_{n-1}^{(2\gamma_{\kappa})}(x)
L_{n}^{(2\gamma_{\kappa})}(x')
\nonumber \\
&-& \sum_{n=0}^{\infty}\frac{n!}{(n+\gamma_{\kappa}-\omega)
\Gamma(n+2\gamma_{\kappa})}L_{n}^{(2\gamma_{\kappa})}(x)
L_{n-1}^{(2\gamma_{\kappa})}(x')
\nonumber \\
&+& \left.
\sum_{n=0}^{\infty}\frac{(n+\gamma_{\kappa}+\varepsilon\zeta
-\kappa)n!}{(n+\gamma_{\kappa}-\omega)\Gamma(n+2\gamma_{\kappa}+1)}
L_{n}^{(2\gamma_{\kappa})}(x)L_{n}^{(2\gamma_{\kappa})}(x')\right]
\nonumber \\*[5mm]
&=& \frac{\alpha}{2\varepsilon}x^{\gamma_{\kappa}}
x'^{\gamma_{\kappa}}{\rm e}^{-(x+x')/2}\left[
\sum_{n=0}^{\infty}\frac{(n+\gamma_{\kappa}+\varepsilon\zeta
+\kappa+1)n!}{(n+\gamma_{\kappa}-\omega+1)
\Gamma(n+2\gamma_{\kappa}+1)}
L_{n}^{(2\gamma_{\kappa})}(x)L_{n}^{(2\gamma_{\kappa})}(x')
\right.
\nonumber \\
&-& \sum_{n=0}^{\infty}\frac{n!}{(n+\gamma_{\kappa}-\omega)
\Gamma(n+2\gamma_{\kappa})}L_{n-1}^{(2\gamma_{\kappa})}(x)
L_{n}^{(2\gamma_{\kappa})}(x')
\nonumber \\
&-& \sum_{n=0}^{\infty}\frac{n!}{(n+\gamma_{\kappa}-\omega)
\Gamma(n+2\gamma_{\kappa})}L_{n}^{(2\gamma_{\kappa})}(x)
L_{n-1}^{(2\gamma_{\kappa})}(x')
\nonumber \\
&+& \left.
\sum_{n=0}^{\infty}\frac{(n+\gamma_{\kappa}+\varepsilon\zeta
-\kappa)n!}{(n+\gamma_{\kappa}-\omega)\Gamma(n+2\gamma_{\kappa}+1)}
L_{n}^{(2\gamma_{\kappa})}(x)L_{n}^{(2\gamma_{\kappa})}(x')\right]
\nonumber \\*[5mm]
&=& \frac{\alpha}{2\varepsilon}x^{\gamma_{\kappa}}
x'^{\gamma_{\kappa}}{\rm e}^{-(x+x')/2}\left\{
\sum_{n=0}^{\infty}\frac{(n+\gamma_{\kappa}+\varepsilon\zeta
+\kappa+1)n!}{(n+\gamma_{\kappa}-\omega+1)
\Gamma(n+2\gamma_{\kappa}+1)}
L_{n}^{(2\gamma_{\kappa})}(x)L_{n}^{(2\gamma_{\kappa})}(x')
\right.
\nonumber \\
&-& \sum_{n=0}^{\infty}\frac{n!}{(n+\gamma_{\kappa}-\omega)
\Gamma(n+2\gamma_{\kappa})}\left[L_{n}^{(2\gamma_{\kappa})}(x)
-L_{n}^{(2\gamma_{\kappa}-1)}(x)\right]L_{n}^{(2\gamma_{\kappa})}(x')
\nonumber \\
&-& \sum_{n=0}^{\infty}\frac{n!}{(n+\gamma_{\kappa}-\omega)
\Gamma(n+2\gamma_{\kappa})}L_{n}^{(2\gamma_{\kappa})}(x)
\left[L_{n}^{(2\gamma_{\kappa})}(x')
-L_{n}^{(2\gamma_{\kappa}-1)}(x')\right]
\nonumber \\
&+& \left.
\sum_{n=0}^{\infty}\frac{(n+\gamma_{\kappa}+\varepsilon\zeta
-\kappa)n!}{(n+\gamma_{\kappa}-\omega)\Gamma(n+2\gamma_{\kappa}+1)}
L_{n}^{(2\gamma_{\kappa})}(x)L_{n}^{(2\gamma_{\kappa})}(x')\right\}
\nonumber \\*[5mm]
&=& \frac{\alpha}{2\varepsilon}x^{\gamma_{\kappa}}
x'^{\gamma_{\kappa}}{\rm e}^{-(x+x')/2}\left[
\sum_{n=0}^{\infty}\frac{(n+\gamma_{\kappa}+\varepsilon\zeta
+\kappa+1)n!}{(n+\gamma_{\kappa}-\omega+1)
\Gamma(n+2\gamma_{\kappa}+1)}
L_{n}^{(2\gamma_{\kappa})}(x)L_{n}^{(2\gamma_{\kappa})}(x')
\right.
\nonumber \\
&-& \sum_{n=0}^{\infty}\frac{n!}{(n+\gamma_{\kappa}-\omega)
\Gamma(n+2\gamma_{\kappa})}L_{n}^{(2\gamma_{\kappa})}(x)
L_{n}^{(2\gamma_{\kappa})}(x')
\nonumber \\
&+& \sum_{n=0}^{\infty}\frac{n!}{(n+\gamma_{\kappa}-\omega)
\Gamma(n+2\gamma_{\kappa})}L_{n}^{(2\gamma_{\kappa}-1)}(x)
L_{n}^{(2\gamma_{\kappa})}(x')
\nonumber \\
&-& \sum_{n=0}^{\infty}\frac{n!}{(n+\gamma_{\kappa}-\omega)
\Gamma(n+2\gamma_{\kappa})}L_{n}^{(2\gamma_{\kappa})}(x)
L_{n}^{(2\gamma_{\kappa})}(x')
\nonumber \\
&+& \sum_{n=0}^{\infty}\frac{n!}{(n+\gamma_{\kappa}-\omega)
\Gamma(n+2\gamma_{\kappa})}L_{n}^{(2\gamma_{\kappa})}(x)
L_{n}^{(2\gamma_{\kappa}-1)}(x')
\nonumber \\
&+& \left.
\sum_{n=0}^{\infty}\frac{(n+\gamma_{\kappa}+\varepsilon\zeta
-\kappa)n!}{(n+\gamma_{\kappa}-\omega)\Gamma(n+2\gamma_{\kappa}+1)}
L_{n}^{(2\gamma_{\kappa})}(x)L_{n}^{(2\gamma_{\kappa})}(x')\right]
\nonumber \\*[5mm]
&=& \frac{\alpha}{2\varepsilon}x^{\gamma_{\kappa}}
x'^{\gamma_{\kappa}}{\rm e}^{-(x+x')/2}\left\{
\sum_{n=0}^{\infty}\frac{(n+\gamma_{\kappa}+\varepsilon\zeta
+\kappa+1)n!}{(n+\gamma_{\kappa}-\omega+1)
\Gamma(n+2\gamma_{\kappa}+1)}
L_{n}^{(2\gamma_{\kappa})}(x)L_{n}^{(2\gamma_{\kappa})}(x')
\right.
\nonumber \\
&+& \sum_{n=0}^{\infty}\frac{n!}{(n+\gamma_{\kappa}-\omega)
\Gamma(n+2\gamma_{\kappa})}\left[L_{n}^{(2\gamma_{\kappa}-1)}(x)
L_{n}^{(2\gamma_{\kappa})}(x')+L_{n}^{(2\gamma_{\kappa})}(x)
L_{n}^{(2\gamma_{\kappa}-1)}(x')\right]
\nonumber \\
&+& \left.
\sum_{n=0}^{\infty}\frac{(-n-3\gamma_{\kappa}+\varepsilon\zeta
-\kappa)n!}{(n+\gamma_{\kappa}-\omega)\Gamma(n+2\gamma_{\kappa}+1)}
L_{n}^{(2\gamma_{\kappa})}(x)L_{n}^{(2\gamma_{\kappa})}(x')\right\}
\nonumber \\*[5mm]
&=& \frac{\alpha}{2\varepsilon}x^{\gamma_{\kappa}}
x'^{\gamma_{\kappa}}{\rm e}^{-(x+x')/2}\left\{
\sum_{n=0}^{\infty}\frac{(n+\gamma_{\kappa}-\omega+1+\varepsilon\zeta
+\omega+\kappa)n!}{(n+\gamma_{\kappa}-\omega+1)
\Gamma(n+2\gamma_{\kappa}+1)}
L_{n}^{(2\gamma_{\kappa})}(x)L_{n}^{(2\gamma_{\kappa})}(x')
\right.
\nonumber \\
&+& \sum_{n=0}^{\infty}\frac{n!}{(n+\gamma_{\kappa}-\omega)
\Gamma(n+2\gamma_{\kappa})}\left[L_{n}^{(2\gamma_{\kappa}-1)}(x)
L_{n}^{(2\gamma_{\kappa})}(x')+L_{n}^{(2\gamma_{\kappa})}(x)
L_{n}^{(2\gamma_{\kappa}-1)}(x')\right]
\nonumber \\
&+& \left.\sum_{n=0}^{\infty}\frac{(-n-\gamma_{\kappa}+\omega
+\varepsilon\zeta-\omega-2\gamma_{\kappa}-\kappa)n!}
{(n+\gamma_{\kappa}-\omega)\Gamma(n+2\gamma_{\kappa}+1)}
L_{n}^{(2\gamma_{\kappa})}(x)L_{n}^{(2\gamma_{\kappa})}(x')\right\}
\nonumber \\*[5mm]
&=& \frac{\alpha}{2\varepsilon}x^{\gamma_{\kappa}}
x'^{\gamma_{\kappa}}{\rm e}^{-(x+x')/2}\left\{
\sum_{n=0}^{\infty}\frac{n!}{\Gamma(n+2\gamma_{\kappa}+1)}
L_{n}^{(2\gamma_{\kappa})}(x)L_{n}^{(2\gamma_{\kappa})}(x')\right.
\nonumber \\
&+&
\sum_{n=0}^{\infty}\frac{(\varepsilon\zeta+\omega+\kappa)n!}
{(n+\gamma_{\kappa}-\omega+1)\Gamma(n+2\gamma_{\kappa}+1)}
L_{n}^{(2\gamma_{\kappa})}(x)L_{n}^{(2\gamma_{\kappa})}(x')
\nonumber \\
&+& \sum_{n=0}^{\infty}\frac{n!}{(n+\gamma_{\kappa}-\omega)
\Gamma(n+2\gamma_{\kappa})}\left[L_{n}^{(2\gamma_{\kappa}-1)}(x)
L_{n}^{(2\gamma_{\kappa})}(x')+L_{n}^{(2\gamma_{\kappa})}(x)
L_{n}^{(2\gamma_{\kappa}-1)}(x')\right]
\nonumber \\
&-& \sum_{n=0}^{\infty}\frac{n!}{\Gamma(n+2\gamma_{\kappa}+1)}
L_{n}^{(2\gamma_{\kappa})}(x)L_{n}^{(2\gamma_{\kappa})}(x')
\nonumber \\
&+& \left.\sum_{n=0}^{\infty}\frac{(\varepsilon\zeta-\omega
-2\gamma_{\kappa}-\kappa)n!}{(n+\gamma_{\kappa}-\omega)
\Gamma(n+2\gamma_{\kappa}+1)}
L_{n}^{(2\gamma_{\kappa})}(x)L_{n}^{(2\gamma_{\kappa})}(x')\right\}
\nonumber \\*[5mm]
&=& \frac{\alpha}{2\varepsilon}x^{\gamma_{\kappa}}
x'^{\gamma_{\kappa}}{\rm e}^{-(x+x')/2}\left\{
(\varepsilon\zeta+\omega+\kappa)\sum_{n=0}^{\infty}\frac{n!}
{(n+\gamma_{\kappa}-\omega+1)\Gamma(n+2\gamma_{\kappa}+1)}
L_{n}^{(2\gamma_{\kappa})}(x)L_{n}^{(2\gamma_{\kappa})}(x')\right.
\nonumber \\
&+& \sum_{n=0}^{\infty}\frac{n!}{(n+\gamma_{\kappa}-\omega)
\Gamma(n+2\gamma_{\kappa})}\left[L_{n}^{(2\gamma_{\kappa}-1)}(x)
L_{n}^{(2\gamma_{\kappa})}(x')+L_{n}^{(2\gamma_{\kappa})}(x)
L_{n}^{(2\gamma_{\kappa}-1)}(x')\right]
\nonumber \\
&+& \left.(\varepsilon\zeta-\omega-2\gamma_{\kappa}-\kappa)
\sum_{n=0}^{\infty}\frac{n!}{(n+\gamma_{\kappa}-\omega)
\Gamma(n+2\gamma_{\kappa}+1)}
L_{n}^{(2\gamma_{\kappa})}(x)L_{n}^{(2\gamma_{\kappa})}(x')\right\}
\nonumber \\*[5mm]
&=& \frac{\alpha}{2\varepsilon}x^{\gamma_{\kappa}}
x'^{\gamma_{\kappa}}{\rm e}^{-(x+x')/2}
\left\{\left(\frac{\zeta}{\sqrt{1-\epsilon^{2}}}+\kappa\right)
\sum_{n=0}^{\infty}\frac{n!}{(n+\gamma_{\kappa}-\omega+1)
\Gamma(n+2\gamma_{\kappa}+1)}L_{n}^{(2\gamma_{\kappa})}(x)
L_{n}^{(2\gamma_{\kappa})}(x')\right.
\nonumber \\
&+& \sum_{n=0}^{\infty}\frac{n!}{(n+\gamma_{\kappa}-\omega)
\Gamma(n+2\gamma_{\kappa})}\left[L_{n}^{(2\gamma_{\kappa}-1)}(x)
L_{n}^{(2\gamma_{\kappa})}(x')+L_{n}^{(2\gamma_{\kappa})}(x)
L_{n}^{(2\gamma_{\kappa}-1)}(x')\right]
\nonumber \\
&+& \left.\left(\frac{\zeta(1-2\epsilon)}{\sqrt{1-\epsilon^{2}}}
-2\gamma_{\kappa}-\kappa\right)\sum_{n=0}^{\infty}
\frac{n!}{(n+\gamma_{\kappa}-\omega)\Gamma(n+2\gamma_{\kappa}+1)}
L_{n}^{(2\gamma_{\kappa})}(x)L_{n}^{(2\gamma_{\kappa})}(x')\right\}
\label{25}
\end{eqnarray}
\begin{eqnarray*}
&& g_{\kappa}^{(12)}(r,r';E)
=\sum_{n=-\infty}^{\infty}\frac{1}{\mu_{n\kappa}-1}
S_{n\kappa}(x)T_{n\kappa}(x')
\nonumber \\*[5mm]
&=& \sum_{n=-\infty}^{\infty}\frac{1}
{\varepsilon\frac{|n|+\gamma_{\kappa}\pm N_{n\kappa}}{\zeta}-1}
\frac{\alpha(|n|+2\gamma_{\kappa})|n|!}
{2N_{n\kappa}(N_{n\kappa}\mp\kappa)\Gamma(|n|+2\gamma_{\kappa})}
x^{\gamma_{\kappa}}x'^{\gamma_{\kappa}}{\rm e}^{-(x+x')/2}
\nonumber \\
&\times& \left[L_{|n|-1}^{(2\gamma_{\kappa})}(x)
+\frac{\kappa\mp N_{n\kappa}}{|n|+2\gamma_{\kappa}}
L_{|n|}^{(2\gamma_{\kappa})}(x)\right]
\left[L_{|n|-1}^{(2\gamma_{\kappa})}(x')
-\frac{\kappa\mp N_{n\kappa}}{|n|+2\gamma_{\kappa}}
L_{|n|}^{(2\gamma_{\kappa})}(x')\right]
\nonumber \\*[5mm]
&=& x^{\gamma_{\kappa}}x'^{\gamma_{\kappa}}{\rm e}^{-(x+x')/2}
\sum_{n=-\infty}^{\infty}
\frac{\alpha\zeta(|n|+2\gamma_{\kappa})|n|!}{2\varepsilon
N_{n\kappa}(N_{n\kappa}\mp\kappa)\Gamma(|n|+2\gamma_{\kappa})}
\frac{|n|+\gamma_{\kappa}-\varepsilon^{-1}\zeta\mp N_{n\kappa}}
{(|n|+\gamma_{\kappa}-\varepsilon^{-1}\zeta)^{2}-N_{n\kappa}^{2}}
\nonumber \\
&\times& \left[L_{|n|-1}^{(2\gamma_{\kappa})}(x)
+\frac{\kappa\mp N_{n\kappa}}{|n|+2\gamma_{\kappa}}
L_{|n|}^{(2\gamma_{\kappa})}(x)\right]
\left[L_{|n|-1}^{(2\gamma_{\kappa})}(x')
-\frac{\kappa\mp N_{n\kappa}}{|n|+2\gamma_{\kappa}}
L_{|n|}^{(2\gamma_{\kappa})}(x')\right]
\nonumber \\*[5mm]
&=& x^{\gamma_{\kappa}}x'^{\gamma_{\kappa}}{\rm e}^{-(x+x')/2}
\sum_{n=-\infty}^{\infty}
\frac{\alpha\zeta(|n|+2\gamma_{\kappa})|n|!}{2\varepsilon
N_{n\kappa}(N_{n\kappa}\mp\kappa)\Gamma(|n|+2\gamma_{\kappa})}
\frac{|n|+\gamma_{\kappa}-\varepsilon^{-1}\zeta\mp N_{n\kappa}}
{(-2)\varepsilon^{-1}\zeta(|n|+\gamma_{\kappa}-\omega)}
\nonumber \\
&\times& \left[L_{|n|-1}^{(2\gamma_{\kappa})}(x)
+\frac{\kappa\mp N_{n\kappa}}{|n|+2\gamma_{\kappa}}
L_{|n|}^{(2\gamma_{\kappa})}(x)\right]
\left[L_{|n|-1}^{(2\gamma_{\kappa})}(x')
-\frac{\kappa\mp N_{n\kappa}}{|n|+2\gamma_{\kappa}}
L_{|n|}^{(2\gamma_{\kappa})}(x')\right]
\nonumber \\
&=& -\frac{\alpha}{4}
x^{\gamma_{\kappa}}x'^{\gamma_{\kappa}}{\rm e}^{-(x+x')/2}
\sum_{n=-\infty}^{\infty}
\frac{(|n|+\gamma_{\kappa}-\varepsilon^{-1}\zeta\mp N_{n\kappa})
(|n|+2\gamma_{\kappa})|n|!}{N_{n\kappa}(N_{n\kappa}\mp\kappa)
(|n|+\gamma_{\kappa}-\omega)\Gamma(|n|+2\gamma_{\kappa})}
\nonumber \\
&\times& \left[L_{|n|-1}^{(2\gamma_{\kappa})}(x)
+\frac{\kappa\mp N_{n\kappa}}{|n|+2\gamma_{\kappa}}
L_{|n|}^{(2\gamma_{\kappa})}(x)\right]
\left[L_{|n|-1}^{(2\gamma_{\kappa})}(x')
-\frac{\kappa\mp N_{n\kappa}}{|n|+2\gamma_{\kappa}}
L_{|n|}^{(2\gamma_{\kappa})}(x')\right]
\nonumber \\*[5mm]
&=& -\frac{\alpha}{4}x^{\gamma_{\kappa}}
x'^{\gamma_{\kappa}}{\rm e}^{-(x+x')/2}\left[
\sum_{n=-\infty}^{\infty}
\frac{(|n|+\gamma_{\kappa}-\varepsilon^{-1}\zeta\mp N_{n\kappa})
(|n|+2\gamma_{\kappa})|n|!}{N_{n\kappa}(N_{n\kappa}\mp\kappa)
(|n|+\gamma_{\kappa}-\omega)\Gamma(|n|+2\gamma_{\kappa})}
L_{|n|-1}^{(2\gamma_{\kappa})}(x)L_{|n|-1}^{(2\gamma_{\kappa})}(x')
\right.
\nonumber \\
&-& \sum_{n=-\infty}^{\infty}
\frac{(|n|+\gamma_{\kappa}-\varepsilon^{-1}\zeta\mp N_{n\kappa})
(|n|+2\gamma_{\kappa})|n|!}{N_{n\kappa}(N_{n\kappa}\mp\kappa)
(|n|+\gamma_{\kappa}-\omega)\Gamma(|n|+2\gamma_{\kappa})}
\frac{\kappa\mp N_{n\kappa}}{|n|+2\gamma_{\kappa}}
L_{|n|-1}^{(2\gamma_{\kappa})}(x)L_{|n|}^{(2\gamma_{\kappa})}(x')
\nonumber \\
&+& \sum_{n=-\infty}^{\infty}
\frac{(|n|+\gamma_{\kappa}-\varepsilon^{-1}\zeta\mp N_{n\kappa})
(|n|+2\gamma_{\kappa})|n|!}{N_{n\kappa}(N_{n\kappa}\mp\kappa)
(|n|+\gamma_{\kappa}-\omega)\Gamma(|n|+2\gamma_{\kappa})}
\frac{\kappa\mp N_{n\kappa}}{|n|+2\gamma_{\kappa}}
L_{|n|}^{(2\gamma_{\kappa})}(x)L_{|n|-1}^{(2\gamma_{\kappa})}(x')
\nonumber \\
&-& \left.\sum_{n=-\infty}^{\infty}
\frac{(|n|+\gamma_{\kappa}-\varepsilon^{-1}\zeta\mp N_{n\kappa})
(|n|+2\gamma_{\kappa})|n|!}{N_{n\kappa}(N_{n\kappa}\mp\kappa)
(|n|+\gamma_{\kappa}-\omega)\Gamma(|n|+2\gamma_{\kappa})}
\left(\frac{\kappa\mp N_{n\kappa}}{|n|+2\gamma_{\kappa}}\right)^{2}
L_{|n|}^{(2\gamma_{\kappa})}(x)L_{|n|}^{(2\gamma_{\kappa})}(x')
\right]
\nonumber \\*[5mm]
&=& -\frac{\alpha}{4}x^{\gamma_{\kappa}}
x'^{\gamma_{\kappa}}{\rm e}^{-(x+x')/2}\left[
\sum_{n=-\infty}^{\infty}
\frac{(|n|+\gamma_{\kappa}-\varepsilon^{-1}\zeta\mp N_{n\kappa})
(|n|+2\gamma_{\kappa})|n|!}{N_{n\kappa}(N_{n\kappa}\mp\kappa)
(|n|+\gamma_{\kappa}-\omega)\Gamma(|n|+2\gamma_{\kappa})}
L_{|n|-1}^{(2\gamma_{\kappa})}(x)L_{|n|-1}^{(2\gamma_{\kappa})}(x')
\right.
\nonumber \\
&-& \sum_{n=-\infty}^{\infty}\frac{\mp(|n|+\gamma_{\kappa}
-\varepsilon^{-1}\zeta\mp N_{n\kappa})|n|!}
{N_{n\kappa}(|n|+\gamma_{\kappa}-\omega)\Gamma(|n|+2\gamma_{\kappa})}
L_{|n|-1}^{(2\gamma_{\kappa})}(x)L_{|n|}^{(2\gamma_{\kappa})}(x')
\nonumber \\
&+& \sum_{n=-\infty}^{\infty}\frac{\mp(|n|+\gamma_{\kappa}
-\varepsilon^{-1}\zeta\mp N_{n\kappa})|n|!}
{N_{n\kappa}(|n|+\gamma_{\kappa}-\omega)\Gamma(|n|+2\gamma_{\kappa})}
L_{|n|}^{(2\gamma_{\kappa})}(x)L_{|n|-1}^{(2\gamma_{\kappa})}(x')
\nonumber \\
&-& \left.\sum_{n=-\infty}^{\infty}
\frac{(|n|+\gamma_{\kappa}-\varepsilon^{-1}\zeta\mp N_{n\kappa})
(N_{n\kappa}\mp\kappa)|n|!}{N_{n\kappa}
(|n|+\gamma_{\kappa}-\omega)\Gamma(|n|+2\gamma_{\kappa}+1)}
L_{|n|}^{(2\gamma_{\kappa})}(x)L_{|n|}^{(2\gamma_{\kappa})}(x')
\right]
\nonumber \\*[5mm]
&=& -\frac{\alpha}{4}x^{\gamma_{\kappa}}
x'^{\gamma_{\kappa}}{\rm e}^{-(x+x')/2}\left[
\sum_{n=1}^{\infty}
\frac{(|n|+\gamma_{\kappa}-\varepsilon^{-1}\zeta-N_{n\kappa})
(|n|+2\gamma_{\kappa})|n|!}{N_{n\kappa}(N_{n\kappa}-\kappa)
(|n|+\gamma_{\kappa}-\omega)\Gamma(|n|+2\gamma_{\kappa})}
L_{|n|-1}^{(2\gamma_{\kappa})}(x)L_{|n|-1}^{(2\gamma_{\kappa})}(x')
\right.
\nonumber \\
&+& \sum_{n=-\infty}^{-1}
\frac{(|n|+\gamma_{\kappa}-\varepsilon^{-1}\zeta+N_{n\kappa})
(|n|+2\gamma_{\kappa})|n|!}{N_{n\kappa}(N_{n\kappa}+\kappa)
(|n|+\gamma_{\kappa}-\omega)\Gamma(|n|+2\gamma_{\kappa})}
L_{|n|-1}^{(2\gamma_{\kappa})}(x)L_{|n|-1}^{(2\gamma_{\kappa})}(x')
\nonumber \\
&+& \sum_{n=1}^{\infty}
\frac{(|n|+\gamma_{\kappa}-\varepsilon^{-1}\zeta-N_{n\kappa})|n|!}
{N_{n\kappa}(|n|+\gamma_{\kappa}-\omega)\Gamma(|n|+2\gamma_{\kappa})}
L_{|n|-1}^{(2\gamma_{\kappa})}(x)L_{|n|}^{(2\gamma_{\kappa})}(x')
\nonumber \\
&-& \sum_{n=-\infty}^{-1}
\frac{(|n|+\gamma_{\kappa}-\varepsilon^{-1}\zeta+N_{n\kappa})|n|!}
{N_{n\kappa}(|n|+\gamma_{\kappa}-\omega)\Gamma(|n|+2\gamma_{\kappa})}
L_{|n|-1}^{(2\gamma_{\kappa})}(x)L_{|n|}^{(2\gamma_{\kappa})}(x')
\nonumber \\
&-& \sum_{n=1}^{\infty}
\frac{(|n|+\gamma_{\kappa}-\varepsilon^{-1}\zeta-N_{n\kappa})|n|!}
{N_{n\kappa}(|n|+\gamma_{\kappa}-\omega)\Gamma(|n|+2\gamma_{\kappa})}
L_{|n|}^{(2\gamma_{\kappa})}(x)L_{|n|-1}^{(2\gamma_{\kappa})}(x')
\nonumber \\
&+& \sum_{n=-\infty}^{-1}
\frac{(|n|+\gamma_{\kappa}-\varepsilon^{-1}\zeta+N_{n\kappa})|n|!}
{N_{n\kappa}(|n|+\gamma_{\kappa}-\omega)\Gamma(|n|+2\gamma_{\kappa})}
L_{|n|}^{(2\gamma_{\kappa})}(x)L_{|n|-1}^{(2\gamma_{\kappa})}(x')
\nonumber \\
&-& \sum_{n=0}^{\infty}
\frac{(|n|+\gamma_{\kappa}-\varepsilon^{-1}\zeta-N_{n\kappa})
(N_{n\kappa}-\kappa)|n|!}{N_{n\kappa}
(|n|+\gamma_{\kappa}-\omega)\Gamma(|n|+2\gamma_{\kappa}+1)}
L_{|n|}^{(2\gamma_{\kappa})}(x)L_{|n|}^{(2\gamma_{\kappa})}(x')
\nonumber \\
&-& \left.\sum_{n=-\infty}^{0}
\frac{(|n|+\gamma_{\kappa}-\varepsilon^{-1}\zeta+N_{n\kappa})
(N_{n\kappa}+\kappa)|n|!}{N_{n\kappa}
(|n|+\gamma_{\kappa}-\omega)\Gamma(|n|+2\gamma_{\kappa}+1)}
L_{|n|}^{(2\gamma_{\kappa})}(x)L_{|n|}^{(2\gamma_{\kappa})}(x')
\right]
\nonumber \\*[5mm]
&=& -\frac{\alpha}{4}x^{\gamma_{\kappa}}
x'^{\gamma_{\kappa}}{\rm e}^{-(x+x')/2}\left[
\sum_{n=1}^{\infty}
\frac{(n+\gamma_{\kappa}-\varepsilon^{-1}\zeta-N_{n\kappa})
(n+2\gamma_{\kappa})n!}{N_{n\kappa}(N_{n\kappa}-\kappa)
(n+\gamma_{\kappa}-\omega)\Gamma(n+2\gamma_{\kappa})}
L_{n-1}^{(2\gamma_{\kappa})}(x)L_{n-1}^{(2\gamma_{\kappa})}(x')
\right.
\nonumber \\
&+& \sum_{n=1}^{\infty}
\frac{(n+\gamma_{\kappa}-\varepsilon^{-1}\zeta+N_{n\kappa})
(n+2\gamma_{\kappa})n!}{N_{n\kappa}(N_{n\kappa}+\kappa)
(n+\gamma_{\kappa}-\omega)\Gamma(n+2\gamma_{\kappa})}
L_{n-1}^{(2\gamma_{\kappa})}(x)L_{n-1}^{(2\gamma_{\kappa})}(x')
\nonumber \\
&+& \sum_{n=1}^{\infty}
\frac{(n+\gamma_{\kappa}-\varepsilon^{-1}\zeta-N_{n\kappa})n!}
{N_{n\kappa}(n+\gamma_{\kappa}-\omega)\Gamma(n+2\gamma_{\kappa})}
L_{n-1}^{(2\gamma_{\kappa})}(x)L_{n}^{(2\gamma_{\kappa})}(x')
\nonumber \\
&-& \sum_{n=1}^{\infty}
\frac{(n+\gamma_{\kappa}-\varepsilon^{-1}\zeta+N_{n\kappa})n!}
{N_{n\kappa}(n+\gamma_{\kappa}-\omega)\Gamma(n+2\gamma_{\kappa})}
L_{n-1}^{(2\gamma_{\kappa})}(x)L_{n}^{(2\gamma_{\kappa})}(x')
\nonumber \\
&-& \sum_{n=1}^{\infty}
\frac{(n+\gamma_{\kappa}-\varepsilon^{-1}\zeta-N_{n\kappa})n!}
{N_{n\kappa}(n+\gamma_{\kappa}-\omega)\Gamma(n+2\gamma_{\kappa})}
L_{n}^{(2\gamma_{\kappa})}(x)L_{n-1}^{(2\gamma_{\kappa})}(x')
\nonumber \\
&+& \sum_{n=1}^{\infty}
\frac{(n+\gamma_{\kappa}-\varepsilon^{-1}\zeta+N_{n\kappa})n!}
{N_{n\kappa}(n+\gamma_{\kappa}-\omega)\Gamma(n+2\gamma_{\kappa})}
L_{n}^{(2\gamma_{\kappa})}(x)L_{n-1}^{(2\gamma_{\kappa})}(x')
\nonumber \\
&-& \sum_{n=0}^{\infty}
\frac{(n+\gamma_{\kappa}-\varepsilon^{-1}\zeta-N_{n\kappa})
(N_{n\kappa}-\kappa)n!}{N_{n\kappa}
(n+\gamma_{\kappa}-\omega)\Gamma(n+2\gamma_{\kappa}+1)}
L_{n}^{(2\gamma_{\kappa})}(x)L_{n}^{(2\gamma_{\kappa})}(x')
\nonumber \\
&-& \left.\sum_{n=0}^{\infty}
\frac{(n+\gamma_{\kappa}-\varepsilon^{-1}\zeta+N_{n\kappa})
(N_{n\kappa}+\kappa)n!}{N_{n\kappa}
(n+\gamma_{\kappa}-\omega)\Gamma(n+2\gamma_{\kappa}+1)}
L_{n}^{(2\gamma_{\kappa})}(x)L_{n}^{(2\gamma_{\kappa})}(x')
\right]
\end{eqnarray*}
\begin{eqnarray}
&=& -\frac{\alpha}{4}x^{\gamma_{\kappa}}
x'^{\gamma_{\kappa}}{\rm e}^{-(x+x')/2}\left[
\sum_{n=1}^{\infty}\frac{(n+\gamma_{\kappa}-\varepsilon^{-1}\zeta
-N_{n\kappa})(N_{n\kappa}+\kappa)(n-1)!}{N_{n\kappa}
(n+\gamma_{\kappa}-\omega)\Gamma(n+2\gamma_{\kappa})}
L_{n-1}^{(2\gamma_{\kappa})}(x)L_{n-1}^{(2\gamma_{\kappa})}(x')
\right.
\nonumber \\
&+& \sum_{n=1}^{\infty}\frac{(n+\gamma_{\kappa}-\varepsilon^{-1}\zeta
+N_{n\kappa})(N_{n\kappa}-\kappa)(n-1)!}{N_{n\kappa}
(n+\gamma_{\kappa}-\omega)\Gamma(n+2\gamma_{\kappa})}
L_{n-1}^{(2\gamma_{\kappa})}(x)L_{n-1}^{(2\gamma_{\kappa})}(x')
\nonumber \\
&-& 2\sum_{n=1}^{\infty}\frac{n!}{(n+\gamma_{\kappa}-\omega)
\Gamma(n+2\gamma_{\kappa})}
L_{n-1}^{(2\gamma_{\kappa})}(x)L_{n}^{(2\gamma_{\kappa})}(x')
\nonumber \\
&+& 2\sum_{n=1}^{\infty}\frac{n!}{(n+\gamma_{\kappa}-\omega)
\Gamma(n+2\gamma_{\kappa})}
L_{n}^{(2\gamma_{\kappa})}(x)L_{n-1}^{(2\gamma_{\kappa})}(x')
\nonumber \\
&-& \sum_{n=0}^{\infty}
\frac{(n+\gamma_{\kappa}-\varepsilon^{-1}\zeta-N_{n\kappa})
(N_{n\kappa}-\kappa)n!}{N_{n\kappa}(n+\gamma_{\kappa}-\omega)
\Gamma(n+2\gamma_{\kappa}+1)}
L_{n}^{(2\gamma_{\kappa})}(x)L_{n}^{(2\gamma_{\kappa})}(x')
\nonumber \\
&-& \left.\sum_{n=0}^{\infty}
\frac{(n+\gamma_{\kappa}-\varepsilon^{-1}\zeta+N_{n\kappa})
(N_{n\kappa}+\kappa)n!}{N_{n\kappa}(n+\gamma_{\kappa}-\omega)
\Gamma(n+2\gamma_{\kappa}+1)}
L_{n}^{(2\gamma_{\kappa})}(x)L_{n}^{(2\gamma_{\kappa})}(x')
\right]
\nonumber \\*[5mm]
&=& -\frac{\alpha}{2}x^{\gamma_{\kappa}}
x'^{\gamma_{\kappa}}{\rm e}^{-(x+x')/2}\left[
\sum_{n=1}^{\infty}\frac{(n+\gamma_{\kappa}-\varepsilon^{-1}\zeta
-\kappa)(n-1)!}{(n+\gamma_{\kappa}-\omega)\Gamma(n+2\gamma_{\kappa})}
L_{n-1}^{(2\gamma_{\kappa})}(x)L_{n-1}^{(2\gamma_{\kappa})}(x')
\right.
\nonumber \\
&-& \sum_{n=0}^{\infty}\frac{n!}{(n+\gamma_{\kappa}-\omega)
\Gamma(n+2\gamma_{\kappa})}L_{n-1}^{(2\gamma_{\kappa})}(x)
L_{n}^{(2\gamma_{\kappa})}(x')
\nonumber \\
&+& \sum_{n=0}^{\infty}\frac{n!}{(n+\gamma_{\kappa}-\omega)
\Gamma(n+2\gamma_{\kappa})}L_{n}^{(2\gamma_{\kappa})}(x)
L_{n-1}^{(2\gamma_{\kappa})}(x')
\nonumber \\
&-& \left.
\sum_{n=0}^{\infty}\frac{(n+\gamma_{\kappa}-\varepsilon^{-1}\zeta
+\kappa)n!}{(n+\gamma_{\kappa}-\omega)\Gamma(n+2\gamma_{\kappa}+1)}
L_{n}^{(2\gamma_{\kappa})}(x)L_{n}^{(2\gamma_{\kappa})}(x')\right]
\nonumber \\*[5mm]
&=& -\frac{\alpha}{2}x^{\gamma_{\kappa}}
x'^{\gamma_{\kappa}}{\rm e}^{-(x+x')/2}\left[
\sum_{n=0}^{\infty}\frac{(n+\gamma_{\kappa}-\varepsilon^{-1}\zeta
-\kappa+1)n!}{(n+\gamma_{\kappa}-\omega+1)
\Gamma(n+2\gamma_{\kappa}+1)}
L_{n}^{(2\gamma_{\kappa})}(x)L_{n}^{(2\gamma_{\kappa})}(x')
\right.
\nonumber \\
&-& \sum_{n=0}^{\infty}\frac{n!}{(n+\gamma_{\kappa}-\omega)
\Gamma(n+2\gamma_{\kappa})}L_{n-1}^{(2\gamma_{\kappa})}(x)
L_{n}^{(2\gamma_{\kappa})}(x')
\nonumber \\
&+& \sum_{n=0}^{\infty}\frac{n!}{(n+\gamma_{\kappa}-\omega)
\Gamma(n+2\gamma_{\kappa})}L_{n}^{(2\gamma_{\kappa})}(x)
L_{n-1}^{(2\gamma_{\kappa})}(x')
\nonumber \\
&-& \left.
\sum_{n=0}^{\infty}\frac{(n+\gamma_{\kappa}-\varepsilon^{-1}\zeta
+\kappa)n!}{(n+\gamma_{\kappa}-\omega)\Gamma(n+2\gamma_{\kappa}+1)}
L_{n}^{(2\gamma_{\kappa})}(x)L_{n}^{(2\gamma_{\kappa})}(x')\right]
\nonumber \\*[5mm]
&=& -\frac{\alpha}{2}x^{\gamma_{\kappa}}
x'^{\gamma_{\kappa}}{\rm e}^{-(x+x')/2}\left\{
\sum_{n=0}^{\infty}\frac{(n+\gamma_{\kappa}-\varepsilon^{-1}\zeta
-\kappa+1)n!}{(n+\gamma_{\kappa}-\omega+1)
\Gamma(n+2\gamma_{\kappa}+1)}
L_{n}^{(2\gamma_{\kappa})}(x)L_{n}^{(2\gamma_{\kappa})}(x')
\right.
\nonumber \\
&-& \sum_{n=0}^{\infty}\frac{n!}{(n+\gamma_{\kappa}-\omega)
\Gamma(n+2\gamma_{\kappa})}\left[L_{n}^{(2\gamma_{\kappa})}(x)
-L_{n}^{(2\gamma_{\kappa}-1)}(x)\right]L_{n}^{(2\gamma_{\kappa})}(x')
\nonumber \\
&+& \sum_{n=0}^{\infty}\frac{n!}{(n+\gamma_{\kappa}-\omega)
\Gamma(n+2\gamma_{\kappa})}L_{n}^{(2\gamma_{\kappa})}(x)
\left[L_{n}^{(2\gamma_{\kappa})}(x')
-L_{n}^{(2\gamma_{\kappa}-1)}(x')\right]
\nonumber \\
&-& \left.
\sum_{n=0}^{\infty}\frac{(n+\gamma_{\kappa}-\varepsilon^{-1}\zeta
+\kappa)n!}{(n+\gamma_{\kappa}-\omega)\Gamma(n+2\gamma_{\kappa}+1)}
L_{n}^{(2\gamma_{\kappa})}(x)L_{n}^{(2\gamma_{\kappa})}(x')\right\}
\nonumber \\*[5mm]
&=& -\frac{\alpha}{2}x^{\gamma_{\kappa}}
x'^{\gamma_{\kappa}}{\rm e}^{-(x+x')/2}\left[
\sum_{n=0}^{\infty}\frac{(n+\gamma_{\kappa}-\varepsilon^{-1}\zeta
-\kappa+1)n!}{(n+\gamma_{\kappa}-\omega+1)
\Gamma(n+2\gamma_{\kappa}+1)}
L_{n}^{(2\gamma_{\kappa})}(x)L_{n}^{(2\gamma_{\kappa})}(x')
\right.
\nonumber \\
&-& \sum_{n=0}^{\infty}\frac{n!}{(n+\gamma_{\kappa}-\omega)
\Gamma(n+2\gamma_{\kappa})}L_{n}^{(2\gamma_{\kappa})}(x)
L_{n}^{(2\gamma_{\kappa})}(x')
\nonumber \\
&+& \sum_{n=0}^{\infty}\frac{n!}{(n+\gamma_{\kappa}-\omega)
\Gamma(n+2\gamma_{\kappa})}L_{n}^{(2\gamma_{\kappa}-1)}(x)
L_{n}^{(2\gamma_{\kappa})}(x')
\nonumber \\
&+& \sum_{n=0}^{\infty}\frac{n!}{(n+\gamma_{\kappa}-\omega)
\Gamma(n+2\gamma_{\kappa})}L_{n}^{(2\gamma_{\kappa})}(x)
L_{n}^{(2\gamma_{\kappa})}(x')
\nonumber \\
&-& \sum_{n=0}^{\infty}\frac{n!}{(n+\gamma_{\kappa}-\omega)
\Gamma(n+2\gamma_{\kappa})}L_{n}^{(2\gamma_{\kappa})}(x)
L_{n}^{(2\gamma_{\kappa}-1)}(x')
\nonumber \\
&-& \left.
\sum_{n=0}^{\infty}\frac{(n+\gamma_{\kappa}-\varepsilon^{-1}\zeta
+\kappa)n!}{(n+\gamma_{\kappa}-\omega)\Gamma(n+2\gamma_{\kappa}+1)}
L_{n}^{(2\gamma_{\kappa})}(x)L_{n}^{(2\gamma_{\kappa})}(x')\right]
\nonumber \\*[5mm]
&=& -\frac{\alpha}{2}x^{\gamma_{\kappa}}
x'^{\gamma_{\kappa}}{\rm e}^{-(x+x')/2}\left\{
\sum_{n=0}^{\infty}\frac{(n+\gamma_{\kappa}-\varepsilon^{-1}\zeta
-\kappa+1)n!}{(n+\gamma_{\kappa}-\omega+1)
\Gamma(n+2\gamma_{\kappa}+1)}
L_{n}^{(2\gamma_{\kappa})}(x)L_{n}^{(2\gamma_{\kappa})}(x')
\right.
\nonumber \\
&+& \sum_{n=0}^{\infty}\frac{n!}{(n+\gamma_{\kappa}-\omega)
\Gamma(n+2\gamma_{\kappa})}\left[L_{n}^{(2\gamma_{\kappa}-1)}(x)
L_{n}^{(2\gamma_{\kappa})}(x')-L_{n}^{(2\gamma_{\kappa})}(x)
L_{n}^{(2\gamma_{\kappa}-1)}(x')\right]
\nonumber \\
&-& \left.
\sum_{n=0}^{\infty}\frac{(n+\gamma_{\kappa}-\varepsilon^{-1}\zeta
+\kappa)n!}{(n+\gamma_{\kappa}-\omega)\Gamma(n+2\gamma_{\kappa}+1)}
L_{n}^{(2\gamma_{\kappa})}(x)L_{n}^{(2\gamma_{\kappa})}(x')\right\}
\nonumber \\*[5mm]
&=& -\frac{\alpha}{2}x^{\gamma_{\kappa}}
x'^{\gamma_{\kappa}}{\rm e}^{-(x+x')/2}\left\{
\sum_{n=0}^{\infty}\frac{(n+\gamma_{\kappa}-\omega+1-\varepsilon^{-1}
\zeta+\omega-\kappa)n!}{(n+\gamma_{\kappa}-\omega+1)
\Gamma(n+2\gamma_{\kappa}+1)}
L_{n}^{(2\gamma_{\kappa})}(x)L_{n}^{(2\gamma_{\kappa})}(x')
\right.
\nonumber \\
&+& \sum_{n=0}^{\infty}\frac{n!}{(n+\gamma_{\kappa}-\omega)
\Gamma(n+2\gamma_{\kappa})}\left[L_{n}^{(2\gamma_{\kappa}-1)}(x)
L_{n}^{(2\gamma_{\kappa})}(x')-L_{n}^{(2\gamma_{\kappa})}(x)
L_{n}^{(2\gamma_{\kappa}-1)}(x')\right]
\nonumber \\
&-& \left.\sum_{n=0}^{\infty}\frac{(n+\gamma_{\kappa}-\omega
-\varepsilon^{-1}\zeta+\omega+\kappa)n!}
{(n+\gamma_{\kappa}-\omega)\Gamma(n+2\gamma_{\kappa}+1)}
L_{n}^{(2\gamma_{\kappa})}(x)L_{n}^{(2\gamma_{\kappa})}(x')\right\}
\nonumber \\*[5mm]
&=& -\frac{\alpha}{2}x^{\gamma_{\kappa}}
x'^{\gamma_{\kappa}}{\rm e}^{-(x+x')/2}\left\{
\sum_{n=0}^{\infty}\frac{n!}{\Gamma(n+2\gamma_{\kappa}+1)}
L_{n}^{(2\gamma_{\kappa})}(x)L_{n}^{(2\gamma_{\kappa})}(x')\right.
\nonumber \\
&-&
\sum_{n=0}^{\infty}\frac{(\varepsilon^{-1}\zeta-\omega+\kappa)n!}
{(n+\gamma_{\kappa}-\omega+1)\Gamma(n+2\gamma_{\kappa}+1)}
L_{n}^{(2\gamma_{\kappa})}(x)L_{n}^{(2\gamma_{\kappa})}(x')
\nonumber \\
&+& \sum_{n=0}^{\infty}\frac{n!}{(n+\gamma_{\kappa}-\omega)
\Gamma(n+2\gamma_{\kappa})}\left[L_{n}^{(2\gamma_{\kappa}-1)}(x)
L_{n}^{(2\gamma_{\kappa})}(x')-L_{n}^{(2\gamma_{\kappa})}(x)
L_{n}^{(2\gamma_{\kappa}-1)}(x')\right]
\nonumber \\
&-& \sum_{n=0}^{\infty}\frac{n!}{\Gamma(n+2\gamma_{\kappa}+1)}
L_{n}^{(2\gamma_{\kappa})}(x)L_{n}^{(2\gamma_{\kappa})}(x')
\nonumber \\
&+& \left.\sum_{n=0}^{\infty}\frac{(\varepsilon^{-1}\zeta-\omega
-\kappa)n!}{(n+\gamma_{\kappa}-\omega)\Gamma(n+2\gamma_{\kappa}+1)}
L_{n}^{(2\gamma_{\kappa})}(x)L_{n}^{(2\gamma_{\kappa})}(x')\right\}
\nonumber \\*[5mm]
&=& \frac{\alpha}{2}x^{\gamma_{\kappa}}
x'^{\gamma_{\kappa}}{\rm e}^{-(x+x')/2}\left\{
(\varepsilon^{-1}\zeta-\omega+\kappa)\sum_{n=0}^{\infty}\frac{n!}
{(n+\gamma_{\kappa}-\omega+1)\Gamma(n+2\gamma_{\kappa}+1)}
L_{n}^{(2\gamma_{\kappa})}(x)L_{n}^{(2\gamma_{\kappa})}(x')\right.
\nonumber \\
&-& \sum_{n=0}^{\infty}\frac{n!}{(n+\gamma_{\kappa}-\omega)
\Gamma(n+2\gamma_{\kappa})}\left[L_{n}^{(2\gamma_{\kappa}-1)}(x)
L_{n}^{(2\gamma_{\kappa})}(x')-L_{n}^{(2\gamma_{\kappa})}(x)
L_{n}^{(2\gamma_{\kappa}-1)}(x')\right]
\nonumber \\
&-& \left.(\varepsilon^{-1}\zeta-\omega-\kappa)
\sum_{n=0}^{\infty}\frac{n!}{(n+\gamma_{\kappa}-\omega)
\Gamma(n+2\gamma_{\kappa}+1)}
L_{n}^{(2\gamma_{\kappa})}(x)L_{n}^{(2\gamma_{\kappa})}(x')\right\}
\nonumber \\*[5mm]
&=& \frac{\alpha}{2}x^{\gamma_{\kappa}}
x'^{\gamma_{\kappa}}{\rm e}^{-(x+x')/2}
\left\{\left(\frac{\zeta}{\sqrt{1-\epsilon^{2}}}+\kappa\right)
\sum_{n=0}^{\infty}\frac{n!}{(n+\gamma_{\kappa}-\omega+1)
\Gamma(n+2\gamma_{\kappa}+1)}L_{n}^{(2\gamma_{\kappa})}(x)
L_{n}^{(2\gamma_{\kappa})}(x')\right.
\nonumber \\
&-& \sum_{n=0}^{\infty}\frac{n!}{(n+\gamma_{\kappa}-\omega)
\Gamma(n+2\gamma_{\kappa})}\left[L_{n}^{(2\gamma_{\kappa}-1)}(x)
L_{n}^{(2\gamma_{\kappa})}(x')-L_{n}^{(2\gamma_{\kappa})}(x)
L_{n}^{(2\gamma_{\kappa}-1)}(x')\right]
\nonumber \\
&-& \left.\left(\frac{\zeta}{\sqrt{1-\epsilon^{2}}}-\kappa\right)
\sum_{n=0}^{\infty}\frac{n!}{(n+\gamma_{\kappa}-\omega)
\Gamma(n+2\gamma_{\kappa}+1)}L_{n}^{(2\gamma_{\kappa})}(x)
L_{n}^{(2\gamma_{\kappa})}(x')\right\},
\label{26}
\end{eqnarray}
\begin{eqnarray*}
&& g_{\kappa}^{(21)}(r,r';E)
=\sum_{n=-\infty}^{\infty}\frac{\mu_{n\kappa}}{\mu_{n\kappa}-1}
T_{n\kappa}(x)S_{n\kappa}(x')
\nonumber \\*[5mm]
&=& \sum_{n=-\infty}^{\infty}
\frac{1}{1+\frac{|n|+\gamma_{\kappa}\mp N_{n\kappa}}
{\varepsilon\zeta}}\:\frac{\alpha(|n|+2\gamma_{\kappa})|n|!}
{2N_{n\kappa}(N_{n\kappa}\mp\kappa)
\Gamma(|n|+2\gamma_{\kappa})}x^{\gamma_{\kappa}}x'^{\gamma_{\kappa}}
{\rm e}^{-(x+x')/2}
\nonumber \\
&\times& \left[L_{|n|-1}^{(2\gamma_{\kappa})}(x)
-\frac{\kappa\mp N_{n\kappa}}{|n|+2\gamma_{\kappa}}
L_{|n|}^{(2\gamma_{\kappa})}(x)\right]
\left[L_{|n|-1}^{(2\gamma_{\kappa})}(x')
+\frac{\kappa\mp N_{n\kappa}}{|n|+2\gamma_{\kappa}}
L_{|n|}^{(2\gamma_{\kappa})}(x')\right]
\nonumber \\*[5mm]
&=& x^{\gamma_{\kappa}}x'^{\gamma_{\kappa}}
{\rm e}^{-(x+x')/2}\sum_{n=-\infty}^{\infty}
\frac{\alpha\varepsilon\zeta(|n|+2\gamma_{\kappa})|n|!}
{2N_{n\kappa}(N_{n\kappa}\mp\kappa)\Gamma(|n|+2\gamma_{\kappa})}
\frac{|n|+\gamma_{\kappa}+\varepsilon\zeta\pm N_{n\kappa}}
{(|n|+\gamma_{\kappa}+\varepsilon\zeta)^{2}-N_{n\kappa}^{2}}
\nonumber \\
&\times& \left[L_{|n|-1}^{(2\gamma_{\kappa})}(x)
-\frac{\kappa\mp N_{n\kappa}}{|n|+2\gamma_{\kappa}}
L_{|n|}^{(2\gamma_{\kappa})}(x)\right]
\left[L_{|n|-1}^{(2\gamma_{\kappa})}(x')
+\frac{\kappa\mp N_{n\kappa}}{|n|+2\gamma_{\kappa}}
L_{|n|}^{(2\gamma_{\kappa})}(x')\right]
\nonumber \\*[5mm]
&=& x^{\gamma_{\kappa}}x'^{\gamma_{\kappa}}
{\rm e}^{-(x+x')/2}\sum_{n=-\infty}^{\infty}
\frac{\alpha\varepsilon\zeta(|n|+2\gamma_{\kappa})|n|!}
{2N_{n\kappa}(N_{n\kappa}\mp\kappa)\Gamma(|n|+2\gamma_{\kappa})}
\frac{|n|+\gamma_{\kappa}+\varepsilon\zeta\pm N_{n\kappa}}
{2\varepsilon\zeta(|n|+\gamma_{\kappa}-\omega)}
\nonumber \\
&\times& \left[L_{|n|-1}^{(2\gamma_{\kappa})}(x)
-\frac{\kappa\mp N_{n\kappa}}{|n|+2\gamma_{\kappa}}
L_{|n|}^{(2\gamma_{\kappa})}(x)\right]
\left[L_{|n|-1}^{(2\gamma_{\kappa})}(x')
+\frac{\kappa\mp N_{n\kappa}}{|n|+2\gamma_{\kappa}}
L_{|n|}^{(2\gamma_{\kappa})}(x')\right]
\nonumber \\*[5mm]
&=& \frac{\alpha}{4}
x^{\gamma_{\kappa}}x'^{\gamma_{\kappa}}
{\rm e}^{-(x+x')/2}\sum_{n=-\infty}^{\infty}
\frac{(|n|+\gamma_{\kappa}+\varepsilon\zeta\pm N_{n\kappa})
(|n|+2\gamma_{\kappa})|n|!}{N_{n\kappa}(N_{n\kappa}\mp\kappa)
(|n|+\gamma_{\kappa}-\omega)\Gamma(|n|+2\gamma_{\kappa})}
\nonumber \\
&\times& \left[L_{|n|-1}^{(2\gamma_{\kappa})}(x)
-\frac{\kappa\mp N_{n\kappa}}{|n|+2\gamma_{\kappa}}
L_{|n|}^{(2\gamma_{\kappa})}(x)\right]
\left[L_{|n|-1}^{(2\gamma_{\kappa})}(x')
+\frac{\kappa\mp N_{n\kappa}}{|n|+2\gamma_{\kappa}}
L_{|n|}^{(2\gamma_{\kappa})}(x')\right]
\nonumber \\*[5mm]
&=& \frac{\alpha}{4}x^{\gamma_{\kappa}}
x'^{\gamma_{\kappa}}{\rm e}^{-(x+x')/2}\left[
\sum_{n=-\infty}^{\infty}
\frac{(|n|+\gamma_{\kappa}+\varepsilon\zeta\pm N_{n\kappa})
(|n|+2\gamma_{\kappa})|n|!}{N_{n\kappa}(N_{n\kappa}\mp\kappa)
(|n|+\gamma_{\kappa}-\omega)\Gamma(|n|+2\gamma_{\kappa})}
L_{|n|-1}^{(2\gamma_{\kappa})}(x)L_{|n|-1}^{(2\gamma_{\kappa})}(x')
\right.
\nonumber \\
&+& \sum_{n=-\infty}^{\infty}
\frac{(|n|+\gamma_{\kappa}+\varepsilon\zeta\pm N_{n\kappa})
(|n|+2\gamma_{\kappa})|n|!}{N_{n\kappa}(N_{n\kappa}\mp\kappa)
(|n|+\gamma_{\kappa}-\omega)\Gamma(|n|+2\gamma_{\kappa})}
\frac{\kappa\mp N_{n\kappa}}{|n|+2\gamma_{\kappa}}
L_{|n|-1}^{(2\gamma_{\kappa})}(x)L_{|n|}^{(2\gamma_{\kappa})}(x')
\nonumber \\
&-& \sum_{n=-\infty}^{\infty}
\frac{(|n|+\gamma_{\kappa}+\varepsilon\zeta\pm N_{n\kappa})
(|n|+2\gamma_{\kappa})|n|!}{N_{n\kappa}(N_{n\kappa}\mp\kappa)
(|n|+\gamma_{\kappa}-\omega)\Gamma(|n|+2\gamma_{\kappa})}
\frac{\kappa\mp N_{n\kappa}}{|n|+2\gamma_{\kappa}}
L_{|n|}^{(2\gamma_{\kappa})}(x)L_{|n|-1}^{(2\gamma_{\kappa})}(x')
\nonumber \\
&-& \left.\sum_{n=-\infty}^{\infty}
\frac{(|n|+\gamma_{\kappa}+\varepsilon\zeta\pm N_{n\kappa})
(|n|+2\gamma_{\kappa})|n|!}{N_{n\kappa}(N_{n\kappa}\mp\kappa)
(|n|+\gamma_{\kappa}-\omega)\Gamma(|n|+2\gamma_{\kappa})}
\left(\frac{\kappa\mp N_{n\kappa}}{|n|+2\gamma_{\kappa}}\right)^{2}
L_{|n|}^{(2\gamma_{\kappa})}(x)L_{|n|}^{(2\gamma_{\kappa})}(x')
\right]
\nonumber \\*[5mm]
&=& \frac{\alpha}{4}x^{\gamma_{\kappa}}
x'^{\gamma_{\kappa}}{\rm e}^{-(x+x')/2}\left[
\sum_{n=-\infty}^{\infty}
\frac{(|n|+\gamma_{\kappa}+\varepsilon\zeta\pm N_{n\kappa})
(|n|+2\gamma_{\kappa})|n|!}{N_{n\kappa}(N_{n\kappa}\mp\kappa)
(|n|+\gamma_{\kappa}-\omega)\Gamma(|n|+2\gamma_{\kappa})}
L_{|n|-1}^{(2\gamma_{\kappa})}(x)L_{|n|-1}^{(2\gamma_{\kappa})}(x')
\right.
\nonumber \\
&+& \sum_{n=-\infty}^{\infty}
\frac{\mp(|n|+\gamma_{\kappa}+\varepsilon\zeta\pm N_{n\kappa})|n|!}
{N_{n\kappa}(|n|+\gamma_{\kappa}-\omega)\Gamma(|n|+2\gamma_{\kappa})}
L_{|n|-1}^{(2\gamma_{\kappa})}(x)L_{|n|}^{(2\gamma_{\kappa})}(x')
\nonumber \\
&-& \sum_{n=-\infty}^{\infty}
\frac{\mp(|n|+\gamma_{\kappa}+\varepsilon\zeta\pm N_{n\kappa})|n|!}
{N_{n\kappa}(|n|+\gamma_{\kappa}-\omega)\Gamma(|n|+2\gamma_{\kappa})}
L_{|n|}^{(2\gamma_{\kappa})}(x)L_{|n|-1}^{(2\gamma_{\kappa})}(x')
\nonumber \\
&-& \left.\sum_{n=-\infty}^{\infty}
\frac{(|n|+\gamma_{\kappa}+\varepsilon\zeta\pm N_{n\kappa})
(N_{n\kappa}\mp\kappa)|n|!}{N_{n\kappa}
(|n|+\gamma_{\kappa}-\omega)\Gamma(|n|+2\gamma_{\kappa}+1)}
L_{|n|}^{(2\gamma_{\kappa})}(x)L_{|n|}^{(2\gamma_{\kappa})}(x')
\right]
\nonumber \\*[5mm]
&=& \frac{\alpha}{4}x^{\gamma_{\kappa}}
x'^{\gamma_{\kappa}}{\rm e}^{-(x+x')/2}\left[
\sum_{n=1}^{\infty}
\frac{(|n|+\gamma_{\kappa}+\varepsilon\zeta+N_{n\kappa})
(|n|+2\gamma_{\kappa})|n|!}{N_{n\kappa}(N_{n\kappa}-\kappa)
(|n|+\gamma_{\kappa}-\omega)\Gamma(|n|+2\gamma_{\kappa})}
L_{|n|-1}^{(2\gamma_{\kappa})}(x)L_{|n|-1}^{(2\gamma_{\kappa})}(x')
\right.
\nonumber \\
&+& \sum_{n=-\infty}^{-1}
\frac{(|n|+\gamma_{\kappa}+\varepsilon\zeta-N_{n\kappa})
(|n|+2\gamma_{\kappa})|n|!}{N_{n\kappa}(N_{n\kappa}+\kappa)
(|n|+\gamma_{\kappa}-\omega)\Gamma(|n|+2\gamma_{\kappa})}
L_{|n|-1}^{(2\gamma_{\kappa})}(x)L_{|n|-1}^{(2\gamma_{\kappa})}(x')
\nonumber \\
&-& \sum_{n=1}^{\infty}
\frac{(|n|+\gamma_{\kappa}+\varepsilon\zeta+N_{n\kappa})|n|!}
{N_{n\kappa}(|n|+\gamma_{\kappa}-\omega)\Gamma(|n|+2\gamma_{\kappa})}
L_{|n|-1}^{(2\gamma_{\kappa})}(x)L_{|n|}^{(2\gamma_{\kappa})}(x')
\nonumber \\
&+& \sum_{n=-\infty}^{-1}
\frac{(|n|+\gamma_{\kappa}+\varepsilon\zeta-N_{n\kappa})|n|!}
{N_{n\kappa}(|n|+\gamma_{\kappa}-\omega)\Gamma(|n|+2\gamma_{\kappa})}
L_{|n|-1}^{(2\gamma_{\kappa})}(x)L_{|n|}^{(2\gamma_{\kappa})}(x')
\nonumber \\
&+& \sum_{n=1}^{\infty}
\frac{(|n|+\gamma_{\kappa}+\varepsilon\zeta+N_{n\kappa})|n|!}
{N_{n\kappa}(|n|+\gamma_{\kappa}-\omega)\Gamma(|n|+2\gamma_{\kappa})}
L_{|n|}^{(2\gamma_{\kappa})}(x)L_{|n|-1}^{(2\gamma_{\kappa})}(x')
\nonumber \\
&-& \sum_{n=-\infty}^{-1}
\frac{(|n|+\gamma_{\kappa}+\varepsilon\zeta-N_{n\kappa})|n|!}
{N_{n\kappa}(|n|+\gamma_{\kappa}-\omega)\Gamma(|n|+2\gamma_{\kappa})}
L_{|n|}^{(2\gamma_{\kappa})}(x)L_{|n|-1}^{(2\gamma_{\kappa})}(x')
\nonumber \\
&-& \sum_{n=0}^{\infty}
\frac{(|n|+\gamma_{\kappa}+\varepsilon\zeta+N_{n\kappa})
(N_{n\kappa}-\kappa)|n|!}{N_{n\kappa}
(|n|+\gamma_{\kappa}-\omega)\Gamma(|n|+2\gamma_{\kappa}+1)}
L_{|n|}^{(2\gamma_{\kappa})}(x)L_{|n|}^{(2\gamma_{\kappa})}(x')
\nonumber \\
&-& \left.\sum_{n=-\infty}^{0}
\frac{(|n|+\gamma_{\kappa}+\varepsilon\zeta-N_{n\kappa})
(N_{n\kappa}+\kappa)|n|!}{N_{n\kappa}
(|n|+\gamma_{\kappa}-\omega)\Gamma(|n|+2\gamma_{\kappa}+1)}
L_{|n|}^{(2\gamma_{\kappa})}(x)L_{|n|}^{(2\gamma_{\kappa})}(x')
\right]
\nonumber \\*[5mm]
&=& \frac{\alpha}{4}x^{\gamma_{\kappa}}
x'^{\gamma_{\kappa}}{\rm e}^{-(x+x')/2}\left[
\sum_{n=1}^{\infty}
\frac{(n+\gamma_{\kappa}+\varepsilon\zeta+N_{n\kappa})
(n+2\gamma_{\kappa})n!}{N_{n\kappa}(N_{n\kappa}-\kappa)
(n+\gamma_{\kappa}-\omega)\Gamma(n+2\gamma_{\kappa})}
L_{n-1}^{(2\gamma_{\kappa})}(x)L_{n-1}^{(2\gamma_{\kappa})}(x')
\right.
\nonumber \\
&+& \sum_{n=1}^{\infty}
\frac{(n+\gamma_{\kappa}+\varepsilon\zeta-N_{n\kappa})
(n+2\gamma_{\kappa})n!}{N_{n\kappa}(N_{n\kappa}+\kappa)
(n+\gamma_{\kappa}-\omega)\Gamma(n+2\gamma_{\kappa})}
L_{n-1}^{(2\gamma_{\kappa})}(x)L_{n-1}^{(2\gamma_{\kappa})}(x')
\nonumber \\
&-& \sum_{n=1}^{\infty}
\frac{(n+\gamma_{\kappa}+\varepsilon\zeta+N_{n\kappa})n!}
{N_{n\kappa}(n+\gamma_{\kappa}-\omega)\Gamma(n+2\gamma_{\kappa})}
L_{n-1}^{(2\gamma_{\kappa})}(x)L_{n}^{(2\gamma_{\kappa})}(x')
\nonumber \\
&+& \sum_{n=1}^{\infty}
\frac{(n+\gamma_{\kappa}+\varepsilon\zeta-N_{n\kappa})n!}
{N_{n\kappa}(n+\gamma_{\kappa}-\omega)\Gamma(n+2\gamma_{\kappa})}
L_{n-1}^{(2\gamma_{\kappa})}(x)L_{n}^{(2\gamma_{\kappa})}(x')
\nonumber \\
&+& \sum_{n=1}^{\infty}
\frac{(n+\gamma_{\kappa}+\varepsilon\zeta+N_{n\kappa})n!}
{N_{n\kappa}(n+\gamma_{\kappa}-\omega)\Gamma(n+2\gamma_{\kappa})}
L_{n}^{(2\gamma_{\kappa})}(x)L_{n-1}^{(2\gamma_{\kappa})}(x')
\nonumber \\
&-& \sum_{n=1}^{\infty}
\frac{(n+\gamma_{\kappa}+\varepsilon\zeta-N_{n\kappa})n!}
{N_{n\kappa}(n+\gamma_{\kappa}-\omega)\Gamma(n+2\gamma_{\kappa})}
L_{n}^{(2\gamma_{\kappa})}(x)L_{n-1}^{(2\gamma_{\kappa})}(x')
\nonumber \\
&-& \sum_{n=0}^{\infty}
\frac{(n+\gamma_{\kappa}+\varepsilon\zeta+N_{n\kappa})
(N_{n\kappa}-\kappa)n!}{N_{n\kappa}
(n+\gamma_{\kappa}-\omega)\Gamma(n+2\gamma_{\kappa}+1)}
L_{n}^{(2\gamma_{\kappa})}(x)L_{n}^{(2\gamma_{\kappa})}(x')
\nonumber \\
&-& \left.\sum_{n=0}^{\infty}
\frac{(n+\gamma_{\kappa}+\varepsilon\zeta-N_{n\kappa})
(N_{n\kappa}+\kappa)n!}{N_{n\kappa}
(n+\gamma_{\kappa}-\omega)\Gamma(n+2\gamma_{\kappa}+1)}
L_{n}^{(2\gamma_{\kappa})}(x)L_{n}^{(2\gamma_{\kappa})}(x')
\right]
\end{eqnarray*}
\begin{eqnarray}
&=& \frac{\alpha}{4}x^{\gamma_{\kappa}}
x'^{\gamma_{\kappa}}{\rm e}^{-(x+x')/2}\left[
\sum_{n=1}^{\infty}\frac{(n+\gamma_{\kappa}+\varepsilon\zeta
+N_{n\kappa})(N_{n\kappa}+\kappa)(n-1)!}{N_{n\kappa}
(n+\gamma_{\kappa}-\omega)\Gamma(n+2\gamma_{\kappa})}
L_{n-1}^{(2\gamma_{\kappa})}(x)L_{n-1}^{(2\gamma_{\kappa})}(x')
\right.
\nonumber \\
&+& \sum_{n=1}^{\infty}\frac{(n+\gamma_{\kappa}+\varepsilon\zeta
-N_{n\kappa})(N_{n\kappa}-\kappa)(n-1)!}{N_{n\kappa}
(n+\gamma_{\kappa}-\omega)\Gamma(n+2\gamma_{\kappa})}
L_{n-1}^{(2\gamma_{\kappa})}(x)L_{n-1}^{(2\gamma_{\kappa})}(x')
\nonumber \\
&-& 2\sum_{n=1}^{\infty}\frac{n!}{(n+\gamma_{\kappa}-\omega)
\Gamma(n+2\gamma_{\kappa})}
L_{n-1}^{(2\gamma_{\kappa})}(x)L_{n}^{(2\gamma_{\kappa})}(x')
\nonumber \\
&+& 2\sum_{n=1}^{\infty}\frac{n!}{(n+\gamma_{\kappa}-\omega)
\Gamma(n+2\gamma_{\kappa})}
L_{n}^{(2\gamma_{\kappa})}(x)L_{n-1}^{(2\gamma_{\kappa})}(x')
\nonumber \\
&-& \sum_{n=0}^{\infty}
\frac{(n+\gamma_{\kappa}+\varepsilon\zeta+N_{n\kappa})
(N_{n\kappa}-\kappa)n!}{N_{n\kappa}(n+\gamma_{\kappa}-\omega)
\Gamma(n+2\gamma_{\kappa}+1)}
L_{n}^{(2\gamma_{\kappa})}(x)L_{n}^{(2\gamma_{\kappa})}(x')
\nonumber \\
&-& \left.\sum_{n=0}^{\infty}
\frac{(n+\gamma_{\kappa}+\varepsilon\zeta-N_{n\kappa})
(N_{n\kappa}+\kappa)n!}{N_{n\kappa}(n+\gamma_{\kappa}-\omega)
\Gamma(n+2\gamma_{\kappa}+1)}
L_{n}^{(2\gamma_{\kappa})}(x)L_{n}^{(2\gamma_{\kappa})}(x')
\right]
\nonumber \\*[5mm]
&=& \frac{\alpha}{2}x^{\gamma_{\kappa}}
x'^{\gamma_{\kappa}}{\rm e}^{-(x+x')/2}\left[
\sum_{n=1}^{\infty}\frac{(n+\gamma_{\kappa}+\varepsilon\zeta+\kappa)
(n-1)!}{(n+\gamma_{\kappa}-\omega)\Gamma(n+2\gamma_{\kappa})}
L_{n-1}^{(2\gamma_{\kappa})}(x)L_{n-1}^{(2\gamma_{\kappa})}(x')
\right.
\nonumber \\
&-& \sum_{n=0}^{\infty}\frac{n!}{(n+\gamma_{\kappa}-\omega)
\Gamma(n+2\gamma_{\kappa})}L_{n-1}^{(2\gamma_{\kappa})}(x)
L_{n}^{(2\gamma_{\kappa})}(x')
\nonumber \\
&+& \sum_{n=0}^{\infty}\frac{n!}{(n+\gamma_{\kappa}-\omega)
\Gamma(n+2\gamma_{\kappa})}L_{n}^{(2\gamma_{\kappa})}(x)
L_{n-1}^{(2\gamma_{\kappa})}(x')
\nonumber \\
&-& \left.
\sum_{n=0}^{\infty}\frac{(n+\gamma_{\kappa}+\varepsilon\zeta
-\kappa)n!}{(n+\gamma_{\kappa}-\omega)\Gamma(n+2\gamma_{\kappa}+1)}
L_{n}^{(2\gamma_{\kappa})}(x)L_{n}^{(2\gamma_{\kappa})}(x')\right]
\nonumber \\*[5mm]
&=& \frac{\alpha}{2}x^{\gamma_{\kappa}}
x'^{\gamma_{\kappa}}{\rm e}^{-(x+x')/2}\left[
\sum_{n=0}^{\infty}\frac{(n+\gamma_{\kappa}+\varepsilon\zeta
+\kappa+1)n!}{(n+\gamma_{\kappa}-\omega+1)
\Gamma(n+2\gamma_{\kappa}+1)}
L_{n}^{(2\gamma_{\kappa})}(x)L_{n}^{(2\gamma_{\kappa})}(x')
\right.
\nonumber \\
&-& \sum_{n=0}^{\infty}\frac{n!}{(n+\gamma_{\kappa}-\omega)
\Gamma(n+2\gamma_{\kappa})}L_{n-1}^{(2\gamma_{\kappa})}(x)
L_{n}^{(2\gamma_{\kappa})}(x')
\nonumber \\
&+& \sum_{n=0}^{\infty}\frac{n!}{(n+\gamma_{\kappa}-\omega)
\Gamma(n+2\gamma_{\kappa})}L_{n}^{(2\gamma_{\kappa})}(x)
L_{n-1}^{(2\gamma_{\kappa})}(x')
\nonumber \\
&-& \left.
\sum_{n=0}^{\infty}\frac{(n+\gamma_{\kappa}+\varepsilon\zeta
-\kappa)n!}{(n+\gamma_{\kappa}-\omega)\Gamma(n+2\gamma_{\kappa}+1)}
L_{n}^{(2\gamma_{\kappa})}(x)L_{n}^{(2\gamma_{\kappa})}(x')\right]
\nonumber \\*[5mm]
&=& \frac{\alpha}{2}x^{\gamma_{\kappa}}
x'^{\gamma_{\kappa}}{\rm e}^{-(x+x')/2}\left\{
\sum_{n=0}^{\infty}\frac{(n+\gamma_{\kappa}+\varepsilon\zeta
+\kappa+1)n!}{(n+\gamma_{\kappa}-\omega+1)
\Gamma(n+2\gamma_{\kappa}+1)}
L_{n}^{(2\gamma_{\kappa})}(x)L_{n}^{(2\gamma_{\kappa})}(x')
\right.
\nonumber \\
&-& \sum_{n=0}^{\infty}\frac{n!}{(n+\gamma_{\kappa}-\omega)
\Gamma(n+2\gamma_{\kappa})}\left[L_{n}^{(2\gamma_{\kappa})}(x)
-L_{n}^{(2\gamma_{\kappa}-1)}(x)\right]L_{n}^{(2\gamma_{\kappa})}(x')
\nonumber \\
&+& \sum_{n=0}^{\infty}\frac{n!}{(n+\gamma_{\kappa}-\omega)
\Gamma(n+2\gamma_{\kappa})}L_{n}^{(2\gamma_{\kappa})}(x)
\left[L_{n}^{(2\gamma_{\kappa})}(x')
-L_{n}^{(2\gamma_{\kappa}-1)}(x')\right]
\nonumber \\
&-& \left.
\sum_{n=0}^{\infty}\frac{(n+\gamma_{\kappa}+\varepsilon\zeta
-\kappa)n!}{(n+\gamma_{\kappa}-\omega)\Gamma(n+2\gamma_{\kappa}+1)}
L_{n}^{(2\gamma_{\kappa})}(x)L_{n}^{(2\gamma_{\kappa})}(x')\right\}
\nonumber \\*[5mm]
&=& \frac{\alpha}{2}x^{\gamma_{\kappa}}
x'^{\gamma_{\kappa}}{\rm e}^{-(x+x')/2}\left[
\sum_{n=0}^{\infty}\frac{(n+\gamma_{\kappa}+\varepsilon\zeta
+\kappa+1)n!}{(n+\gamma_{\kappa}-\omega+1)
\Gamma(n+2\gamma_{\kappa}+1)}
L_{n}^{(2\gamma_{\kappa})}(x)L_{n}^{(2\gamma_{\kappa})}(x')
\right.
\nonumber \\
&-& \sum_{n=0}^{\infty}\frac{n!}{(n+\gamma_{\kappa}-\omega)
\Gamma(n+2\gamma_{\kappa})}L_{n}^{(2\gamma_{\kappa})}(x)
L_{n}^{(2\gamma_{\kappa})}(x')
\nonumber \\
&+& \sum_{n=0}^{\infty}\frac{n!}{(n+\gamma_{\kappa}-\omega)
\Gamma(n+2\gamma_{\kappa})}L_{n}^{(2\gamma_{\kappa}-1)}(x)
L_{n}^{(2\gamma_{\kappa})}(x')
\nonumber \\
&+& \sum_{n=0}^{\infty}\frac{n!}{(n+\gamma_{\kappa}-\omega)
\Gamma(n+2\gamma_{\kappa})}L_{n}^{(2\gamma_{\kappa})}(x)
L_{n}^{(2\gamma_{\kappa})}(x')
\nonumber \\
&-& \sum_{n=0}^{\infty}\frac{n!}{(n+\gamma_{\kappa}-\omega)
\Gamma(n+2\gamma_{\kappa})}L_{n}^{(2\gamma_{\kappa})}(x)
L_{n}^{(2\gamma_{\kappa}-1)}(x')
\nonumber \\
&-& \left.
\sum_{n=0}^{\infty}\frac{(n+\gamma_{\kappa}+\varepsilon\zeta
-\kappa)n!}{(n+\gamma_{\kappa}-\omega)\Gamma(n+2\gamma_{\kappa}+1)}
L_{n}^{(2\gamma_{\kappa})}(x)L_{n}^{(2\gamma_{\kappa})}(x')\right]
\nonumber \\*[5mm]
&=& \frac{\alpha}{2}x^{\gamma_{\kappa}}
x'^{\gamma_{\kappa}}{\rm e}^{-(x+x')/2}\left\{
\sum_{n=0}^{\infty}\frac{(n+\gamma_{\kappa}+\varepsilon\zeta
+\kappa+1)n!}{(n+\gamma_{\kappa}-\omega+1)
\Gamma(n+2\gamma_{\kappa}+1)}
L_{n}^{(2\gamma_{\kappa})}(x)L_{n}^{(2\gamma_{\kappa})}(x')
\right.
\nonumber \\
&+& \sum_{n=0}^{\infty}\frac{n!}{(n+\gamma_{\kappa}-\omega)
\Gamma(n+2\gamma_{\kappa})}\left[L_{n}^{(2\gamma_{\kappa}-1)}(x)
L_{n}^{(2\gamma_{\kappa})}(x')-L_{n}^{(2\gamma_{\kappa})}(x)
L_{n}^{(2\gamma_{\kappa}-1)}(x')\right]
\nonumber \\
&-& \left.
\sum_{n=0}^{\infty}\frac{(n+\gamma_{\kappa}+\varepsilon\zeta
-\kappa)n!}{(n+\gamma_{\kappa}-\omega)\Gamma(n+2\gamma_{\kappa}+1)}
L_{n}^{(2\gamma_{\kappa})}(x)L_{n}^{(2\gamma_{\kappa})}(x')\right\}
\nonumber \\*[5mm]
&=& \frac{\alpha}{2}x^{\gamma_{\kappa}}
x'^{\gamma_{\kappa}}{\rm e}^{-(x+x')/2}\left\{
\sum_{n=0}^{\infty}\frac{(n+\gamma_{\kappa}-\omega+1+\varepsilon\zeta
+\omega+\kappa)n!}{(n+\gamma_{\kappa}-\omega+1)
\Gamma(n+2\gamma_{\kappa}+1)}
L_{n}^{(2\gamma_{\kappa})}(x)L_{n}^{(2\gamma_{\kappa})}(x')
\right.
\nonumber \\
&+& \sum_{n=0}^{\infty}\frac{n!}{(n+\gamma_{\kappa}-\omega)
\Gamma(n+2\gamma_{\kappa})}\left[L_{n}^{(2\gamma_{\kappa}-1)}(x)
L_{n}^{(2\gamma_{\kappa})}(x')-L_{n}^{(2\gamma_{\kappa})}(x)
L_{n}^{(2\gamma_{\kappa}-1)}(x')\right]
\nonumber \\
&-& \left.\sum_{n=0}^{\infty}\frac{(n+\gamma_{\kappa}-\omega
+\varepsilon\zeta+\omega-\kappa)n!}
{(n+\gamma_{\kappa}-\omega)\Gamma(n+2\gamma_{\kappa}+1)}
L_{n}^{(2\gamma_{\kappa})}(x)L_{n}^{(2\gamma_{\kappa})}(x')\right\}
\nonumber \\*[5mm]
&=& \frac{\alpha}{2}x^{\gamma_{\kappa}}
x'^{\gamma_{\kappa}}{\rm e}^{-(x+x')/2}\left\{
\sum_{n=0}^{\infty}\frac{n!}{\Gamma(n+2\gamma_{\kappa}+1)}
L_{n}^{(2\gamma_{\kappa})}(x)L_{n}^{(2\gamma_{\kappa})}(x')\right.
\nonumber \\
&+&
\sum_{n=0}^{\infty}\frac{(\varepsilon\zeta+\omega+\kappa)n!}
{(n+\gamma_{\kappa}-\omega+1)\Gamma(n+2\gamma_{\kappa}+1)}
L_{n}^{(2\gamma_{\kappa})}(x)L_{n}^{(2\gamma_{\kappa})}(x')
\nonumber \\
&+& \sum_{n=0}^{\infty}\frac{n!}{(n+\gamma_{\kappa}-\omega)
\Gamma(n+2\gamma_{\kappa})}\left[L_{n}^{(2\gamma_{\kappa}-1)}(x)
L_{n}^{(2\gamma_{\kappa})}(x')-L_{n}^{(2\gamma_{\kappa})}(x)
L_{n}^{(2\gamma_{\kappa}-1)}(x')\right]
\nonumber \\
&-& \sum_{n=0}^{\infty}\frac{n!}{\Gamma(n+2\gamma_{\kappa}+1)}
L_{n}^{(2\gamma_{\kappa})}(x)L_{n}^{(2\gamma_{\kappa})}(x')
\nonumber \\
&-& \left.\sum_{n=0}^{\infty}\frac{(\varepsilon\zeta+\omega
-\kappa)n!}{(n+\gamma_{\kappa}-\omega)\Gamma(n+2\gamma_{\kappa}+1)}
L_{n}^{(2\gamma_{\kappa})}(x)L_{n}^{(2\gamma_{\kappa})}(x')\right\}
\nonumber \\*[5mm]
&=& \frac{\alpha}{2}x^{\gamma_{\kappa}}
x'^{\gamma_{\kappa}}{\rm e}^{-(x+x')/2}\left\{
(\varepsilon\zeta+\omega+\kappa)\sum_{n=0}^{\infty}\frac{n!}
{(n+\gamma_{\kappa}-\omega+1)\Gamma(n+2\gamma_{\kappa}+1)}
L_{n}^{(2\gamma_{\kappa})}(x)L_{n}^{(2\gamma_{\kappa})}(x')\right.
\nonumber \\
&+& \sum_{n=0}^{\infty}\frac{n!}{(n+\gamma_{\kappa}-\omega)
\Gamma(n+2\gamma_{\kappa})}\left[L_{n}^{(2\gamma_{\kappa}-1)}(x)
L_{n}^{(2\gamma_{\kappa})}(x')-L_{n}^{(2\gamma_{\kappa})}(x)
L_{n}^{(2\gamma_{\kappa}-1)}(x')\right]
\nonumber \\
&-& \left.(\varepsilon\zeta+\omega-\kappa)
\sum_{n=0}^{\infty}\frac{n!}{(n+\gamma_{\kappa}-\omega)
\Gamma(n+2\gamma_{\kappa}+1)}
L_{n}^{(2\gamma_{\kappa})}(x)L_{n}^{(2\gamma_{\kappa})}(x')\right\}
\nonumber \\*[5mm]
&=& \frac{\alpha}{2}x^{\gamma_{\kappa}}
x'^{\gamma_{\kappa}}{\rm e}^{-(x+x')/2}
\left\{\left(\frac{\zeta}{\sqrt{1-\epsilon^{2}}}+\kappa\right)
\sum_{n=0}^{\infty}\frac{n!}{(n+\gamma_{\kappa}-\omega+1)
\Gamma(n+2\gamma_{\kappa}+1)}L_{n}^{(2\gamma_{\kappa})}(x)
L_{n}^{(2\gamma_{\kappa})}(x')\right.
\nonumber \\
&+& \sum_{n=0}^{\infty}\frac{n!}{(n+\gamma_{\kappa}-\omega)
\Gamma(n+2\gamma_{\kappa})}\left[L_{n}^{(2\gamma_{\kappa}-1)}(x)
L_{n}^{(2\gamma_{\kappa})}(x')-L_{n}^{(2\gamma_{\kappa})}(x)
L_{n}^{(2\gamma_{\kappa}-1)}(x')\right]
\nonumber \\
&-& \left.\left(\frac{\zeta}{\sqrt{1-\epsilon^{2}}}
-\kappa\right)\sum_{n=0}^{\infty}
\frac{n!}{(n+\gamma_{\kappa}-\omega)\Gamma(n+2\gamma_{\kappa}+1)}
L_{n}^{(2\gamma_{\kappa})}(x)L_{n}^{(2\gamma_{\kappa})}(x')\right\}
\label{27}
\end{eqnarray}
\begin{eqnarray*}
&& g_{\kappa}^{(22)}(r,r';E)
=\sum_{n=-\infty}^{\infty}\frac{1}{\mu_{n\kappa}-1}
T_{n\kappa}(x)T_{n\kappa}(x')
\nonumber \\*[5mm]
&=& \sum_{n=-\infty}^{\infty}\frac{1}
{\varepsilon\frac{|n|+\gamma_{\kappa}\pm N_{n\kappa}}{\zeta}-1}
\frac{\alpha\varepsilon(|n|+2\gamma_{\kappa})|n|!}
{2N_{n\kappa}(N_{n\kappa}\mp\kappa)\Gamma(|n|+2\gamma_{\kappa})}
x^{\gamma_{\kappa}}x'^{\gamma_{\kappa}}{\rm e}^{-(x+x')/2}
\nonumber \\
&\times& \left[L_{|n|-1}^{(2\gamma_{\kappa})}(x)
-\frac{\kappa\mp N_{n\kappa}}{|n|+2\gamma_{\kappa}}
L_{|n|}^{(2\gamma_{\kappa})}(x)\right]
\left[L_{|n|-1}^{(2\gamma_{\kappa})}(x')
-\frac{\kappa\mp N_{n\kappa}}{|n|+2\gamma_{\kappa}}
L_{|n|}^{(2\gamma_{\kappa})}(x')\right]
\nonumber \\*[5mm]
&=& x^{\gamma_{\kappa}}x'^{\gamma_{\kappa}}{\rm e}^{-(x+x')/2}
\sum_{n=-\infty}^{\infty}
\frac{\alpha\zeta(|n|+2\gamma_{\kappa})|n|!}
{2N_{n\kappa}(N_{n\kappa}\mp\kappa)\Gamma(|n|+2\gamma_{\kappa})}
\frac{|n|+\gamma_{\kappa}-\varepsilon^{-1}\zeta\mp N_{n\kappa}}
{(|n|+\gamma_{\kappa}-\varepsilon^{-1}\zeta)^{2}-N_{n\kappa}^{2}}
\nonumber \\
&\times& \left[L_{|n|-1}^{(2\gamma_{\kappa})}(x)
-\frac{\kappa\mp N_{n\kappa}}{|n|+2\gamma_{\kappa}}
L_{|n|}^{(2\gamma_{\kappa})}(x)\right]
\left[L_{|n|-1}^{(2\gamma_{\kappa})}(x')
-\frac{\kappa\mp N_{n\kappa}}{|n|+2\gamma_{\kappa}}
L_{|n|}^{(2\gamma_{\kappa})}(x')\right]
\nonumber \\*[5mm]
&=& x^{\gamma_{\kappa}}x'^{\gamma_{\kappa}}{\rm e}^{-(x+x')/2}
\sum_{n=-\infty}^{\infty}
\frac{\alpha\zeta(|n|+2\gamma_{\kappa})|n|!}
{2N_{n\kappa}(N_{n\kappa}\mp\kappa)\Gamma(|n|+2\gamma_{\kappa})}
\frac{|n|+\gamma_{\kappa}-\varepsilon^{-1}\zeta\mp N_{n\kappa}}
{(-2)\varepsilon^{-1}\zeta(|n|+\gamma_{\kappa}-\omega)}
\nonumber \\
&\times& \left[L_{|n|-1}^{(2\gamma_{\kappa})}(x)
-\frac{\kappa\mp N_{n\kappa}}{|n|+2\gamma_{\kappa}}
L_{|n|}^{(2\gamma_{\kappa})}(x)\right]
\left[L_{|n|-1}^{(2\gamma_{\kappa})}(x')
-\frac{\kappa\mp N_{n\kappa}}{|n|+2\gamma_{\kappa}}
L_{|n|}^{(2\gamma_{\kappa})}(x')\right]
\nonumber \\
&=& -\frac{\alpha\varepsilon}{4}
x^{\gamma_{\kappa}}x'^{\gamma_{\kappa}}{\rm e}^{-(x+x')/2}
\sum_{n=-\infty}^{\infty}
\frac{(|n|+\gamma_{\kappa}-\varepsilon^{-1}\zeta\mp N_{n\kappa})
(|n|+2\gamma_{\kappa})|n|!}{N_{n\kappa}(N_{n\kappa}\mp\kappa)
(|n|+\gamma_{\kappa}-\omega)\Gamma(|n|+2\gamma_{\kappa})}
\nonumber \\
&\times& \left[L_{|n|-1}^{(2\gamma_{\kappa})}(x)
-\frac{\kappa\mp N_{n\kappa}}{|n|+2\gamma_{\kappa}}
L_{|n|}^{(2\gamma_{\kappa})}(x)\right]
\left[L_{|n|-1}^{(2\gamma_{\kappa})}(x')
-\frac{\kappa\mp N_{n\kappa}}{|n|+2\gamma_{\kappa}}
L_{|n|}^{(2\gamma_{\kappa})}(x')\right]
\nonumber \\*[5mm]
&=& -\frac{\alpha\varepsilon}{4}x^{\gamma_{\kappa}}
x'^{\gamma_{\kappa}}{\rm e}^{-(x+x')/2}\left[
\sum_{n=-\infty}^{\infty}
\frac{(|n|+\gamma_{\kappa}-\varepsilon^{-1}\zeta\mp N_{n\kappa})
(|n|+2\gamma_{\kappa})|n|!}{N_{n\kappa}(N_{n\kappa}\mp\kappa)
(|n|+\gamma_{\kappa}-\omega)\Gamma(|n|+2\gamma_{\kappa})}
L_{|n|-1}^{(2\gamma_{\kappa})}(x)L_{|n|-1}^{(2\gamma_{\kappa})}(x')
\right.
\nonumber \\
&-& \sum_{n=-\infty}^{\infty}
\frac{(|n|+\gamma_{\kappa}-\varepsilon^{-1}\zeta\mp N_{n\kappa})
(|n|+2\gamma_{\kappa})|n|!}{N_{n\kappa}(N_{n\kappa}\mp\kappa)
(|n|+\gamma_{\kappa}-\omega)\Gamma(|n|+2\gamma_{\kappa})}
\frac{\kappa\mp N_{n\kappa}}{|n|+2\gamma_{\kappa}}
L_{|n|-1}^{(2\gamma_{\kappa})}(x)L_{|n|}^{(2\gamma_{\kappa})}(x')
\nonumber \\
&-& \sum_{n=-\infty}^{\infty}
\frac{(|n|+\gamma_{\kappa}-\varepsilon^{-1}\zeta\mp N_{n\kappa})
(|n|+2\gamma_{\kappa})|n|!}{N_{n\kappa}(N_{n\kappa}\mp\kappa)
(|n|+\gamma_{\kappa}-\omega)\Gamma(|n|+2\gamma_{\kappa})}
\frac{\kappa\mp N_{n\kappa}}{|n|+2\gamma_{\kappa}}
L_{|n|}^{(2\gamma_{\kappa})}(x)L_{|n|-1}^{(2\gamma_{\kappa})}(x')
\nonumber \\
&+& \left.\sum_{n=-\infty}^{\infty}
\frac{(|n|+\gamma_{\kappa}-\varepsilon^{-1}\zeta\mp N_{n\kappa})
(|n|+2\gamma_{\kappa})|n|!}{N_{n\kappa}(N_{n\kappa}\mp\kappa)
(|n|+\gamma_{\kappa}-\omega)\Gamma(|n|+2\gamma_{\kappa})}
\left(\frac{\kappa\mp N_{n\kappa}}{|n|+2\gamma_{\kappa}}\right)^{2}
L_{|n|}^{(2\gamma_{\kappa})}(x)L_{|n|}^{(2\gamma_{\kappa})}(x')
\right]
\nonumber \\*[5mm]
&=& -\frac{\alpha\varepsilon}{4}x^{\gamma_{\kappa}}
x'^{\gamma_{\kappa}}{\rm e}^{-(x+x')/2}\left[
\sum_{n=-\infty}^{\infty}
\frac{(|n|+\gamma_{\kappa}-\varepsilon^{-1}\zeta\mp N_{n\kappa})
(|n|+2\gamma_{\kappa})|n|!}{N_{n\kappa}(N_{n\kappa}\mp\kappa)
(|n|+\gamma_{\kappa}-\omega)\Gamma(|n|+2\gamma_{\kappa})}
L_{|n|-1}^{(2\gamma_{\kappa})}(x)L_{|n|-1}^{(2\gamma_{\kappa})}(x')
\right.
\nonumber \\
&-& \sum_{n=-\infty}^{\infty}\frac{\mp(|n|+\gamma_{\kappa}
-\varepsilon^{-1}\zeta\mp N_{n\kappa})|n|!}
{N_{n\kappa}(|n|+\gamma_{\kappa}-\omega)\Gamma(|n|+2\gamma_{\kappa})}
L_{|n|-1}^{(2\gamma_{\kappa})}(x)L_{|n|}^{(2\gamma_{\kappa})}(x')
\nonumber \\
&-& \sum_{n=-\infty}^{\infty}\frac{\mp(|n|+\gamma_{\kappa}
-\varepsilon^{-1}\zeta\mp N_{n\kappa})|n|!}
{N_{n\kappa}(|n|+\gamma_{\kappa}-\omega)\Gamma(|n|+2\gamma_{\kappa})}
L_{|n|}^{(2\gamma_{\kappa})}(x)L_{|n|-1}^{(2\gamma_{\kappa})}(x')
\nonumber \\
&+& \left.\sum_{n=-\infty}^{\infty}
\frac{(|n|+\gamma_{\kappa}-\varepsilon^{-1}\zeta\mp N_{n\kappa})
(N_{n\kappa}\mp\kappa)|n|!}{N_{n\kappa}
(|n|+\gamma_{\kappa}-\omega)\Gamma(|n|+2\gamma_{\kappa}+1)}
L_{|n|}^{(2\gamma_{\kappa})}(x)L_{|n|}^{(2\gamma_{\kappa})}(x')
\right]
\nonumber \\*[5mm]
&=& -\frac{\alpha\varepsilon}{4}x^{\gamma_{\kappa}}
x'^{\gamma_{\kappa}}{\rm e}^{-(x+x')/2}\left[
\sum_{n=1}^{\infty}
\frac{(|n|+\gamma_{\kappa}-\varepsilon^{-1}\zeta-N_{n\kappa})
(|n|+2\gamma_{\kappa})|n|!}{N_{n\kappa}(N_{n\kappa}-\kappa)
(|n|+\gamma_{\kappa}-\omega)\Gamma(|n|+2\gamma_{\kappa})}
L_{|n|-1}^{(2\gamma_{\kappa})}(x)L_{|n|-1}^{(2\gamma_{\kappa})}(x')
\right.
\nonumber \\
&+& \sum_{n=-\infty}^{-1}
\frac{(|n|+\gamma_{\kappa}-\varepsilon^{-1}\zeta+N_{n\kappa})
(|n|+2\gamma_{\kappa})|n|!}{N_{n\kappa}(N_{n\kappa}+\kappa)
(|n|+\gamma_{\kappa}-\omega)\Gamma(|n|+2\gamma_{\kappa})}
L_{|n|-1}^{(2\gamma_{\kappa})}(x)L_{|n|-1}^{(2\gamma_{\kappa})}(x')
\nonumber \\
&+& \sum_{n=1}^{\infty}
\frac{(|n|+\gamma_{\kappa}-\varepsilon^{-1}\zeta-N_{n\kappa})|n|!}
{N_{n\kappa}(|n|+\gamma_{\kappa}-\omega)\Gamma(|n|+2\gamma_{\kappa})}
L_{|n|-1}^{(2\gamma_{\kappa})}(x)L_{|n|}^{(2\gamma_{\kappa})}(x')
\nonumber \\
&-& \sum_{n=-\infty}^{-1}
\frac{(|n|+\gamma_{\kappa}-\varepsilon^{-1}\zeta+N_{n\kappa})|n|!}
{N_{n\kappa}(|n|+\gamma_{\kappa}-\omega)\Gamma(|n|+2\gamma_{\kappa})}
L_{|n|-1}^{(2\gamma_{\kappa})}(x)L_{|n|}^{(2\gamma_{\kappa})}(x')
\nonumber \\
&+& \sum_{n=1}^{\infty}
\frac{(|n|+\gamma_{\kappa}-\varepsilon^{-1}\zeta-N_{n\kappa})|n|!}
{N_{n\kappa}(|n|+\gamma_{\kappa}-\omega)\Gamma(|n|+2\gamma_{\kappa})}
L_{|n|}^{(2\gamma_{\kappa})}(x)L_{|n|-1}^{(2\gamma_{\kappa})}(x')
\nonumber \\
&-& \sum_{n=-\infty}^{-1}
\frac{(|n|+\gamma_{\kappa}-\varepsilon^{-1}\zeta+N_{n\kappa})|n|!}
{N_{n\kappa}(|n|+\gamma_{\kappa}-\omega)\Gamma(|n|+2\gamma_{\kappa})}
L_{|n|}^{(2\gamma_{\kappa})}(x)L_{|n|-1}^{(2\gamma_{\kappa})}(x')
\nonumber \\
&+& \sum_{n=0}^{\infty}
\frac{(|n|+\gamma_{\kappa}-\varepsilon^{-1}\zeta-N_{n\kappa})
(N_{n\kappa}-\kappa)|n|!}{N_{n\kappa}
(|n|+\gamma_{\kappa}-\omega)\Gamma(|n|+2\gamma_{\kappa}+1)}
L_{|n|}^{(2\gamma_{\kappa})}(x)L_{|n|}^{(2\gamma_{\kappa})}(x')
\nonumber \\
&+& \left.\sum_{n=-\infty}^{0}
\frac{(|n|+\gamma_{\kappa}-\varepsilon^{-1}\zeta+N_{n\kappa})
(N_{n\kappa}+\kappa)|n|!}{N_{n\kappa}
(|n|+\gamma_{\kappa}-\omega)\Gamma(|n|+2\gamma_{\kappa}+1)}
L_{|n|}^{(2\gamma_{\kappa})}(x)L_{|n|}^{(2\gamma_{\kappa})}(x')
\right]
\nonumber \\*[5mm]
&=& -\frac{\alpha\varepsilon}{4}x^{\gamma_{\kappa}}
x'^{\gamma_{\kappa}}{\rm e}^{-(x+x')/2}\left[
\sum_{n=1}^{\infty}
\frac{(n+\gamma_{\kappa}-\varepsilon^{-1}\zeta-N_{n\kappa})
(n+2\gamma_{\kappa})n!}{N_{n\kappa}(N_{n\kappa}-\kappa)
(n+\gamma_{\kappa}-\omega)\Gamma(n+2\gamma_{\kappa})}
L_{n-1}^{(2\gamma_{\kappa})}(x)L_{n-1}^{(2\gamma_{\kappa})}(x')
\right.
\nonumber \\
&+& \sum_{n=1}^{\infty}
\frac{(n+\gamma_{\kappa}-\varepsilon^{-1}\zeta+N_{n\kappa})
(n+2\gamma_{\kappa})n!}{N_{n\kappa}(N_{n\kappa}+\kappa)
(n+\gamma_{\kappa}-\omega)\Gamma(n+2\gamma_{\kappa})}
L_{n-1}^{(2\gamma_{\kappa})}(x)L_{n-1}^{(2\gamma_{\kappa})}(x')
\nonumber \\
&+& \sum_{n=1}^{\infty}
\frac{(n+\gamma_{\kappa}-\varepsilon^{-1}\zeta-N_{n\kappa})n!}
{N_{n\kappa}(n+\gamma_{\kappa}-\omega)\Gamma(n+2\gamma_{\kappa})}
L_{n-1}^{(2\gamma_{\kappa})}(x)L_{n}^{(2\gamma_{\kappa})}(x')
\nonumber \\
&-& \sum_{n=1}^{\infty}
\frac{(n+\gamma_{\kappa}-\varepsilon^{-1}\zeta+N_{n\kappa})n!}
{N_{n\kappa}(n+\gamma_{\kappa}-\omega)\Gamma(n+2\gamma_{\kappa})}
L_{n-1}^{(2\gamma_{\kappa})}(x)L_{n}^{(2\gamma_{\kappa})}(x')
\nonumber \\
&+& \sum_{n=1}^{\infty}
\frac{(n+\gamma_{\kappa}-\varepsilon^{-1}\zeta-N_{n\kappa})n!}
{N_{n\kappa}(n+\gamma_{\kappa}-\omega)\Gamma(n+2\gamma_{\kappa})}
L_{n}^{(2\gamma_{\kappa})}(x)L_{n-1}^{(2\gamma_{\kappa})}(x')
\nonumber \\
&-& \sum_{n=1}^{\infty}
\frac{(n+\gamma_{\kappa}-\varepsilon^{-1}\zeta+N_{n\kappa})n!}
{N_{n\kappa}(n+\gamma_{\kappa}-\omega)\Gamma(n+2\gamma_{\kappa})}
L_{n}^{(2\gamma_{\kappa})}(x)L_{n-1}^{(2\gamma_{\kappa})}(x')
\nonumber \\
&+& \sum_{n=0}^{\infty}
\frac{(n+\gamma_{\kappa}-\varepsilon^{-1}\zeta-N_{n\kappa})
(N_{n\kappa}-\kappa)n!}{N_{n\kappa}
(n+\gamma_{\kappa}-\omega)\Gamma(n+2\gamma_{\kappa}+1)}
L_{n}^{(2\gamma_{\kappa})}(x)L_{n}^{(2\gamma_{\kappa})}(x')
\nonumber \\
&+& \left.\sum_{n=0}^{\infty}
\frac{(n+\gamma_{\kappa}-\varepsilon^{-1}\zeta+N_{n\kappa})
(N_{n\kappa}+\kappa)n!}{N_{n\kappa}
(n+\gamma_{\kappa}-\omega)\Gamma(n+2\gamma_{\kappa}+1)}
L_{n}^{(2\gamma_{\kappa})}(x)L_{n}^{(2\gamma_{\kappa})}(x')
\right]
\nonumber \\*[5mm]
&=& -\frac{\alpha\varepsilon}{4}x^{\gamma_{\kappa}}
x'^{\gamma_{\kappa}}{\rm e}^{-(x+x')/2}\left[
\sum_{n=1}^{\infty}\frac{(n+\gamma_{\kappa}-\varepsilon^{-1}\zeta
-N_{n\kappa})(N_{n\kappa}+\kappa)(n-1)!}{N_{n\kappa}
(n+\gamma_{\kappa}-\omega)\Gamma(n+2\gamma_{\kappa})}
L_{n-1}^{(2\gamma_{\kappa})}(x)L_{n-1}^{(2\gamma_{\kappa})}(x')
\right.
\nonumber \\
&+& \sum_{n=0}^{\infty}\frac{(n+\gamma_{\kappa}-\varepsilon^{-1}\zeta
+N_{n\kappa})(N_{n\kappa}-\kappa)(n-1)!}{N_{n\kappa}
(n+\gamma_{\kappa}-\omega)\Gamma(n+2\gamma_{\kappa})}
L_{n-1}^{(2\gamma_{\kappa})}(x)L_{n-1}^{(2\gamma_{\kappa})}(x')
\nonumber \\
&-& 2\sum_{n=1}^{\infty}\frac{n!}{(n+\gamma_{\kappa}-\omega)
\Gamma(n+2\gamma_{\kappa})}
L_{n-1}^{(2\gamma_{\kappa})}(x)L_{n}^{(2\gamma_{\kappa})}(x')
\nonumber \\
&-& 2\sum_{n=1}^{\infty}\frac{n!}{(n+\gamma_{\kappa}-\omega)
\Gamma(n+2\gamma_{\kappa})}
L_{n}^{(2\gamma_{\kappa})}(x)L_{n-1}^{(2\gamma_{\kappa})}(x')
\nonumber \\
&+& \sum_{n=0}^{\infty}
\frac{(n+\gamma_{\kappa}-\varepsilon^{-1}\zeta-N_{n\kappa})
(N_{n\kappa}-\kappa)n!}{N_{n\kappa}(n+\gamma_{\kappa}-\omega)
\Gamma(n+2\gamma_{\kappa}+1)}
L_{n}^{(2\gamma_{\kappa})}(x)L_{n}^{(2\gamma_{\kappa})}(x')
\nonumber \\
&+& \left.\sum_{n=0}^{\infty}
\frac{(n+\gamma_{\kappa}-\varepsilon^{-1}\zeta+N_{n\kappa})
(N_{n\kappa}+\kappa)n!}{N_{n\kappa}(n+\gamma_{\kappa}-\omega)
\Gamma(n+2\gamma_{\kappa}+1)}
L_{n}^{(2\gamma_{\kappa})}(x)L_{n}^{(2\gamma_{\kappa})}(x')
\right]
\end{eqnarray*}
\begin{eqnarray}
&=& -\frac{\alpha\varepsilon}{2}x^{\gamma_{\kappa}}
x'^{\gamma_{\kappa}}{\rm e}^{-(x+x')/2}\left[
\sum_{n=0}^{\infty}\frac{(n+\gamma_{\kappa}-\varepsilon^{-1}\zeta
-\kappa)(n-1)!}{(n+\gamma_{\kappa}-\omega)\Gamma(n+2\gamma_{\kappa})}
L_{n-1}^{(2\gamma_{\kappa})}(x)L_{n-1}^{(2\gamma_{\kappa})}(x')
\right.
\nonumber \\
&-& \sum_{n=0}^{\infty}\frac{n!}{(n+\gamma_{\kappa}-\omega)
\Gamma(n+2\gamma_{\kappa})}L_{n-1}^{(2\gamma_{\kappa})}(x)
L_{n}^{(2\gamma_{\kappa})}(x')
\nonumber \\
&-& \sum_{n=0}^{\infty}\frac{n!}{(n+\gamma_{\kappa}-\omega)
\Gamma(n+2\gamma_{\kappa})}L_{n}^{(2\gamma_{\kappa})}(x)
L_{n-1}^{(2\gamma_{\kappa})}(x')
\nonumber \\
&+& \left.
\sum_{n=0}^{\infty}\frac{(n+\gamma_{\kappa}-\varepsilon^{-1}\zeta
+\kappa)n!}{(n+\gamma_{\kappa}-\omega)\Gamma(n+2\gamma_{\kappa}+1)}
L_{n}^{(2\gamma_{\kappa})}(x)L_{n}^{(2\gamma_{\kappa})}(x')\right]
\nonumber \\*[5mm]
&=& -\frac{\alpha\varepsilon}{2}x^{\gamma_{\kappa}}
x'^{\gamma_{\kappa}}{\rm e}^{-(x+x')/2}\left[
\sum_{n=0}^{\infty}\frac{(n+\gamma_{\kappa}-\varepsilon^{-1}\zeta
-\kappa+1)n!}{(n+\gamma_{\kappa}-\omega+1)
\Gamma(n+2\gamma_{\kappa}+1)}
L_{n}^{(2\gamma_{\kappa})}(x)L_{n}^{(2\gamma_{\kappa})}(x')
\right.
\nonumber \\
&-& \sum_{n=0}^{\infty}\frac{n!}{(n+\gamma_{\kappa}-\omega)
\Gamma(n+2\gamma_{\kappa})}L_{n-1}^{(2\gamma_{\kappa})}(x)
L_{n}^{(2\gamma_{\kappa})}(x')
\nonumber \\
&-& \sum_{n=0}^{\infty}\frac{n!}{(n+\gamma_{\kappa}-\omega)
\Gamma(n+2\gamma_{\kappa})}L_{n}^{(2\gamma_{\kappa})}(x)
L_{n-1}^{(2\gamma_{\kappa})}(x')
\nonumber \\
&+& \left.
\sum_{n=0}^{\infty}\frac{(n+\gamma_{\kappa}-\varepsilon^{-1}\zeta
+\kappa)n!}{(n+\gamma_{\kappa}-\omega)\Gamma(n+2\gamma_{\kappa}+1)}
L_{n}^{(2\gamma_{\kappa})}(x)L_{n}^{(2\gamma_{\kappa})}(x')\right]
\nonumber \\*[5mm]
&=& -\frac{\alpha\varepsilon}{2}x^{\gamma_{\kappa}}
x'^{\gamma_{\kappa}}{\rm e}^{-(x+x')/2}\left\{
\sum_{n=0}^{\infty}\frac{(n+\gamma_{\kappa}-\varepsilon^{-1}\zeta
-\kappa+1)n!}{(n+\gamma_{\kappa}-\omega+1)
\Gamma(n+2\gamma_{\kappa}+1)}
L_{n}^{(2\gamma_{\kappa})}(x)L_{n}^{(2\gamma_{\kappa})}(x')
\right.
\nonumber \\
&-& \sum_{n=0}^{\infty}\frac{n!}{(n+\gamma_{\kappa}-\omega)
\Gamma(n+2\gamma_{\kappa})}\left[L_{n}^{(2\gamma_{\kappa})}(x)
-L_{n}^{(2\gamma_{\kappa}-1)}(x)\right]L_{n}^{(2\gamma_{\kappa})}(x')
\nonumber \\
&-& \sum_{n=0}^{\infty}\frac{n!}{(n+\gamma_{\kappa}-\omega)
\Gamma(n+2\gamma_{\kappa})}L_{n}^{(2\gamma_{\kappa})}(x)
\left[L_{n}^{(2\gamma_{\kappa})}(x')
-L_{n}^{(2\gamma_{\kappa}-1)}(x')\right]
\nonumber \\
&+& \left.
\sum_{n=0}^{\infty}\frac{(n+\gamma_{\kappa}-\varepsilon^{-1}\zeta
+\kappa)n!}{(n+\gamma_{\kappa}-\omega)\Gamma(n+2\gamma_{\kappa}+1)}
L_{n}^{(2\gamma_{\kappa})}(x)L_{n}^{(2\gamma_{\kappa})}(x')\right\}
\nonumber \\*[5mm]
&=& -\frac{\alpha\varepsilon}{2}x^{\gamma_{\kappa}}
x'^{\gamma_{\kappa}}{\rm e}^{-(x+x')/2}\left[
\sum_{n=0}^{\infty}\frac{(n+\gamma_{\kappa}-\varepsilon^{-1}\zeta
-\kappa+1)n!}{(n+\gamma_{\kappa}-\omega+1)
\Gamma(n+2\gamma_{\kappa}+1)}
L_{n}^{(2\gamma_{\kappa})}(x)L_{n}^{(2\gamma_{\kappa})}(x')
\right.
\nonumber \\
&-& \sum_{n=0}^{\infty}\frac{n!}{(n+\gamma_{\kappa}-\omega)
\Gamma(n+2\gamma_{\kappa})}L_{n}^{(2\gamma_{\kappa})}(x)
L_{n}^{(2\gamma_{\kappa})}(x')
\nonumber \\
&+& \sum_{n=0}^{\infty}\frac{n!}{(n+\gamma_{\kappa}-\omega)
\Gamma(n+2\gamma_{\kappa})}L_{n}^{(2\gamma_{\kappa}-1)}(x)
L_{n}^{(2\gamma_{\kappa})}(x')
\nonumber \\
&-& \sum_{n=0}^{\infty}\frac{n!}{(n+\gamma_{\kappa}-\omega)
\Gamma(n+2\gamma_{\kappa})}L_{n}^{(2\gamma_{\kappa})}(x)
L_{n}^{(2\gamma_{\kappa})}(x')
\nonumber \\
&+& \sum_{n=0}^{\infty}\frac{n!}{(n+\gamma_{\kappa}-\omega)
\Gamma(n+2\gamma_{\kappa})}L_{n}^{(2\gamma_{\kappa})}(x)
L_{n}^{(2\gamma_{\kappa}-1)}(x')
\nonumber \\
&+& \left.
\sum_{n=0}^{\infty}\frac{(n+\gamma_{\kappa}-\varepsilon^{-1}\zeta
+\kappa)n!}{(n+\gamma_{\kappa}-\omega)\Gamma(n+2\gamma_{\kappa}+1)}
L_{n}^{(2\gamma_{\kappa})}(x)L_{n}^{(2\gamma_{\kappa})}(x')\right]
\nonumber \\*[5mm]
&=& -\frac{\alpha\varepsilon}{2}x^{\gamma_{\kappa}}
x'^{\gamma_{\kappa}}{\rm e}^{-(x+x')/2}\left\{
\sum_{n=0}^{\infty}\frac{(n+\gamma_{\kappa}-\varepsilon^{-1}\zeta
-\kappa+1)n!}{(n+\gamma_{\kappa}-\omega+1)
\Gamma(n+2\gamma_{\kappa}+1)}
L_{n}^{(2\gamma_{\kappa})}(x)L_{n}^{(2\gamma_{\kappa})}(x')
\right.
\nonumber \\
&+& \sum_{n=0}^{\infty}\frac{n!}{(n+\gamma_{\kappa}-\omega)
\Gamma(n+2\gamma_{\kappa})}\left[L_{n}^{(2\gamma_{\kappa}-1)}(x)
L_{n}^{(2\gamma_{\kappa})}(x')+L_{n}^{(2\gamma_{\kappa})}(x)
L_{n}^{(2\gamma_{\kappa}-1)}(x')\right]
\nonumber \\
&+& \left.
\sum_{n=0}^{\infty}\frac{(-n-3\gamma_{\kappa}-\varepsilon^{-1}\zeta
+\kappa)n!}{(n+\gamma_{\kappa}-\omega)\Gamma(n+2\gamma_{\kappa}+1)}
L_{n}^{(2\gamma_{\kappa})}(x)L_{n}^{(2\gamma_{\kappa})}(x')\right\}
\nonumber \\*[5mm]
&=& -\frac{\alpha\varepsilon}{2}x^{\gamma_{\kappa}}
x'^{\gamma_{\kappa}}{\rm e}^{-(x+x')/2}\left\{
\sum_{n=0}^{\infty}\frac{(n+\gamma_{\kappa}-\omega+1-\varepsilon^{-1}
\zeta+\omega-\kappa)n!}{(n+\gamma_{\kappa}-\omega+1)
\Gamma(n+2\gamma_{\kappa}+1)}
L_{n}^{(2\gamma_{\kappa})}(x)L_{n}^{(2\gamma_{\kappa})}(x')
\right.
\nonumber \\
&+& \sum_{n=0}^{\infty}\frac{n!}{(n+\gamma_{\kappa}-\omega)
\Gamma(n+2\gamma_{\kappa})}\left[L_{n}^{(2\gamma_{\kappa}-1)}(x)
L_{n}^{(2\gamma_{\kappa})}(x')+L_{n}^{(2\gamma_{\kappa})}(x)
L_{n}^{(2\gamma_{\kappa}-1)}(x')\right]
\nonumber \\
&+& \left.\sum_{n=0}^{\infty}\frac{(-n-\gamma_{\kappa}+\omega
-\varepsilon^{-1}\zeta-\omega-2\gamma_{\kappa}+\kappa)n!}
{(n+\gamma_{\kappa}-\omega)\Gamma(n+2\gamma_{\kappa}+1)}
L_{n}^{(2\gamma_{\kappa})}(x)L_{n}^{(2\gamma_{\kappa})}(x')\right\}
\nonumber \\*[5mm]
&=& -\frac{\alpha\varepsilon}{2}x^{\gamma_{\kappa}}
x'^{\gamma_{\kappa}}{\rm e}^{-(x+x')/2}\left\{
\sum_{n=0}^{\infty}\frac{n!}{\Gamma(n+2\gamma_{\kappa}+1)}
L_{n}^{(2\gamma_{\kappa})}(x)L_{n}^{(2\gamma_{\kappa})}(x')\right.
\nonumber \\
&-&
\sum_{n=0}^{\infty}\frac{(\varepsilon^{-1}\zeta-\omega+\kappa)n!}
{(n+\gamma_{\kappa}-\omega+1)\Gamma(n+2\gamma_{\kappa}+1)}
L_{n}^{(2\gamma_{\kappa})}(x)L_{n}^{(2\gamma_{\kappa})}(x')
\nonumber \\
&+& \sum_{n=0}^{\infty}\frac{n!}{(n+\gamma_{\kappa}-\omega)
\Gamma(n+2\gamma_{\kappa})}\left[L_{n}^{(2\gamma_{\kappa}-1)}(x)
L_{n}^{(2\gamma_{\kappa})}(x')+L_{n}^{(2\gamma_{\kappa})}(x)
L_{n}^{(2\gamma_{\kappa}-1)}(x')\right]
\nonumber \\
&-& \sum_{n=0}^{\infty}\frac{n!}{\Gamma(n+2\gamma_{\kappa}+1)}
L_{n}^{(2\gamma_{\kappa})}(x)L_{n}^{(2\gamma_{\kappa})}(x')
\nonumber \\
&-& \left.\sum_{n=0}^{\infty}\frac{(\varepsilon^{-1}\zeta+\omega
+2\gamma_{\kappa}-\kappa)n!}{(n+\gamma_{\kappa}-\omega)
\Gamma(n+2\gamma_{\kappa}+1)}
L_{n}^{(2\gamma_{\kappa})}(x)L_{n}^{(2\gamma_{\kappa})}(x')\right\}
\nonumber \\*[5mm]
&=& \frac{\alpha\varepsilon}{2}x^{\gamma_{\kappa}}
x'^{\gamma_{\kappa}}{\rm e}^{-(x+x')/2}\left\{
(\varepsilon^{-1}\zeta-\omega+\kappa)\sum_{n=0}^{\infty}\frac{n!}
{(n+\gamma_{\kappa}-\omega+1)\Gamma(n+2\gamma_{\kappa}+1)}
L_{n}^{(2\gamma_{\kappa})}(x)L_{n}^{(2\gamma_{\kappa})}(x')\right.
\nonumber \\
&-& \sum_{n=0}^{\infty}\frac{n!}{(n+\gamma_{\kappa}-\omega)
\Gamma(n+2\gamma_{\kappa})}\left[L_{n}^{(2\gamma_{\kappa}-1)}(x)
L_{n}^{(2\gamma_{\kappa})}(x')+L_{n}^{(2\gamma_{\kappa})}(x)
L_{n}^{(2\gamma_{\kappa}-1)}(x')\right]
\nonumber \\
&+& \left.(\varepsilon^{-1}\zeta+\omega+2\gamma_{\kappa}-\kappa)
\sum_{n=0}^{\infty}\frac{n!}{(n+\gamma_{\kappa}-\omega)
\Gamma(n+2\gamma_{\kappa}+1)}
L_{n}^{(2\gamma_{\kappa})}(x)L_{n}^{(2\gamma_{\kappa})}(x')\right\}
\nonumber \\*[5mm]
&=& \frac{\alpha\varepsilon}{2}x^{\gamma_{\kappa}}
x'^{\gamma_{\kappa}}{\rm e}^{-(x+x')/2}
\left\{\left(\frac{\zeta}{\sqrt{1-\epsilon^{2}}}+\kappa\right)
\sum_{n=0}^{\infty}\frac{n!}{(n+\gamma_{\kappa}-\omega+1)
\Gamma(n+2\gamma_{\kappa}+1)}L_{n}^{(2\gamma_{\kappa})}(x)
L_{n}^{(2\gamma_{\kappa})}(x')\right.
\nonumber \\
&-& \sum_{n=0}^{\infty}\frac{n!}{(n+\gamma_{\kappa}-\omega)
\Gamma(n+2\gamma_{\kappa})}\left[L_{n}^{(2\gamma_{\kappa}-1)}(x)
L_{n}^{(2\gamma_{\kappa})}(x')+L_{n}^{(2\gamma_{\kappa})}(x)
L_{n}^{(2\gamma_{\kappa}-1)}(x')\right]
\nonumber \\
&+& \left.\left(\frac{\zeta(1+2\epsilon)}{\sqrt{1-\epsilon^{2}}}
+2\gamma_{\kappa}-\kappa\right)\sum_{n=0}^{\infty}
\frac{n!}{(n+\gamma_{\kappa}-\omega)\Gamma(n+2\gamma_{\kappa}+1)}
L_{n}^{(2\gamma_{\kappa})}(x)L_{n}^{(2\gamma_{\kappa})}(x')\right\}.
\label{28}
\end{eqnarray}

\begin{thebibliography}{99}
\bibitem{Szmy97} R.~Szmytkowski
   {\em J.\ Phys.\ B:\ At.\ Mol.\ Opt.\ Phys.\/} {\bf 30} (1997)
   825--61 [Erratum: {\bf 30} (1997) 2747]
\bibitem{Szmy98} R.~Szmytkowski
   {\em J.~Phys.~A:~Math.~Gen.\/} {\bf 31} (1998) 4963--90 
   [Erratum: {\bf 31} (1998) 7415--6]
\end{thebibliography}
\end{document}